\pdfoutput=1
\documentclass[10pt,a4paper]{article}

\usepackage[T1]{fontenc}
\usepackage{amsmath,amssymb}
\usepackage{stix}                          
\usepackage[scaled=.92,varqu]{inconsolata} 
\usepackage[a4paper,hmargin=22.7mm,vmargin=25mm]{geometry}
\usepackage{microtype}
\usepackage{graphicx}
\usepackage{booktabs,makecell,multirow,array}
\usepackage[svgnames,dvipsnames]{xcolor}
\usepackage[numbers,sort&compress]{natbib}
\usepackage{pifont}      
\usepackage{listings}    
\usepackage[font=small,labelfont=bf,labelsep=period]{caption}
\usepackage{tikz}
\usepackage[colorlinks,linkcolor=blue!60!black,citecolor=blue!60!black,urlcolor=blue!60!black]{hyperref}
\usepackage{orcidlink}
\hypersetup{pdftitle={A code-level survey of CPU scheduling mechanisms in open-source embedded hypervisors},pdfauthor={Lisa Mignerey, Anthony Ferrari, Damien Sauveron, Karim Tamine}}

\newcommand{\tblwidth}{\textwidth}
\newcolumntype{L}{@{\extracolsep{\fill}}l}
\newcolumntype{R}{@{\extracolsep{\fill}}r}
\newcolumntype{C}{@{\extracolsep{\fill}}c}

\newcommand*{\code}[1]{\texttt{#1}}

\usepackage{tikz}
\usetikzlibrary{arrows.meta}
\definecolor{tikzcat}{HTML}{E6E8FA}   
\definecolor{tikzfam}{HTML}{F2F3FC}   
\definecolor{tikzleaf}{HTML}{DFFADF}  
\definecolor{spgreen}{HTML}{2DB03F}   
\definecolor{spgreenln}{HTML}{4CAF50}
\definecolor{spgreenbg}{HTML}{DCF2D3}
\definecolor{spblue}{HTML}{5FC9F3}
\definecolor{spblueln}{HTML}{1E9AD6}
\definecolor{spbluebg}{HTML}{CCEDFB}
\definecolor{sporange}{HTML}{E97132}
\tikzset{
  box/.style  = {draw=black!75, rounded corners=2pt, align=center,
                 inner sep=3pt, font=\footnotesize},
  cat/.style  = {box, fill=tikzcat},
  fam/.style  = {box, fill=tikzfam},
  leaf/.style = {box, fill=tikzleaf},
  arr/.style  = {-{Stealth[length=4.5pt,width=3.5pt]}, semithick},
  ans/.style  = {fill=white, inner sep=1pt, font=\scriptsize\bfseries},
}

\newcommand{\rh}[1]{\rotatebox{90}{\makecell[l]{#1}}}

\let\origunderscore\_
\renewcommand{\_}{\origunderscore\allowbreak}

\title{A code-level survey of CPU scheduling mechanisms in open-source embedded hypervisors}
\author{%
  Lisa Mignerey$^{1,2,}$\thanks{Corresponding author: \href{mailto:lisa.mignerey@thalesgroup.com}{lisa.mignerey@thalesgroup.com}}\,\orcidlink{0009-0007-2203-3726},\quad
  Anthony Ferrari$^{1}$,\quad
  Damien Sauveron$^{2}$\,\orcidlink{0000-0001-7948-6143},\quad
  Karim Tamine$^{2}$\,\orcidlink{0000-0001-7613-699X}\\[2ex]
  \small $^{1}$Thales Cybersecurity \& Digital Identity, La Ciotat, France\\
  \small $^{2}$XLIM, UMR CNRS 7252, University of Limoges, Limoges, France\\[1ex]
  \small\texttt{\{lisa.mignerey, anthony.ferrari\}@thalesgroup.com},
  \texttt{\{damien.sauveron, karim.tamine\}@unilim.fr}}
\date{}

\begin{document}

\maketitle

\begin{abstract}
Virtualization is increasingly used in embedded systems to consolidate mixed-criticality applications onto a single platform, relying on the hypervisor’s CPU scheduler to share physical CPU (pCPU) time among virtual CPUs (vCPUs). This paper surveys ten CPU scheduling algorithms implemented in four open-source hypervisors for resource-constrained systems (Bao, Xen, ACRN, and Xvisor): static partitioning, ARINC653, Credit, Credit2, RTDS, Priority-based, BVT, IORR, PRR, and PRM. Analyzed at the source-code level without runtime measurements, each algorithm is characterized by its decision rule, parameters, selection criterion, triggers, and admission control. They are evaluated against nine criteria (temporal isolation, real-time guarantees, I/O responsiveness, resource utilization, fairness, mixed-criticality support, efficiency/scalability, configurability, and multicore support) and illustrated on a common single-pCPU workload. Source-code inspection reveals key deviations from documented designs: ACRN’s default BVT scheduler lacks the warp mechanism, its priority-based scheduler exposes only two priority levels, and apart from ARINC653’s static frame check, only Xen’s development-branch RTDS performs utilization-based admission control. Furthermore, no policy natively supports criticality attributes or mode changes, requiring integrators to encode criticality via partitions, budgets, caps, or priorities. Finally, a qualitative matrix and a decision tree guide system integrators.
\end{abstract}


\section{Introduction}\label{sec:introduction}

Modern embedded systems are undergoing a major transformation, moving from simple, single-purpose devices to complex multi-functional platforms. This evolution is driven by the need to run several applications with different levels of importance, known as mixed-criticality workloads, on a single hardware platform. To manage this transition, virtualization has emerged as a widely adopted technology, using a hypervisor to create several isolated virtual machines (VMs). In these resource-constrained environments, the CPU scheduler is considered one of the most critical software components because it manages how the CPU time is divided among all active vCPUs.

Consequently, designing a reliable scheduler for mixed-criticality support is a complex multi-objective problem: a difficult balance must be achieved between guaranteeing strict temporal isolation and real-time predictability for critical workloads, while simultaneously providing fair CPU sharing, good I/O responsiveness, and work-conserving resource utilization for non-critical applications; all while maintaining low scheduling overhead and high scalability and configurability for the overall system. If the scheduler is not reliable, the vCPUs of a non-essential VM, such as a user interface for instance, could accidentally consume too many resources and delay the vCPUs of a critical VM, like a control loop in a car. Therefore, understanding the logic behind different scheduling strategies is essential for building safe, secure, and efficient embedded systems that can handle unpredictable real-world demands. While many open-source hypervisors are now available for embedded development, they implement widely different scheduling methods, ranging from extremely simple static designs to complex dynamic algorithms.

A fundamental assumption throughout this paper is that the embedded system is designed and integrated by a single entity (the system architect). Unlike cloud computing environments where VMs belong to independent, agnostic users, this holistic design implies that the architect possesses complete knowledge of the behavior, criticality, and resource requirements of every VM. Consequently, it is assumed that this central authority is fully capable of determining and assigning the appropriate scheduling parameters required by each algorithm.

To clarify the scope of the analysis and address potential concerns regarding hardware contention, explicit boundaries for this evaluation are established. This study conducts a conceptual and algorithmic analysis of the scheduling mechanisms, focusing entirely on the mathematical logic behind the CPU allocation decision-making process. Consequently, hardware-induced limitations, such as micro-architectural contention (including cache, bus, and DRAM interference), are excluded from the scope. Nevertheless, this exclusion does not prevent such a baseline survey from providing meaningful insights to the reader: understanding the core mathematical logic and behavioral properties of these algorithms remains an essential prerequisite and serves as a practical preliminary guide for system architects before conducting hardware-specific tuning.

\subsection{Research questions}\label{sec:research-questions}

This survey addresses three research questions:

\textbf{RQ1.} Which families of scheduling rules do the selected open-source embedded hypervisors implement, and how can these rules be classified?

\textbf{RQ2.} Which temporal-isolation, wake-up handling, and admission-control mechanisms are present in the source code, and how do they differ from their description in the literature and in the hypervisors' documentation?

\textbf{RQ3.} Which structural properties of these mechanisms condition their use for mixed-criticality workloads, and which configuration assumptions do they impose on the system integrator?

\subsection{Contributions}\label{sec:contributions}

In answering these research questions, this paper makes the following three contributions:

\begin{enumerate}

\item \textbf{(RQ1)} A classification of the ten CPU allocation policies by decision rule (Figure~\ref{fig:1}), which places policies of different hypervisors into directly comparable algorithmic families.

\item \textbf{(RQ2)} A uniform, code-level description of each policy at a frozen repository state (parameters, selection criterion, triggers, admission control, and a pseudo-code faithful to the implementation) illustrated by schedules derived by hand on a common reference workload, with the deviations between kernel implementation and published descriptions made explicit (Table~\ref{tab:rq2_deviations}).

\item \textbf{(RQ3)} A qualitative assessment of every policy against nine explicitly defined criteria (Table~\ref{tab:4}), in which each favorable property is stated together with the assumption it places on the system integrator, and a decision tree (Figure~\ref{fig:12}) that guides the selection of a policy from system requirements.

\end{enumerate}

\subsection{Structure of the paper}\label{sec:structure}

The remainder of this paper is organized as follows. Section~\ref{sec:algorithms} presents and analyzes the scheduling algorithms. It first explains how the four hypervisors were selected (Section~\ref{sec:hypervisor-selection}), defines the nine evaluation criteria (Section~\ref{sec:criteria-definitions}), and introduces the common reference workload (Section~\ref{sec:reference-workload}). It then analyzes the ten algorithms one by one (Sections~\ref{sec:static-partitioning} to~\ref{sec:prm}): for each of them, it describes the parameters, selection criterion, triggers, and admission control, gives the pseudo-code, discusses advantages and limitations, and presents a schedule timeline on the common reference workload. Section~\ref{sec:synthesis} synthesizes and compares the reference schedules derived across all algorithms. Section~\ref{sec:qualitative-comparison} generalizes these findings by comparing all algorithms against the qualitative criteria and presents a decision tree that helps the system architect choose a scheduler according to application requirements. Section~\ref{sec:related-work} reviews related work on embedded virtualization surveys, mixed-criticality scheduling models, and evaluation and customization of hypervisor schedulers. Finally, Section~\ref{sec:conclusion} concludes the paper.

\section{Scheduling algorithms analysis}\label{sec:algorithms}

This section explores ten scheduling algorithms implemented in at least one open-source hypervisor suitable for resource-constrained systems. For each algorithm, the scheduling policy is described, then the main advantages and disadvantages are analyzed. Identifiers in typewriter font refer to the source code at the analyzed commit (Table~\ref{tab:2}). In the pseudo-code, bold names denote generic handlers shared by all algorithms (pick\_next, on\_wake, on\_sleep, ...), and the trailing comment gives the function that implements each of them.

\subsection{Hypervisor selection}\label{sec:hypervisor-selection}

To provide a representative analysis of the current virtualization landscape for resource-constrained systems, this study focuses on four hypervisors selected through a purposive, criteria-based process: Bao, Xen, ACRN~\cite{li2019acrn}, and Xvisor. To ensure a rigorous selection, a filtering process was defined based on inclusion criteria. Specifically, the selected hypervisors must feature a public scheduler source code under an open-source license. Their official documentation must explicitly target embedded or real-time systems, excluding generic server-oriented solutions. Furthermore, they must support at least one typical embedded hardware architecture (e.g., ARMv8-A, RISC-V, or embedded x86). Active maintenance is also required, defined as having at least one commit modifying the hypervisor source code on the main branch within the 24 months preceding the cut-off date (documentation-only commits are not counted). Finally, the public repository must be the project's development tree and support at least one physical embedded platform, so that the analyzed code is the code an integrator can deploy. When several candidates implement the same scheduling decision rule, one representative per rule is retained: Bao for static partitioning, Xen's ARINC653 scheduler for offline cyclic tables, and Xvisor's PRR for time-sliced round-robin.

The initial candidate list was compiled using exact queries on major academic databases (Google Scholar, IEEE Xplore, ACM Digital Library), combined with a methodical exploration of GitHub and GitLab repositories. References from major existing surveys (notably Gu and Zhao~\cite{gu2012survey}) are also included. Out of 24 initial candidates identified, the filtering process isolated the 4 hypervisors evaluated in this paper.

During the selection process, 20 hypervisors were excluded for failing to meet one or more of the established criteria. These exclusions are grouped into the following categories:

\textbf{Inactive projects and redundant decision rules:} Minos has had no source-code commit since September 2022 (only a roadmap update in June 2026), and Jailhouse~\cite{ramsauer2017jailhouse} has had no commit since January 2023. Jailhouse also relies on static partitioning, already represented by Bao. PRTOS and the Muen separation kernel~\cite{buerki2013muen} rely on offline cyclic scheduling tables, already represented by Xen's ARINC653 scheduler. Rust-Shyper pins vCPUs to pCPUs and multiplexes vCPUs sharing a pCPU with a plain round-robin scheduler driven by a 10~ms tick (\texttt{src/kernel/sched/sched\_rr.rs}, without priority or weight, where waking vCPUs wait for the next tick), while its real-time scheduler (\texttt{sched\_rt.rs}) is an unimplemented stub (\texttt{todo!()}); plain round-robin is the single-priority case of Xvisor's PRR.

\textbf{Proprietary or closed-source:} PikeOS, QNX Hypervisor, Wind River Helix Virtualization Platform, Green Hills INTEGRITY Multivisor, Microsoft Hyper-V, OKL4 (proprietary since 2012), Quest-V (hypervisor distributed only as an SDK), and XtratuM NG were excluded. Their closed-source nature prevents public algorithmic analysis of their scheduling logic.

\textbf{Generic and server-oriented:} KVM/QEMU and bhyve target general-purpose or server hosting and delegate vCPU scheduling to the host kernel thread scheduler.

\textbf{Microkernels and delegated scheduling:} seL4, Fiasco.OC, NOVA and Hafnium were excluded. These solutions operate as microkernels where virtualization is a user-space layer, or they deliberately delegate scheduling decisions to a primary guest OS rather than providing a native vCPU scheduler at the hypervisor level.

\textbf{Simulator-only open-source tree:} Gunyah implements a native fixed-priority round-robin vCPU scheduler (\texttt{hyp/core/scheduler\_fprr}: 64 priority levels, per-vCPU time slice from 0.1~ms to 100~ms, default 5~ms), but its public repository receives periodic code drops from Qualcomm's internal development and supports only the QEMU and Arm FVP simulators. It is left for a future extension of this survey. Notably, both excluded native schedulers (Gunyah's FPRR and Rust-Shyper's plain round-robin) belong to the fixed-priority round-robin family already analyzed, so no decision rule is missing from the taxonomy of RQ1.

To provide a clear overview of the resulting landscape after this filtering process, Table~\ref{tab:1} maps the ten evaluated scheduling algorithms to the four retained open-source hypervisors that implement them.

The survey targets dedicated embedded hypervisors whose vCPU scheduler is implemented in the hypervisor itself. Two families of systems are therefore outside its scope although they host real-time virtual machines. In KVM, a vCPU is a thread of the host Linux kernel, scheduled by the Linux scheduling classes: the mechanisms are those of the Linux scheduler, which the real-time literature analyzes separately~\cite{abeni2019hierarchical, abeni2024virtualized}. In microkernel-based systems (seL4, NOVA, Fiasco.OC), vCPUs are kernel threads scheduled by the ordinary thread scheduler, and the virtual machine monitor runs in user space. In Hafnium, the primary VM schedules the vCPUs of the secondary VMs itself. Two of these systems are nevertheless useful reference points, because they offer what most surveyed policies lack: Linux \texttt{SCHED\_DEADLINE} implements Constant Bandwidth Servers with a utilization-based admission test, and the MCS configuration of seL4 implements scheduling contexts (budget, period) with sporadic-server budget enforcement~\cite{lyons2018scheduling}. Neither performs a sufficient schedulability test on multiprocessors: the admission test of \texttt{SCHED\_DEADLINE} checks total utilization, as RTDS does.

\begin{table}[!htbp]
\caption{Summary of the evaluated scheduling algorithms}\label{tab:1}
\begin{tabular*}{\tblwidth}{@{}LCCCC@{}}
\toprule
 & \multicolumn{4}{c}{Hypervisor} \\
\cmidrule(l){2-5}
Algorithm & Bao & Xen & ACRN & Xvisor \\
\midrule
Static partitioning & \ding{51} & \ding{51} & \ding{51} &  \\
ARINC653 &  & \ding{51} &  &  \\
Credit &  & \ding{51} &  &  \\
Credit2 &  & \ding{51} &  &  \\
RTDS &  & \ding{51} &  &  \\
Priority-based &  &  & \ding{51} &  \\
BVT &  &  & \ding{51} &  \\
IORR &  &  & \ding{51} &  \\
PRR &  &  &  & \ding{51} \\
PRM &  &  &  & \ding{51} \\
\bottomrule
\end{tabular*}
\end{table}

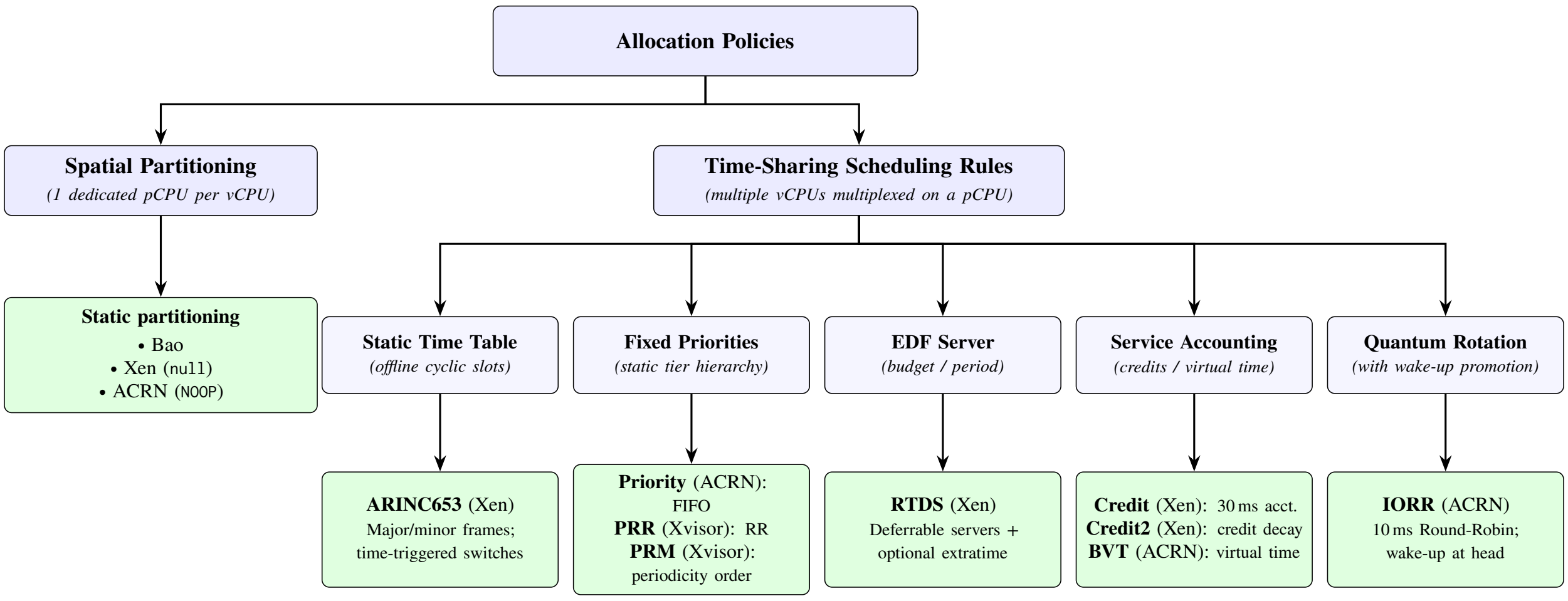
\begin{figure}[!htbp]
\centering
\resizebox{\textwidth}{!}{%
\begin{tikzpicture}[
    >=Stealth,
    cat/.style={
        draw,
        rounded corners=3pt,
        fill=blue!8,
        align=center,
        text width=4.2cm,
        minimum height=1.0cm,
        inner sep=4pt,
        font=\small\bfseries
    },
    subcat/.style={
        draw,
        rounded corners=3pt,
        fill=blue!4,
        align=center,
        text width=3.1cm,
        minimum height=1.1cm,
        inner sep=4pt,
        font=\footnotesize\bfseries
    },
    pol/.style={
        draw,
        rounded corners=3pt,
        fill=green!12,
        align=center,
        text width=3.1cm,
        minimum height=1.65cm,
        inner sep=4pt,
        font=\footnotesize
    },
    edge/.style={->, thick}
]

\node[cat, text width=5.8cm] (root) at (0, 0)
    {Allocation Policies};

\node[cat, text width=4.2cm] (spatial) at (-7.8, -2.0)
    {Spatial Partitioning\\{\scriptsize\normalfont\itshape (1 dedicated pCPU per vCPU)}};

\node[cat, text width=5.6cm] (temporal) at (2.2, -2.0)
    {Time-Sharing Scheduling Rules\\{\scriptsize\normalfont\itshape (multiple vCPUs multiplexed on a pCPU)}};

\node[pol, text width=4.2cm] (p_static) at (-7.8, -4.5)
    {\textbf{Static partitioning}\\[2pt]
    $\bullet$ Bao\\
    $\bullet$ Xen (\texttt{null})\\
    $\bullet$ ACRN (\texttt{NOOP})};

\node[subcat] (f_table) at (-3.8, -4.5)
    {Static Time Table\\{\scriptsize\normalfont\itshape (offline cyclic slots)}};

\node[subcat] (f_prio) at (-0.2, -4.5)
    {Fixed Priorities\\{\scriptsize\normalfont\itshape (static tier hierarchy)}};

\node[subcat] (f_edf) at (3.4, -4.5)
    {EDF Server\\{\scriptsize\normalfont\itshape (budget / period)}};

\node[subcat] (f_acct) at (7.0, -4.5)
    {Service Accounting\\{\scriptsize\normalfont\itshape (credits / virtual time)}};

\node[subcat] (f_rr) at (10.6, -4.5)
    {Quantum Rotation\\{\scriptsize\normalfont\itshape (with wake-up promotion)}};

\node[pol] (p_arinc) at (-3.8, -7.0)
    {\textbf{ARINC653} (Xen)\\[2pt]
    {\scriptsize Major/minor frames;\\time-triggered switches}};

\node[pol] (p_prio) at (-0.2, -7.0)
    {\textbf{Priority} (ACRN):\\{\scriptsize FIFO}\\
    \textbf{PRR} (Xvisor): {\scriptsize RR}\\
    \textbf{PRM} (Xvisor): {\scriptsize periodicity order}};

\node[pol] (p_rtds) at (3.4, -7.0)
    {\textbf{RTDS} (Xen)\\[2pt]
    {\scriptsize Deferrable servers +\\optional extratime}};

\node[pol] (p_acct) at (7.0, -7.0)
    {\textbf{Credit} (Xen): {\scriptsize 30\,ms acct.}\\
    \textbf{Credit2} (Xen): {\scriptsize credit decay}\\
    \textbf{BVT} (ACRN): {\scriptsize virtual time}};

\node[pol] (p_iorr) at (10.6, -7.0)
    {\textbf{IORR} (ACRN)\\[2pt]
    {\scriptsize 10\,ms Round-Robin;\\wake-up at head}};

\draw[edge] (root.south) -- ++(0,-0.4) -| (spatial.north);
\draw[edge] (root.south) -- ++(0,-0.4) -| (temporal.north);

\draw[edge] (spatial.south) -- (p_static.north);

\draw[edge] (temporal.south) -- ++(0,-0.4) -| (f_table.north);
\draw[edge] (temporal.south) -- ++(0,-0.4) -| (f_prio.north);
\draw[edge] (temporal.south) -- ++(0,-0.4) -| (f_edf.north);
\draw[edge] (temporal.south) -- ++(0,-0.4) -| (f_acct.north);
\draw[edge] (temporal.south) -- ++(0,-0.4) -| (f_rr.north);

\draw[edge] (f_table.south) -- (p_arinc.north);
\draw[edge] (f_prio.south)  -- (p_prio.north);
\draw[edge] (f_edf.south)   -- (p_rtds.north);
\draw[edge] (f_acct.south)  -- (p_acct.north);
\draw[edge] (f_rr.south)    -- (p_iorr.north);

\end{tikzpicture}%
}
\caption{Taxonomy of the ten CPU allocation policies across the four selected open-source embedded hypervisors, classified by their primary allocation paradigm and dispatch decision rule (addressing \textbf{RQ1}).}
\label{fig:1}
\end{figure}

While Table~\ref{tab:1} maps the ten policies to their respective hypervisors, answering \textbf{RQ1} requires classifying them by their underlying scheduling decision rule so that policies from different codebases can be compared directly. As shown in the taxonomy of Figure~\ref{fig:1}, the policies first divide into spatial partitioning (which dedicates a physical core to each vCPU and eliminates runtime scheduling (Static partitioning in Bao, Xen \code{null}, and ACRN \code{NOOP})) and time-sharing scheduling rules, which multiplex multiple vCPUs onto shared pCPUs. The nine time-sharing policies further fall into five algorithmic families according to how they allocate CPU time and order the run-queue: (1) static time-table lookup over a cyclic major frame (Xen ARINC653); (2) fixed-priority hierarchies, which differ by their intra-level tie-breaking policy (FIFO without time-slicing in ACRN Priority-based, Round-Robin in Xvisor PRR, and Rate Monotonic periodicity ordering in Xvisor PRM); (3) Earliest Deadline First (EDF) over budget–period deferrable servers (Xen RTDS); (4) proportional service accounting, tracked either via decaying credits (Xen Credit and Credit2) or virtual time (ACRN BVT); and (5) egalitarian quantum rotation with wake-up promotion to the head of the queue (ACRN IORR).

The analysis is based on the state of each repository at the cut-off date (20 October 2026): Xen \code{7e020a4} (15/10/2026), ACRN \code{37b1c13} (01/12/2025), Xvisor \code{ddbfcd6} (01/10/2026) and Bao release v2.0.0 (latest release before cut-off) (\code{0af4a1a}, 04/03/2026). Table~\ref{tab:2} additionally lists, for each policy, the last commit that modified its algorithm file; the generic scheduling code on which each policy relies is analyzed at the repository state given above. All dates are written as DD/MM/YYYY.

\begin{table}[!htbp]
\caption{Analyzed source code state (last commit modifying each scheduler implementation prior to the 20/10/2026 cut-off date).}\label{tab:2}
\begin{tabular*}{\tblwidth}{@{}LLL@{}}
\toprule
Hypervisor & Algorithm & Last commit modifying the scheduler file \\
\midrule
Xen & ARINC653 & \code{58db743} (12/08/2026) \\
Xen & Credit & \code{58db743} (12/08/2026) \\
Xen & Credit2 & \code{e4da182} (01/10/2026) \\
Xen & RTDS & \code{04584da} (21/09/2026) \\
ACRN & Priority & \code{946a927} (08/05/2024) \\
ACRN & BVT & \code{3d6ca84} (27/06/2024) \\
ACRN & IORR & \code{3d6ca84} (27/06/2024) \\
Xvisor & PRR & \code{3379e5c} (26/01/2020) \\
Xvisor & PRM & \code{123df7b} (22/09/2015) \\
Bao & Static & \code{0af4a1a} (04/03/2026) \\
\bottomrule
\end{tabular*}
\end{table}

The analyzed repository states differ from the latest official release of each hypervisor at the cut-off date as follows. Xen release 4.22.0 (29/07/2026) does not include the RTDS admission control added to the staging development branch in September 2026 (commits \texttt{5772eaa} and \texttt{04584da}), nor the budget-refill fix applied when \texttt{extratime} is toggled on a depleted vCPU (commit \texttt{37b49cb}). All other scheduler commits since release 4.22.0 are code refactorings or concern multicore tickling in Credit2. The analyzed scheduler files of ACRN are identical to release v3.3 (02/08/2024), and those of Xvisor are identical to release v0.3.2 (01/01/2023). Bao is analyzed directly at its release v2.0.0. Notably, the algorithm files of PRR and PRM in Xvisor have not changed since 2020 and 2015, respectively: while the Xvisor repository is actively maintained, these two scheduling policies are dormant, which is in itself a relevant consideration for system integrators.

\subsection{Criteria definitions}\label{sec:criteria-definitions}

To effectively evaluate and compare the scheduling algorithms, a specific set of qualitative criteria is established. To clarify the evaluation framework, the optimization objective for each criterion is explicitly stated (i.e., whether the property should be maximized, minimized, or if the optimal configuration is context-dependent). The multi-objective problem of CPU scheduling is analyzed based on the following dimensions:

\textit{Temporal isolation (Objective: To be maximized)}: This is defined as the scheduler’s ability to guarantee that each vCPU receives its allocated pCPU capacity, specified by a temporal service model such as a budget allocated within a period~\cite{abeni1998integrating,rajkumar1998resource}, regardless of the execution time, overload, or misbehavior of other entities~\cite{mercer1994reserves,rajkumar1998resource}. By enforcing pCPU usage bounds~\cite{abeni1998integrating,rajkumar1998resource}, an execution overrun in one vCPU is prevented from compromising the guaranteed pCPU bandwidth or scheduling deadlines of others~\cite{mercer1994reserves}. Note that temporal isolation at the scheduler level strictly applies to pCPU time allocation and scheduling guarantees, explicitly excluding microarchitectural hardware contention (e.g., shared cache, DRAM, or memory bus contention).

\textit{Real-time guarantees} (Objective: To be maximized): This refers to the scheduler’s capacity to ensure that critical vCPUs meet their specified temporal constraints and deadlines (e.g., completing execution budget C within period T, or before deadline D)~\cite{liu1973scheduling,stankovic1988misconceptions} through bounded response times~\cite{axer2014building}.

\textit{I/O responsiveness} (Objective: Responsiveness to be maximized / Latency to be minimized): This criterion evaluates the scheduler's ability to minimize dispatch latency and quickly assign pCPU time when interactive or event-driven I/O vCPUs wake up~\cite{duda1999bvt,ongaro2008scheduling}. General I/O workloads can also include non-interactive data streams that mainly require sustained bandwidth~\cite{ongaro2008scheduling}. However, this evaluation focuses on how well the scheduler handles aperiodic events. This criterion is evaluated strictly at the scheduler level: it measures the dispatch latency of a vCPU that has already been notified of an event. The end-to-end I/O latency experienced by a guest also depends on the interrupt virtualization path (e.g., virtual interrupt controller or posted interrupts versus direct injection in static partitioning) and the hypervisor's I/O architecture (e.g., device drivers running in \texttt{dom0} or driver domains in Xen, or in the Service VM in ACRN), which are outside the scope of this scheduler-focused survey. A low dispatch latency is therefore a necessary, but not sufficient, condition for low end-to-end I/O latency.

\textit{Resource utilization (work-conserving vs. non-work-conserving) (Objective: Context-dependent)}: Work-conserving support refers to the scheduler’s ability to use the pCPU efficiently by ensuring that a pCPU does not remain idle when a runnable vCPU is waiting to run~\cite{kleinrock1975queueing}. Under this policy, unused pCPU time or time left by an early task completion can be given to other active workloads. In contrast, a non-work-conserving scheduler may leave a pCPU idle even when runnable vCPUs are waiting.

\textit{Fair CPU sharing} (Objective: To be maximized): This refers to the scheduler’s capacity to allocate pCPU time among active vCPUs in proportion to their assigned weights or shares~\cite{waldspurger1994lottery,stoica1996proportional}. Each runnable vCPU $i$ receives an allocation fraction $f_{i} = w_{i} / \sum_{j} w_{j}$ of the total pCPU capacity over active execution intervals~\cite{parekh1993generalized}, ensuring equitable resource distribution without long-term disparity or starvation.

\textit{Mixed-criticality support} (Objective: To be maximized): This refers to the scheduler’s ability to co-locate workloads assigned different safety assurance levels on shared pCPU while guaranteeing freedom from temporal interference~\cite{burns2017survey,ernst2016mixed}. To favor safety-critical entities during resource contention or overruns, the scheduler maps these criticality levels onto preferential execution parameters, ensuring that lower-assurance workloads cannot compromise the timing guarantees of critical vCPUs~\cite{ernst2016mixed,baruah2012preemptive}. Because none of the surveyed hypervisor policies carries an explicit criticality attribute or performs a Vestal-style mode change, this criterion specifically evaluates criticality-agnostic co-location, i.e., whether the scheduler bounds the temporal interference of lower-assurance vCPUs on the pCPU time allocated to critical vCPUs.

\textit{Scheduler efficiency and scalability} (Objective: Efficiency to be maximized / Algorithmic complexity to be minimized): This criterion evaluates the structural and algorithmic complexity of the scheduling algorithm as determined directly from its design and source code. Specifically, it assesses: (i) the algorithmic time complexity of core scheduling operations (e.g., $\mathcal{O}(1)$, $\mathcal{O}(\log N)$, or $\mathcal{O}(N)$ run-queue insertion and selection as the number of vCPUs N grows)~\cite{nieh2001virtual,brandenburg2011scheduling}, (ii) the invocation trigger mechanism (i.e., event-driven preemption versus tick-driven periodic timer interrupts)~\cite{katcher1993engineering}, and (iii) the resulting scalability under expanding vCPU counts and pCPU topologies without introducing severe queue lock contention or structural traversal bottlenecks~\cite{brandenburg2011scheduling}.

\textit{Configurability} (Objective: Context-dependent): This criterion characterizes: (a) the controllable scheduling parameters; (b) their binding time (build, boot, or runtime); (c) the reconfiguration interface and online admission checks; and (d) dynamic VM/vCPU lifecycle support. Following~\cite{real2004mode}, runtime changes are temporally safe only if mode transitions guarantee deadline preservation without hypervisor restarts. No assumption that dynamic reconfigurability is superior is made: static configuration remains a core safety requirement (ARINC 653~\cite{vanderleest2010arinc}, AUTOSAR Classic~\cite{autosar2022}), whereas dynamic adaptation addresses Software-Defined Vehicle (SDV) and Over-The-Air (OTA) demands (UN R156~\cite{unece2020r156}, AUTOSAR Adaptive~\cite{autosar2022}).

\textit{Multicore support} (Objective: Context-dependent): This criterion evaluates how the scheduler manages and scales across multiprocessor and multicore hardware topologies. It categorizes schedulers into three primary structural paradigms: (i) Uniprocessor-only scheduling, designed strictly for single-core execution without multicore awareness; (ii) Partitioned multiprocessor scheduling, which instantiates a scheduler per pCPU~\cite{brandenburg2011scheduling,carpenter2004categorization}, eliminating cross-core migration overheads and run-queue lock contention at the expense of bin-packing utilization loss; and (iii) global multiprocessor scheduling, which employs shared run-queues across all pCPUs~\cite{brandenburg2011scheduling,ongaro2008scheduling} or utilizes active inter-core load balancing~\cite{duda1999bvt,brandenburg2011scheduling} to maximize hardware capacity utilization and operational flexibility.

These criteria are applied to each algorithm in Sections~\ref{sec:static-partitioning} to~\ref{sec:prm}. They are then used to build the global qualitative comparison and the decision tree in Section~\ref{sec:qualitative-comparison}, which help the system architect select a scheduler according to the requirements of the application.

\subsection{Reference workload}\label{sec:reference-workload}

To illustrate and compare the behavior of the schedulers, a common reference workload on a single pCPU was scheduled manually over 20\,ms. The workload consists of four distinct vCPUs: vCPU\_RT, a periodic real-time vCPU; vCPU\_IO, an interrupt-driven I/O vCPU; and two background compute-intensive vCPUs, vCPU\_CPU1 and vCPU\_CPU2. Each vCPU is the single vCPU of its own VM, so VM-level parameters apply directly to it. Five key events occur during execution. vCPU\_RT is released periodically at $t=0$ and $t=10$, each release requiring 3\,ms of execution time. Asynchronous I/O wake-up interrupts for vCPU\_IO occur at $t=2$, $t=12$ and $t=16$, each requiring 1\,ms of execution time. vCPU\_CPU1 and vCPU\_CPU2 remain continuously ready for execution. For BVT, however, the same workload is observed over the $t=20$ to $t=40$ window, because its virtual times only reach a representative state after the first cycles (see Section~\ref{sec:bvt}).

Time is expressed in ms. At $t=0$, vCPU\_RT, vCPU\_CPU1 and vCPU\_CPU2 are queued in this order and vCPU\_IO is blocked. Implicit deadlines are equal to periods. An interrupt delivered to a vCPU that is already runnable does not wake it and only adds to its pending work. When several events occur at the same instant, they are all processed before the next scheduling decision, as in Xen, where the timer softirq is handled before the schedule softirq.

\subsection{Static partitioning}\label{sec:static-partitioning}

\subsubsection{Definition}

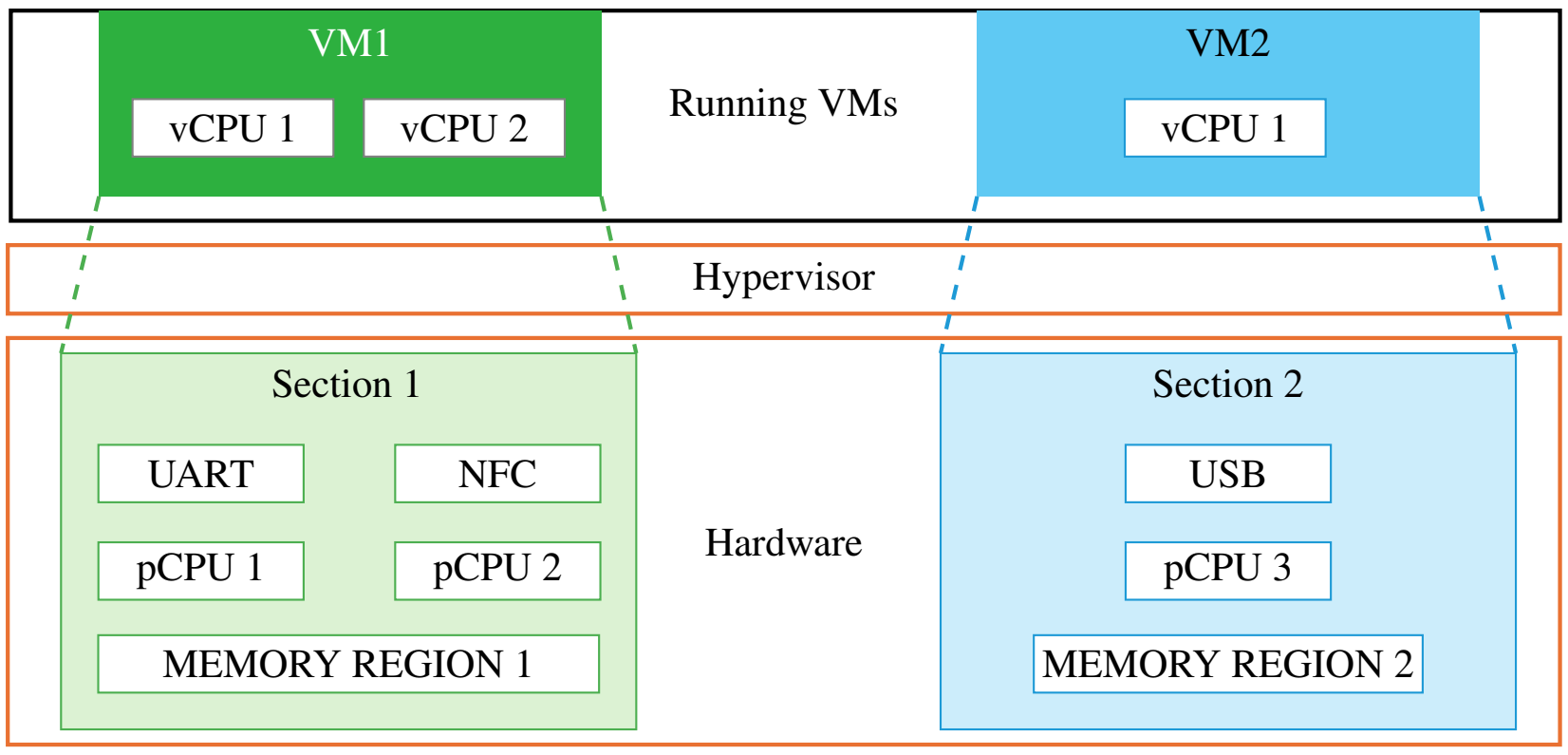
\begin{figure}[!htbp]
\centering
\begin{tikzpicture}[x=0.0106cm, y=-0.0106cm, font=\footnotesize,
    part/.style={draw, minimum height=0.40cm, inner sep=1pt, fill=white}]
  \draw[black, thick]     (10,25)  rectangle (1028,163);
  \draw[sporange, thick]  (8,179)  rectangle (1028,224);
  \draw[sporange, thick]  (8,240)  rectangle (1028,507);
  \node at (518,88)  {Running VMs};
  \node at (518,202) {Hypervisor};
  \node at (518,374) {Hardware};
  \fill[spgreen] (68,25)  rectangle (398,147);
  \fill[spblue]  (645,25) rectangle (975,147);
  \node[text=white] at (233,46) {VM1};
  \node at (810,46) {VM2};
  \node[part, draw=gray, minimum width=1.40cm] at (156,102) {vCPU 1};
  \node[part, draw=gray, minimum width=1.40cm] at (308,102) {vCPU 2};
  \node[part, draw=spblueln, minimum width=1.40cm] at (808,102) {vCPU 1};
  \draw[spgreenln, fill=spgreenbg] (43,250)  rectangle (421,497);
  \draw[spblueln,  fill=spbluebg]  (621,250) rectangle (999,497);
  \node at (231,270) {Section 1};
  \node at (810,270) {Section 2};
  \foreach \x/\y/\t in {135/329/UART, 330/329/NFC, 135/393/pCPU 1, 330/393/pCPU 2}
    \node[part, draw=spgreenln, minimum width=1.43cm] at (\x,\y) {\t};
  \node[part, draw=spgreenln, minimum width=3.49cm] at (232,454) {MEMORY REGION 1};
  \node[part, draw=spblueln,  minimum width=1.43cm] at (810,329) {USB};
  \node[part, draw=spblueln,  minimum width=1.43cm] at (810,393) {pCPU 3};
  \node[part, draw=spblueln,  minimum width=2.71cm] at (810,454) {MEMORY REGION 2};
  \draw[spgreenln, dashed, thick] (68,147)  -- (43,250);
  \draw[spgreenln, dashed, thick] (398,147) -- (421,250);
  \draw[spblueln,  dashed, thick] (645,147) -- (621,250);
  \draw[spblueln,  dashed, thick] (975,147) -- (999,250);
\end{tikzpicture}
\caption{Example of static hardware resource split}\label{fig:2}
\end{figure}

Static partitioning is a resource management method where hardware resources are split into predetermined, non-overlapping sections, with each section attributed to a VM, as Figure~\ref{fig:2} describes. For most systems, this mapping is established during the system’s configuration phase and remains unchanged at runtime. Typically, each vCPU is associated with a pCPU, meaning there is no sharing of pCPUs between VMs and eliminating the need for CPU scheduling. This unchangeable type of configuration is used by hypervisors such as Bao~\cite{martins2020bao,bao2026repo} and ACRN (using the \code{NOOP} scheduler~\cite{acrn2026noop}). However, Xen (using the \code{null} scheduler~\cite{xen2026null}) provides a notable exception to this rule: while it operates on the same static partitioning principle to minimize overhead, it retains the hypervisor's dynamic capabilities, allowing administrators to insert, remove, or migrate VMs at runtime without requiring a system reboot.

\subsubsection{Pros}

By using Static partitioning, strict temporal isolation is achieved at the scheduler level, since no vCPU competes for a shared pCPU. Micro-architectural interference through shared caches and memory buses remains and must be bounded separately~\cite{martins2023shedding}. Furthermore, \textit{real-time guarantees} are maintained due to the absence of dynamic scheduling. Because no scheduling decision is taken at runtime ($\mathcal{O}(1)$), this approach results in a minimal hypervisor codebase and reduced memory footprint. Because vCPUs access their assigned resources directly without the indirection of a scheduler, immediate \textit{I/O responsiveness} and minimal latency are ensured. Finally, regarding \textit{mixed-criticality support}, safety certification is greatly simplified and the system attack surface is minimized by this architectural simplicity, while its partitioned multiprocessor design eliminates cross-core lock contention.

\subsubsection{Cons}

On the other hand, the system's \textit{configurability} is bound to build or boot time, making runtime adjustments impossible without full hypervisor reboots (except in Xen's \code{null} scheduler). Additionally, \textit{fair CPU sharing} and dynamic redistribution of resources are prevented, rendering the platform incapable of adapting to fluctuating workloads. Because the system is inherently \textit{non-work-conserving}, the inability to reclaim unused time from idle VMs leads to significant resource underutilization.

\subsection{ARINC653 scheduler}\label{sec:arinc653}

The ARINC653 scheduler represents a deterministic, cyclic scheduling model designed for real-time and safety-critical virtualized environments. This scheduling paradigm is implemented in Xen through its ARINC653 scheduler~\cite{vanderleest2010arinc,xen2026arinc653wiki,xen2026arinc653code}, which enforces temporal partitioning according to the ARINC 653 standard~\cite{arinc653}.

\subsubsection{Parameters}

\textbf{Major frame (}\code{major\_frame}\textbf{)}: The~fixed time interval~over which the~scheduling cycle repeats.

\noindent\textbf{Minor frame (}\code{runtime}\textbf{)}: A fixed interval of the major frame assigned to a specific vCPU.

\noindent\textbf{Schedule allocation table (}\code{schedule}\textbf{)}: A static array~specifying the ordered~sequence of minor~frames, their durations, time offsets, and~assigned vCPUs.

\subsubsection{Selection criterion}

The ARINC653 scheduler applies a time-based static lookup selection criterion (\code{a653sched\_do\_schedule}). At any point during system execution, the scheduler selects the vCPU that is statically mapped to the active minor frame based on current elapsed time within the major frame.

If the assigned vCPU is in a non-runnable state (e.g., halted, sleeping, or yielded), or if an execution slot is intentionally left unassigned, the scheduler selects the system IDLE vCPU for the remainder of that time slot.

\subsubsection{Triggers}

Scheduling decisions and context switches in an ARINC653 scheduler are invoked by three triggers:
\begin{itemize}
    \item \textbf{Slot or major frame expiration:} A hardware timer interrupt fires when the current minor frame (\code{next\_switch\_time}) or the overall major frame (\code{next\_major\_frame}) expires, forcing a switch to the next slot.

\item\textbf{Voluntary yield / block / sleep (}\code{a653sched\_unit\_sleep}\textbf{):} When a running vCPU completes its task early or blocks waiting for an I/O operation, it cedes the pCPU. The scheduler switches execution to the IDLE vCPU until the current slot duration elapses.

\item\textbf{Host maintenance operations (Xen-specific):} Urgent hypervisor-level tasklets preempt the current minor frame to execute essential host management routines.
\end{itemize}

\subsubsection{Admission control}

The ARINC653 scheduler enforces admission control whenever a candidate schedule is submitted to the hypervisor. Prior to committing a configuration, the scheduler validates that the major frame duration is positive and that the entry count does not exceed the maximum storage capacity (\code{ARINC653\_MAX\_DOMAINS\_PER\_SCHEDULE}). Additionally, it verifies that each time slot receives a non-zero runtime allocation and that the cumulative runtime across all windows does not exceed the major frame boundary. If any condition is violated or if the combined window durations overcommit the major frame budget, the hypervisor rejects the schedule and returns an error (\code{-EINVAL}).

\subsubsection{Pseudo-code}

\begin{lstlisting}
"""
(*\textbf{Parameters}*):
    schedule_table: Schedule allocation table of minor frames (vcpu, start_offset, duration).
    major_frame: Fixed time interval over which the scheduling cycle repeats.
"""

(*\textbf{pick\_next}*)(current_time):   # a653sched_do_schedule()
    time_in_cycle = current_time % major_frame
    for slot in schedule_table:
        if slot.start_offset <= time_in_cycle < (slot.start_offset + slot.duration):
            return slot.vcpu if (slot.vcpu and slot.vcpu.is_runnable()) else IDLE_VCPU
    return IDLE_VCPU

(*\textbf{on\_timer\_interrupt}*)(current_time):   # a653sched_do_schedule()
    """Trigger: Minor or major frame expiration."""
    return pick_next(current_time)

(*\textbf{on\_sleep}*)(vcpu, current_time):   # a653sched_unit_sleep()
    """Trigger: Voluntary yield / block / sleep."""
    vcpu.state = SLEEP
    return pick_next(current_time)

(*\textbf{on\_wake}*)(vcpu):   # a653sched_unit_wake()
    """Trigger: Asynchronous wake-up event."""
    vcpu.state = RUNNABLE
\end{lstlisting}

\subsubsection{Pros}

The ARINC653 scheduler provides complete \textit{temporal isolation} by confining each vCPU to its pre-allocated minor frame, which prevents timing interference between vCPUs. This structure delivers deterministic real-time guarantees, making the supply of pCPU time to each vCPU known offline and enabling compositional schedulability analysis of the guest workload~\cite{shin2003periodic}. Consequently, \textit{mixed-criticality support} is robustly enforced, as safety-critical vCPUs are shielded from less-critical workloads. Furthermore, the algorithm achieves high \textit{scheduler efficiency} with $\mathcal{O}(1)$ time complexity, relying on simple static lookup operations and hardware timer comparisons. Finally, system \textit{configurability} is enhanced in Xen's implementation through online reconfiguration hypercalls, enabling administrators to dynamically update the global schedule allocation table at runtime without requiring system reboots.

\subsubsection{Cons}

Conversely, its primary drawback lies in its resource utilization, as its \textit{non-work-conserving} design forces pCPUs to remain idle when a vCPU yields or completes early, preventing the dynamic redistribution of unused time. The algorithm also exhibits poor \textit{I/O responsiveness}, as asynchronous events cannot trigger preemption and must await the next assigned slot. Furthermore, it lacks \textit{fair CPU sharing} mechanisms, requiring manual offline slot sizing. Finally, the implementation is restricted to uniprocessor execution (a single pCPU per scheduling pool), lacking native \textit{multicore} scaling.

\subsubsection{Timeline}

\begin{figure}[!htbp]
\centering
\includegraphics[width=\linewidth]{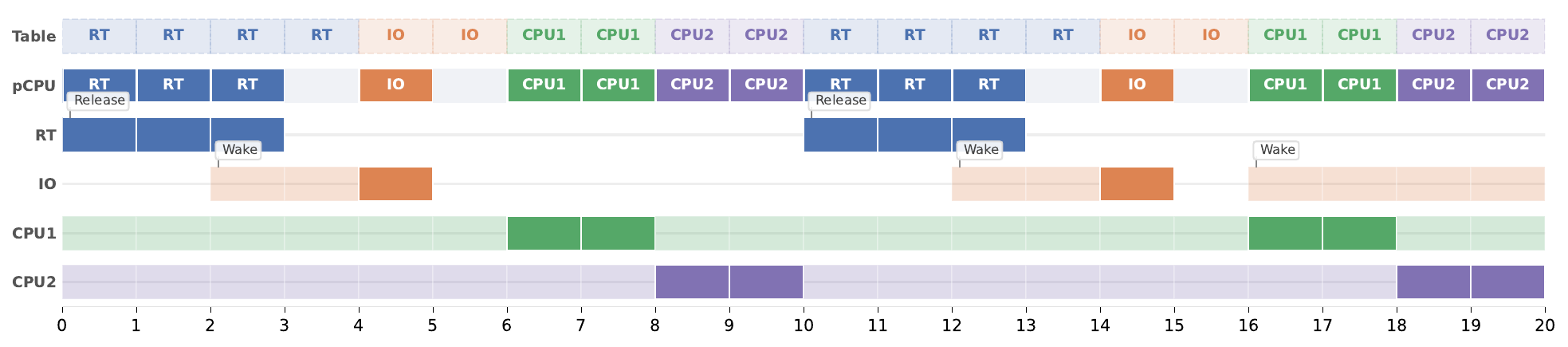}
\caption{Example timeline for ARINC653 scheduler}\label{fig:3}
\end{figure}

The schedule under the ARINC653 scheduler was derived from the rules presented in Section~\ref{sec:arinc653}. The global schedule table (Major Frame) spans a total duration of 10\,ms and is partitioned into 4 time windows (Minor Frames): Minor Frame 1 allocates 4\,ms to vCPU\_RT, Minor Frame 2 allocates 2\,ms to vCPU\_IO, Minor Frame 3 allocates 2\,ms to vCPU\_CPU1, and Minor Frame 4 allocates 2\,ms to vCPU\_CPU2. The scheduling plan is shown in the top track of Figure~\ref{fig:3}.

At $t=0$, the first major frame begins, and Minor Frame 1 activates vCPU\_RT, which starts executing on the pCPU. At $t=2$, an I/O interrupt wakes vCPU\_IO. However, because the time-triggered paradigm prohibits asynchronous preemption, vCPU\_IO cannot interrupt vCPU\_RT and must wait for its designated minor frame.

At $t=3$, vCPU\_RT completes its 3\,ms required execution time and transitions to a sleeping state. Even though vCPU\_IO is ready and waiting, the non-work-conserving policy prevents the hypervisor from reallocating the remaining time slice to any other runnable vCPU, forcing the pCPU into an idle state for 1\,ms and demonstrating the inherent resource underutilization of this model.

At $t=4$, the hypervisor transitions to Minor Frame 2, allowing vCPU\_IO to take control of the pCPU and execute from $t=4$ to $t=5$. Having finished its workload at $t=5$, vCPU\_IO goes back to sleep, causing the remaining time to remain unallocated in an idle state.

Subsequently, Minor Frames 3 and 4 execute sequentially from $t=6$ to $t=10$, where vCPU\_CPU1 consumes its full 2\,ms allocation, followed by vCPU\_CPU2, which also consumes its full 2\,ms allocation.

Finally, at $t=10$, the first Major Frame completes and the schedule rolls over to the next cycle.

The second Major Frame, however, does not simply replay the first one, because the interrupt arrivals are not aligned with the corresponding windows. The I/O wake-up at $t=12$ must wait for its window and only executes from $t=14$ to $t=15$. The third interrupt, which arrives at $t=16$, falls just after the I/O window has closed and must wait for the next one, starting at $t=24$, which results in a latency of 8\,ms.

\subsection{Credit scheduler}\label{sec:credit}

The Credit scheduler~\cite{xen2026creditwiki,xen2026creditcode} is a proportional fair-share CPU scheduling algorithm designed for symmetric multiprocessing (SMP) virtualized environments. This scheduling model is implemented in Xen.

\subsubsection{Parameters}

\textbf{Weight (}\code{sdom->weight}\textbf{)}: A VM-level parameter (ranging from 1 to 65,535, with a default value of \code{CSCHED\_DEFAULT\_WEIGHT} = 256) that defines the relative proportion of pCPU time allocated to a VM relative to competing VMs.

\noindent\textbf{Cap (}\code{sdom->cap}\textbf{):} An optional VM-level ceiling (expressed as a percentage of a single pCPU core, where 0 denotes uncapped execution) that limits the maximum pCPU time a VM can consume even when host cycles are idle.

\noindent\textbf{Time slice (}\code{tslice}\textbf{):} The global continuous execution quantum (configured by default to 30\,ms) allowed for a vCPU before preemption is evaluated.

\noindent\textbf{Accounting period:} The global periodic system interval (30\,ms), managed by \code{master\_ticker}, during which the hypervisor recalculates credit balances (\code{csched\_acct}) and updates priority for all active VMs.

\noindent\textbf{Context switch rate limit (}\code{prv->ratelimit}\textbf{):} A minimum execution threshold (configured by default to 1\,ms) designed to prevent excessive context switching caused by rapid interrupt wake-ups.

\noindent\textbf{Priority states (}\code{svc->pri}\textbf{):} Discrete scheduling priority levels assigned to vCPUs based on credit balance and state, defined as BOOST (0, assigned upon I/O wake-up only if the vCPU is currently in the UNDER state), UNDER (-1, positive credit balance), OVER (-2, exhausted credit balance), and IDLE (-64).

Weight and cap are set per VM (\code{sdom}); at each accounting period, the credit share of a VM is divided among its active vCPUs.

\subsubsection{Selection criterion}

The Credit scheduler applies a priority-based run-queue lookup with work-conserving task stealing. On each pCPU, the scheduler inspects its local run-queue and selects the runnable vCPU with the highest priority state, adhering to a strict hierarchy: BOOST > UNDER > OVER > IDLE. vCPUs belonging to the same priority are serviced in a First-In-First-Out (FIFO) sequence.

If a pCPU's local run-queue contains no UNDER or BOOST vCPUs, or becomes entirely idle, the scheduler executes a work-conserving load balancing routine. It scans the run-queues of other pCPUs to steal runnable vCPUs, thereby redistributing execution workloads across cores before falling back to the system IDLE vCPU.

\subsubsection{Triggers}

Scheduling decisions and context switches in the Credit scheduler are driven by five triggers:

\begin{itemize}

\item \textbf{Wake-up event (}\code{csched\_unit\_wake}\textbf{):} An asynchronous hardware interrupt or event wakes a sleeping vCPU, prompting the scheduler to temporarily elevate its priority to BOOST and insert it into the local run-queue.

\item \textbf{Voluntary yield / sleep / block (}\code{csched\_unit\_sleep} / \code{csched\_unit\_yield}\textbf{):} When a running vCPU completes its task, blocks on I/O, or explicitly yields execution, it cedes the pCPU and triggers a rescheduling event.

\item \textbf{Evaluation tick (}\code{csched\_tick}\textbf{, 10\,ms):} A local per-pCPU periodic hardware timer interrupt burns credits consumed by the currently running vCPU, resets BOOST vCPUs back to UNDER if they continue running and resorts the local run-queue.

\item \textbf{Global accounting timer (}\code{csched\_acct}\textbf{, 30\,ms):} A system timer recalculates credit allocations globally for all active VMs based on their assigned weights and caps, updating priority states between UNDER and OVER.

\item \textbf{Load balancing / task stealing (}\code{csched\_load\_balance}\textbf{):} When a pCPU becomes idle or its top candidate is in the OVER state, the scheduler triggers an active scan across peer pCPUs to steal runnable vCPUs.

\end{itemize}

\subsubsection{Admission control}

The Credit scheduler does not enforce workload admission control or CPU capacity bounds. As a proportional-share scheduler, it permits an arbitrary number of VMs and vCPUs to be instantiated without verifying whether total execution demand exceeds physical capacity. Rather than rejecting configurations or raising errors under overcommitted scenarios, the hypervisor relies on parameter sanitization during initialization and system management operations. Out-of-bounds configuration parameters, such as time-slice duration or rate-limiting thresholds, are automatically clamped to valid operational boundaries alongside a kernel warning, maintaining system stability without restricting workload entry.

\subsubsection{Pseudo-code}

\begin{lstlisting}
"""
(*\textbf{Parameters}*):
    weight: Proportional share weight assigned to each VM (1 to 65535, default 256).
    cap: Maximum pCPU time ceiling percentage for a VM (0 for uncapped).
    time_slice: Continuous execution quantum before preemption evaluation (30 ms).
    accounting_period: Global interval to recalculate credits and priorities (30 ms).
    ratelimit: Minimum execution threshold enforced before context switching (1 ms).
"""

(*\textbf{pick\_next}*)(cpu, now):   # csched_schedule()
    # Priority lookup: BOOST > UNDER > OVER > IDLE with work-conserving task stealing
    cur = cpu.current
    if cur.is_runnable and (now - cur.start_time) < rate_limit:
        return cur
    vcpu = runq_get_highest_priority(cpu)
    if not vcpu or vcpu.priority <= PRI_OVER:
        stolen_vcpu = load_balance_steal(cpu)   # csched_load_balance()
        if stolen_vcpu:
            return stolen_vcpu   # csched_runq_steal()
    return vcpu if (vcpu and vcpu.is_runnable()) else IDLE_VCPU

(*\textbf{on\_tick}*)(vcpu, delta_time):   # csched_tick()
    """Trigger: Per-pCPU accounting tick (10 ms)."""
    vcpu.credit -= delta_time * CREDITS_PER_MSEC   # burn_credits()
    if vcpu.priority == PRI_BOOST:
        vcpu.priority = PRI_UNDER

(*\textbf{on\_accounting\_timer}*)():   # csched_acct()
    """Trigger: Global credit accounting timer (30 ms)."""
    for vcpu in active_vcpus:
        vcpu.credit += compute_fair_share(vcpu.vm.weight, vcpu.vm.cap)/vcpu.vm.nr_active_vcpus
        vcpu.priority = PRI_UNDER if vcpu.credit >= 0 else PRI_OVER
    runq_resort_all()

(*\textbf{on\_wake}*)(vcpu):   # csched_unit_wake()
    """Trigger: Asynchronous wake-up event."""
    vcpu.state = RUNNABLE
    if vcpu.priority == PRI_UNDER:
        vcpu.priority = PRI_BOOST
    runq_insert_by_priority(vcpu)
    return pick_next(cpu, now)

(*\textbf{on\_sleep}*)(vcpu):   # csched_unit_sleep()
    """Trigger: Voluntary yield / block / sleep."""
    vcpu.state = SLEEP
    runq_remove(vcpu)
    return pick_next(cpu, now)
\end{lstlisting}

\subsubsection{Pros}

The Credit scheduler achieves effective \textit{fair CPU sharing} by distributing pCPU time among active vCPUs in direct proportion to their assigned weights. Its \textit{configurability} and runtime reconfiguration are flexible, allowing administrators to dynamically adjust weights and caps at runtime without requiring system reboots. As a \textit{work-conserving} policy, it maximizes resource utilization by immediately reallocating idle cycles. Furthermore, \textit{I/O responsiveness} is enhanced by its dedicated wake-up mechanism, which grants waking vCPUs a BOOST priority that minimizes wake-up latency. Finally, it provides \textit{multicore support} with active work-stealing load balancing across per-pCPU run-queues.

\subsubsection{Cons}

Conversely, the Credit scheduler exhibits limitations regarding \textit{temporal isolation}, as its BOOST mechanism grants BOOST to any waking vCPU in the UNDER state, regardless of its criticality, allowing interactive background vCPUs to preempt safety-critical vCPUs. This lack of timing control compromises \textit{real-time guarantees}, as the algorithm lacks execution deadline tracking, exposing critical workloads to execution jitter. Consequently, its support for \textit{mixed-criticality} workloads offers limited protection against cross-VM interference. Finally, \textit{scheduler efficiency} is bounded by the $\mathcal{O}(N)$ complexity of periodic credit accounting (every 30\,ms over all active vCPUs), the 10\,ms evaluation tick, and the cross-core run-queue scan during task stealing.

\subsubsection{Timeline}

The schedule under the Credit scheduler was derived from the rules presented in Section~\ref{sec:credit}. The Credit scheduler manages vCPU priorities across three classes: BOOST, UNDER, and OVER. Since global credit accounting is updated only every 30\,ms, all non-boosted vCPUs remain in the UNDER state throughout the 20\,ms observation window. As no vCPU transitions to the OVER state, the weights assigned to each vCPU do not affect scheduling decisions in this trace and are therefore not specified.

\begin{figure}[!htbp]
\centering
\includegraphics[width=\linewidth]{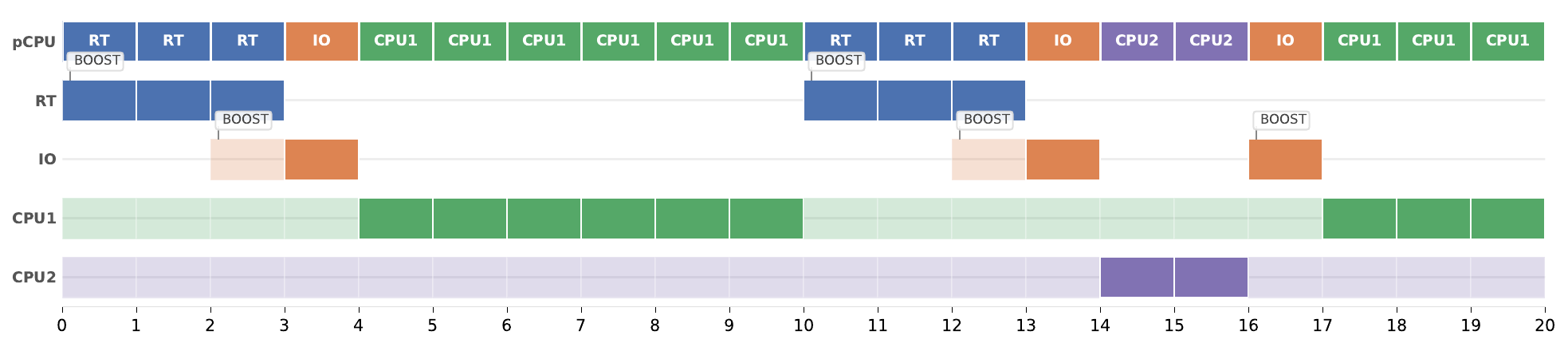}
\caption{Example timeline for Credit scheduler}\label{fig:4}
\end{figure}

At $t=0$, vCPU\_RT wakes up and is granted a BOOST priority allocation, taking control of the pCPU.

At $t=2$, an I/O interrupt wakes vCPU\_IO, which also receives a BOOST priority boost. However, because vCPU\_RT is already executing at the same BOOST priority level, no preemption occurs, and vCPU\_IO is placed at the tail of the BOOST tier in FIFO order.

At $t=3$, vCPU\_RT completes its 3\,ms execution requirement and goes back to sleep. vCPU\_IO immediately takes the pCPU and executes from $t=3$ to $t=4$ before returning to sleep.

At $t=4$, with no BOOST vCPUs remaining runnable, the scheduler selects from the UNDER priority runqueue. vCPU\_CPU1 is scheduled and executes from $t=4$ to $t=10$.

At $t=10$, vCPU\_RT is selected in the BOOST state and preempts vCPU\_CPU1. \code{csched\_schedule()} reinserts the preempted vCPU\_CPU1 behind all UNDER vCPUs (behind vCPU\_CPU2).

At $t=12$, an I/O interrupt wakes vCPU\_IO, elevating its priority to BOOST. Since vCPU\_RT is already running in the BOOST state, no preemption occurs: vCPU\_RT finishes its execution until $t=13$, after which vCPU\_IO runs from $t=13$ to $t=14$ and blocks.

At $t=14$, the scheduler falls back to the UNDER tier and selects vCPU\_CPU2 (now at the head of the UNDER queue following the $t=10$ rotation), which runs from $t=14$ to $t=16$.

At $t=16$, the third I/O interrupt wakes vCPU\_IO in the BOOST state. Because vCPU\_CPU2 is running at UNDER priority, vCPU\_IO immediately preempts it and executes from $t=16$ to $t=17$.

Upon being preempted at $t=16$, vCPU\_CPU2 is reinserted at the tail of the UNDER tier, placing it behind vCPU\_CPU1. Consequently, when vCPU\_IO goes back to sleep at $t=17$, the scheduler selects vCPU\_CPU1, which executes for the remaining $3\text{ms}$ until $t=20$.

\subsection{Credit2 scheduler}\label{sec:credit2}

The Credit2 scheduler~\cite{xen2026credit2wiki,xen2026credit2code} is a proportional fair-share scheduling algorithm designed to improve upon the original Credit scheduler by focusing on fairness, responsiveness, and multicore scalability. This scheduling model is implemented in Xen.

\subsubsection{Parameters}

\textbf{Weight (}\code{sdom->weight}\textbf{):} A VM-level parameter (default value of 256) defining the relative proportion of pCPU time allocated to a VM. Each vCPU inherits this weight, which dictates the rate at which its individual credit balance decays during execution (credits are expressed in time units: a vCPU holding the maximum weight burns one unit of credit per unit of execution time, with an initial allocation \texttt{CSCHED2\_CREDIT\_INIT} of 10\,ms. In the timeline of Section~2.7.8, credits are expressed in milliseconds of execution at the maximum weight).

\noindent\textbf{Cap (}\code{sdom->cap}\textbf{):} An optional VM-level execution ceiling expressed as a percentage of a single pCPU. The hypervisor converts this percentage into a periodic time budget. VM vCPUs consume execution time from this budget and are “parked” (blocked from running) once it is exhausted until the next replenishment cycle.

\noindent\textbf{Context switch rate limit (}\code{prv->ratelimit\_us}\textbf{):} A minimum execution threshold (default 1\,ms) enforced to prevent excessive context switching caused by rapid interrupt wake-ups.

\noindent\textbf{EWMA load tracking parameters (}\code{opt\_load\_window\_shift}, \code{opt\_load\_precision\_shift}\textbf{):} Parameters controlling the Exponentially Weighted Moving Average (EWMA) algorithm used to track~the changing load~of run-queues~and vCPUs.

\noindent\textbf{Migration resistance threshold (}\code{opt\_migrate\_resist}\textbf{):} A credit bonus threshold (default 500\textmu{}s) required for a vCPU to be migrated to a different pCPU, minimizing cache invalidation overheads.

\subsubsection{Selection criterion}

The Credit2 scheduler orders vCPUs by their credit values and balances the workload across pCPUs while considering the hardware topology. On each shared run-queue, vCPUs are inserted in descending order of their remaining credits. When a pCPU needs to select a new vCPU, it picks the runnable vCPU with the highest remaining credits at the head of the queue.

For capped VMs, a vCPU must also have an available time budget in its VM pool. If no budget is available, the vCPU is “parked” and excluded from selection. Additionally, the scheduler dynamically balances the load across run-queues by migrating vCPUs when significant load imbalances occur.

When the vCPU at the head of the run-queue exhausts its credits (reaching 0 or less), the scheduler triggers a credit reset (\code{reset\_credit()}): every vCPU associated with the run-queue, including sleeping and parked ones, receives a fresh allocation of credits added to a carry-over fraction of its remaining balance, capped at a maximum ceiling to prevent credit hoarding.

\subsubsection{Triggers}

Scheduling decisions and context switches in the Credit2 scheduler are driven by five triggers:

\begin{itemize}

\item \textbf{Wake-up event (}\code{csched2\_unit\_wake}\textbf{):} An asynchronous hardware interrupt or event wakes a sleeping vCPU. Because sleeping vCPUs preserve their credits, they are re-inserted near the head of the run-queue according to their credit values.

\item \textbf{Voluntary yield / sleep / block (}\code{csched2\_unit\_sleep} / \code{csched2\_unit\_yield}\textbf{):} When a running vCPU completes its execution, yields, or blocks on I/O, it returns any unconsumed budget (if capped), updates run-queue load statistics and invokes a rescheduling decision.

\item \textbf{Dynamic time slice timer expiration (}\code{csched2\_runtime}\textbf{):} The hypervisor calculates and sets a dynamic hardware timer based on when the current vCPU will exhaust its credits relative to competing vCPUs, hit its cap budget, or reach timer bounds.

\item \textbf{Budget exhaustion \& replenishment timer (}\code{replenish\_domain\_budget}\textbf{):} If a capped VM depletes its budget pool, its running vCPUs are parked. A periodic replenishment timer (default 10\,ms) restores the VM budget and unparks waiting vCPUs.

\item \textbf{Load balancing evaluation (}\code{balance\_load}\textbf{):} Periodic and event-driven EWMA load calculations monitor load disparities across run-queues, triggering dynamic vCPU migrations when imbalances are detected.

\end{itemize}

\subsubsection{Admission control}

The Credit2 scheduler does not restrict vCPU instantiation based on total processor capacity. However, it enforces targeted parameter validation for VM-level execution caps. Specifically, if an administrator configures a utilization cap exceeding the maximum physical limits of the domain's allocated vCPUs, the request is explicitly rejected with an error code (\code{-EINVAL}). System-wide initialization settings are verified against operational limits, issuing warnings and falling back to safe default configurations whenever invalid inputs are detected.

\subsubsection{Pseudo-code}

\begin{lstlisting}
"""
(*\textbf{Parameters}*):
    weight: Proportional share weight assigned to each VM (default 256).
    cap: Maximum pCPU percentage ceiling converted into a periodic time budget (0 for uncapped).
    ratelimit: Minimum execution quantum enforced before context switching (default 1 ms).
    load_window: Window length for Exponentially Weighted Moving Average (EWMA) load tracking.
    migrate_resist: Credit resistance threshold required before cross-pCPU migration (default 500us).
"""

(*\textbf{pick\_next}*)(runq):   # csched2_schedule()
    # Select highest credit runnable vCPU from credit-ordered head of runqueue
    for vcpu in runq:
        if vcpu.is_runnable():
            if has_cap(vcpu) and not grab_budget(vcpu):   # unit_grab_budget()
                park_unit(vcpu)   # park_unit()
                continue
            return vcpu
    return IDLE_VCPU

(*\textbf{on\_schedule}*)(vcpu, current_time):   # csched2_schedule()
    """Trigger: Dynamic time slice expiration."""
    burn_credits(vcpu, current_time)
    if top_runq_credit(vcpu.runq) <= 0:
        reset_credits(vcpu.runq)
        balance_load_ewma()
    return pick_next(vcpu.runq)

(*\textbf{on\_wake}*)(vcpu):   # csched2_unit_wake()
    """Trigger: Asynchronous wake-up event."""
    vcpu.state = RUNNABLE
    update_load_ewma(vcpu, +1)
    runq_insert_by_credit(vcpu)
    return pick_next(vcpu.runq)

(*\textbf{on\_sleep}*)(vcpu):   # csched2_unit_sleep()
    """Trigger: Voluntary yield / block / sleep."""
    vcpu.state = SLEEP
    if has_cap(vcpu):
        return_budget(vcpu)   # unit_return_budget()
    update_load_ewma(vcpu, -1)   # update_load()
    runq_remove(vcpu)
    return pick_next(vcpu.runq)

(*\textbf{on\_replenish\_timer}*)(vm):   # replenish_domain_budget()
    """Trigger: Periodic cap budget replenishment timer (default 10 ms)."""
    vm.budget = vm.tot_budget
    unpark_units(vm.parked_units)   # unpark_parked_units()
\end{lstlisting}

\subsubsection{Pros}

The Credit2 scheduler delivers strong \textit{fair CPU sharing} through a credit decay model based on vCPU weights. Its \textit{configurability} is well supported, enabling dynamic runtime adjustments of weights, caps, and rate limits while grouping pCPUs into topology-aware run-queues. By preserving credits during sleep, it maintains high \textit{I/O responsiveness}, enabling waking vCPUs to naturally preempt lower-credit workloads. As a \textit{work-conserving} algorithm, it maintains high resource utilization by dynamically redistributing idle pCPU cycles through active EWMA load balancing. Finally, it provides \textit{global} \textit{multicore support}, leveraging migration resistance thresholds to reduce lock contention and cache invalidation overhead across multiple pCPUs.

\subsubsection{Cons}

The Credit2 scheduler incurs some limitations in \textit{scheduler efficiency}, due to the $\mathcal{O}(N)$ insertion operations required to maintain run queues ordered by credit, as well as the computational overhead of frequent EWMA load calculations. Because priority relies on credit decay rather than explicit deadlines, it cannot provide hard \textit{real-time guarantees}. Consequently, its \textit{temporal isolation} and \textit{mixed-criticality support} remain limited under uncapped configurations, as non-critical vCPUs can introduce timing jitter into critical workloads.

\subsubsection{Timeline}

The schedule under the Credit2 scheduler was derived from the rules presented in Section~\ref{sec:credit2}. Credit2 manages vCPUs using credit balances (initially 10 credits), where vCPUs burn credits proportionally to their weights and are ordered within run-queues by remaining credits. To reflect a realistic proportional-share allocation, higher weights are assigned to vCPU\_RT and vCPU\_IO (weight = 256, burning 1 credit/ms), while the background vCPUs receive lower weights (weight = 64, burning 4 credits/ms). In addition, to highlight budget enforcement, a 20\% cap is configured for vCPU\_CPU1, limiting its execution time to a maximum of 2\,ms within each 10\,ms replenishment period. vCPU\_CPU1 and vCPU\_CPU2 remain continuously ready, subject to their respective credit and budget constraints.

\begin{figure}[!htbp]
\centering
\includegraphics[width=\linewidth]{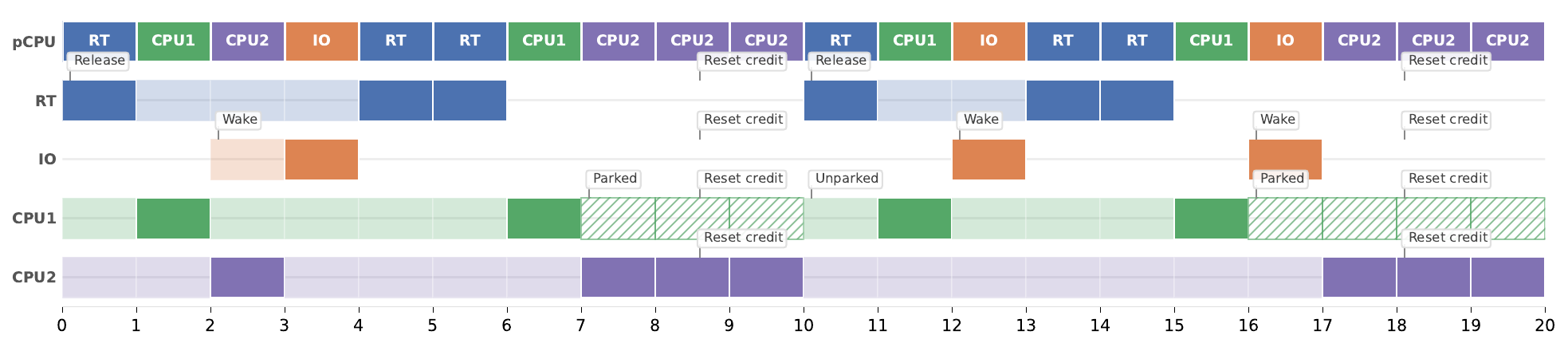}
\caption{Example timeline for Credit2 scheduler}\label{fig:5}
\end{figure}

At $t=0$, vCPU\_RT, vCPU\_CPU1, and vCPU\_CPU2 are all ready with an equal initial balance of $10\text{ credits}$. Because \code{runq\_insert()} enforces FIFO ordering among vCPUs with identical credit balances, vCPU\_RT occupies the head of the run-queue and is selected to execute. However, since the next candidate in the run-queue (vCPU\_CPU1) holds the exact same credit balance, \code{csched2\_runtime()} bounds vCPU\_RT's allocated runtime to the minimum timer slice, set to $1\text{ms}$. vCPU\_RT therefore executes from $t=0$ to $t=1$, burning $1\text{ credit}$.

At $t=1$, vCPU\_CPU1 ($10\text{ credits}$) moves to the head of the run-queue ahead of vCPU\_CPU2 ($10\text{ credits}$) and vCPU\_RT ($9\text{ credits}$). Due to the credit tie with vCPU\_CPU2, vCPU\_CPU1 is also granted a $1\text{ms}$ minimum slice and executes from $t=1$ to $t=2$, consuming $4\text{ credits}$ (down to $6\text{ credits}$) and $1\text{ms}$ of its $2\text{ms}$ cap budget.

At $t=2$, an interrupt wakes vCPU\_IO with its untouched initial balance of $10\text{ credits}$. Because \code{runq\_insert()} requires a strictly greater credit balance to overtake existing entries, vCPU\_IO ($10\text{ credits}$) is inserted behind vCPU\_CPU2 ($10\text{ credits}$) in FIFO order. Consequently, vCPU\_CPU2 executes from $t=2$ to $t=3$ (burning $4\text{ credits}$ down to $6\text{ credits}$), after which vCPU\_IO becomes the sole leader and executes from $t=3$ to $t=4$ (burning $1\text{ credit}$ down to $9\text{ credits}$) before returning to sleep.

At $t=4$, vCPU\_RT holds the highest remaining balance among runnable vCPUs ($9\text{ credits}$ vs. $6\text{ credits}$ for vCPU\_CPU1 and vCPU\_CPU2). \code{csched2\_runtime()} allocates a $3\text{ms}$ slice (the time required for vCPU\_RT to burn down to $6\text{ credits}$) allowing it to complete its remaining $2\text{ms}$ execution requirement uninterrupted from $t=4$ to $t=6$ and go to sleep (at $7\text{ credits}$).

At $t=6$, vCPU\_CPU1 ($6\text{ credits}$) precedes vCPU\_CPU2 in FIFO order and executes for $1\text{ms}$.

At $t=7$, vCPU\_CPU1 exhausts its 2\,ms budget and is parked. Alone in the run-queue, vCPU\_CPU2 receives the time needed to exhaust its own 6 credits at weight 64 ($6 \times 64/256 = 1.5\,\text{ms}$). At $t=8.5$ its credit reaches zero, which triggers \code{reset\_credit()}: every vCPU of the run-queue, including blocked and parked ones, receives 10 credits, clipped to 10.5. vCPU\_CPU2 (now 10) keeps running.

At $t=10$, vCPU\_RT is woken and vCPU\_CPU1 is unparked within the same timer softirq; both hold 10.5 credits whereas vCPU\_CPU2 has burned down to 4. Assuming the timer of vCPU\_RT is processed first, vCPU\_RT runs from $t=10$ to $t=11$, followed by vCPU\_CPU1 from $t=11$ to $t=12$.

At $t=12$, the second I/O interrupt wakes vCPU\_IO with $10.5\text{ credits}$, more than vCPU\_RT (9.5), vCPU\_CPU1 (6.5) and vCPU\_CPU2 (4), and runs until $t=13$.

Then, vCPU\_RT completes its job from $t=13$ to $t=15$, and vCPU\_CPU1 runs from $t=15$ to $t=16$, when its budget is exhausted and it is parked.

At $t=16$, vCPU\_IO wakes with 9.5 credits and runs until $t=17$. vCPU\_CPU2 then runs until $t=20$; its credit reaches zero at $t=18$, which triggers another reset without changing the running vCPU.

\subsection{Real-time Deferrable Server (RTDS) scheduler}\label{sec:rtds}

The RTDS scheduler~\cite{xi2014realtime,xen2026rtdswiki,xen2026rtdscode} is a real-time scheduling algorithm designed to provide deterministic timing guarantees for latency-sensitive workloads. This scheduling model is implemented in Xen through its RTDS module, which evolved from a scheduler based on fixed time slices in Xen 4.5 to a more precise event-based model in Xen 4.7.

\subsubsection{Parameters}

\textbf{Period (}\code{svc->period}\textbf{):} A per-vCPU temporal window (configured by default to 10\,ms) over which execution time reservations are periodically renewed.

\noindent\textbf{Budget (}\code{svc->budget}\textbf{):} The maximum execution time (configured by default to 4\,ms) allocated to a vCPU within each period.

\noindent\textbf{Extratime flag (}\code{RTDS\_extratime}\textbf{):} An optional per-vCPU boolean parameter that allows a vCPU whose nominal budget is exhausted to continue executing during idle pCPU cycles at a lower priority tier.

\noindent\textbf{Absolute deadline (}\code{svc->cur\_deadline}\textbf{):} A dynamic timestamp computed at the beginning of each period that represents the execution deadline used for EDF ordering.

\noindent\textbf{Priority level (}\code{svc->priority\_level}\textbf{):} An integer tracking tier (0 for real-time reserved status, incremented upon budget exhaustion if extratime is enabled) used to separate real-time reserved execution from deprioritized background execution.

\subsubsection{Selection criterion}

The RTDS scheduler applies a Preemptive Global EDF policy with priority demotion. The hypervisor maintains a global run-queue containing all runnable vCPUs with available budget, sorted primarily by priority level (tier 0 first) and secondarily by absolute deadline in ascending order. The scheduler inspects the global run-queue and selects the vCPU with the earliest deadline to execute on an available pCPU.

If a vCPU exhausts its budget and lacks the extratime flag, it is moved to an unsorted global depleted queue and excluded from execution until its next period replenishment. If the extratime flag is enabled, the vCPU's priority level is demoted by 1, its budget is refilled, and it remains in the run-queue to harvest leftover pCPU cycles without interfering with real-time vCPUs.

\subsubsection{Triggers}

Scheduling decisions and context switches in the RTDS scheduler are driven by four triggers:

\begin{itemize}

\item \textbf{Wake-up event (}\code{rt\_unit\_wake}\textbf{):} An asynchronous event or interrupt wakes a sleeping vCPU. If the vCPU slept past its deadline, its deadline is updated and its budget replenished; otherwise, its unconsumed budget is preserved, and it is inserted into the global run-queue.

\item \textbf{Voluntary sleep / block / yield (}\code{rt\_unit\_sleep}\textbf{):} When a running vCPU completes its task or blocks waiting for I/O, it yields the pCPU, triggering a rescheduling event to pick the next earliest deadline candidate.

\item \textbf{Budget exhaustion (}\code{burn\_budget}\textbf{):} A dynamic hardware timer fires when the running vCPU consumes its remaining budget, forcing preemption. The vCPU is either transferred to the depleted queue or demoted in priority if extratime is active.

\item \textbf{Replenishment timer expiration (}\code{repl\_timer\_handler}\textbf{):} A global event-driven replenishment timer, programmed to fire at the most imminent deadline among active vCPUs, restores depleted budgets, recalculates deadlines, and moves depleted vCPUs back to the global run-queue.

\end{itemize}

\subsubsection{Admission control}

The RTDS scheduler incorporates a strict admission control mechanism based on total pCPU capacity. During vCPU instantiation or online reservation adjustments, the hypervisor first validates candidate period and budget parameters via \code{rt\_validate\_params()}, rejecting malformed requests with an error code (\code{-EINVAL}). Next, RTDS computes the aggregate fixed-point utilization demand $(\sum \mathit{budget}/\mathit{period})$ across all admitted workloads. If a proposed allocation or reservation increase causes total demand to exceed the processing capacity limit, the hypervisor explicitly denies the change, issuing a warning log and returning an error (\code{-ENOSPC}). System administrators can also dynamically query or toggle this enforcement mechanism via hypervisor system management operations.

\subsubsection{Pseudo-code}

\begin{lstlisting}
"""
(*\textbf{Parameters}*):
    period: Per-vCPU temporal window over which execution time is renewed (default 10 ms).
    budget: Maximum execution time allocated to a vCPU within each period (default 4 ms).
    extratime: Boolean flag allowing execution during idle cycles at a demoted priority tier.
    cur_deadline: Dynamic absolute deadline used for Earliest Deadline First (EDF) ordering.
    priority_level: Priority tier (0 for real-time reserved; incremented when budget is exhausted with extratime).
"""

(*\textbf{pick\_next}*)(runq):   # rt_schedule()
    # Preemptive Global EDF: pick runnable vCPU with lowest priority_level, then earliest cur_deadline
    return min(runq, key=lambda v: (v.priority_level, v.cur_deadline), default=IDLE_VCPU)

(*\textbf{on\_tick}*)(vcpu, delta_time):
    """Trigger: Per-pCPU timer / budget burn."""
    vcpu.cur_budget -= delta_time   # burn_budget()
    if vcpu.cur_budget <= 0:
        if vcpu.extratime:
            vcpu.priority_level += 1
            vcpu.cur_budget = vcpu.budget
        else:
            vcpu.state = DEPLETED
            runq_move_to_depletedq(vcpu)

(*\textbf{on\_replenish\_timer}*)():   # repl_timer_handler()
    """Trigger: Global replenishment timer firing at the most imminent deadline."""
    for vcpu in replq_get_expired():
        vcpu.cur_deadline += vcpu.period
        vcpu.cur_budget = vcpu.budget
        vcpu.priority_level = 0
        vcpu.state = RUNNABLE
        depletedq_move_to_runq(vcpu)

(*\textbf{on\_wake}*)(vcpu, current_time):   # rt_unit_wake()
    """Trigger: Asynchronous wake-up event."""
    vcpu.state = RUNNABLE
    if current_time >= vcpu.cur_deadline:
        vcpu.cur_deadline = current_time + vcpu.period   # rt_update_deadline()
        vcpu.cur_budget = vcpu.budget
        vcpu.priority_level = 0
    runq_insert_by_edf(vcpu)   # runq_insert()
    return pick_next(runq)

(*\textbf{on\_sleep}*)(vcpu):   # rt_unit_sleep()
    """Trigger: Voluntary yield / block / sleep."""
    vcpu.state = SLEEP
    runq_remove(vcpu)
    return pick_next(runq)
\end{lstlisting}

\subsubsection{Pros}

The RTDS scheduler delivers \textit{real-time guarantees} through its preemptive global EDF policy, provided the system integrator verifies offline that the reservation set is schedulable. Its per-vCPU budget and period reservations enforce \textit{temporal isolation} by preventing misbehaving or computationally heavy vCPUs from exceeding their assigned allocations. Furthermore, \textit{mixed-criticality} co-location is supported thanks to the extratime mechanism, which allows less-critical vCPUs to harvest idle cycles at a demoted priority tier without violating tier-0 reservations. This feature also enhances \textit{resource utilization}. For interactive workloads, \textit{I/O responsiveness} is high because sleeping vCPUs preserve their unconsumed budget upon blocking. It also exhibits rich \textit{runtime reconfigurability}, allowing administrators to dynamically adjust per-vCPU budgets, periods, and extratime flags at runtime without hypervisor reboots. Finally, it provides \textit{global} \textit{multicore support}, as it is specifically designed for SMP architectures using a global EDF run-queue and inter-processor tickle interrupts across pCPUs.

\subsubsection{Cons}

\textit{Scheduler efficiency} represents a key constraint for RTDS. The scheduler must track remaining execution budgets at nanosecond granularity, reprogram hardware timers at every budget event, and perform $\mathcal{O}(N)$ insertions into the deadline-ordered run-queue and replenishment queue. Furthermore, the algorithm lacks \textit{fair CPU sharing} for background vCPUs. While the extratime flag reclaims unreserved pCPU cycles, it lacks proportional ratios, forcing deprioritized vCPUs to compete for remaining cycles without guaranteed fairness.

\subsubsection{Timeline}

The schedule under the RTDS scheduler was derived from the rules presented in Section~\ref{sec:rtds}. Each vCPU is defined by a period and an execution budget: vCPU\_RT (period = 10, budget = 3, extratime = False), vCPU\_IO (period = 6, budget = 1, extratime = False), vCPU\_CPU1 (period = 10, budget = 3, extratime = True), and vCPU\_CPU2 (period = 15, budget = 3, extratime = True).

\begin{figure}[!htbp]
\centering
\includegraphics[width=\linewidth]{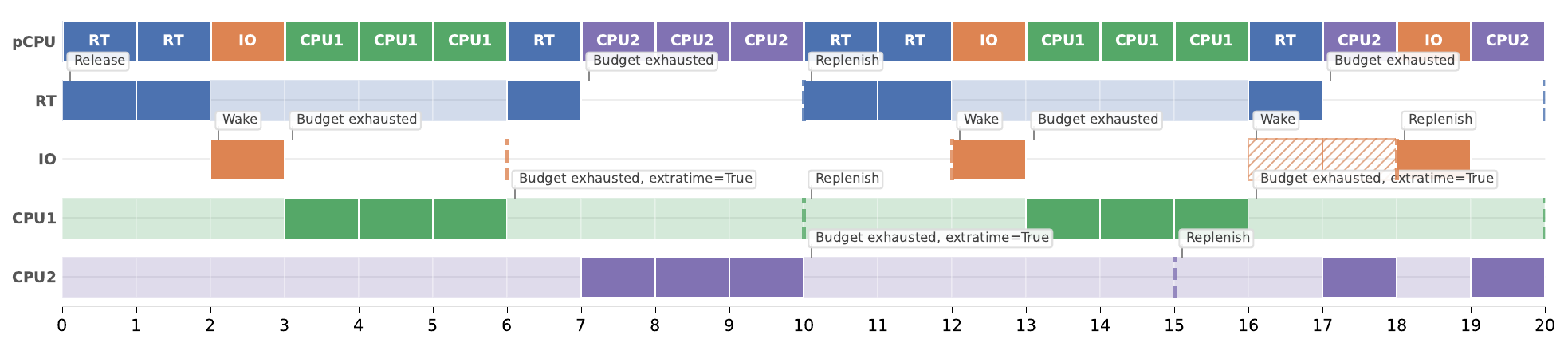}
\caption{Example timeline for RTDS scheduler}\label{fig:6}
\end{figure}

At $t=0$, vCPU\_RT releases its first period (\code{cur\_deadline} = 10, \code{cur\_budget} = 3, \code{priority\_level} = 0) and takes the pCPU.

At $t=2$, an I/O interrupt wakes vCPU\_IO, with its parameters set to \code{cur\_deadline} = 6, \code{cur\_budget} = 1, and \code{priority\_level} = 0. The scheduler invokes \code{runq\_tickle()} to compare deadlines. Because vCPU\_IO has an earlier deadline than the running vCPU\_RT, vCPU\_IO preempts vCPU\_RT at $t=2$ and executes from $t=2$ to $t=3$.

At $t=3$, vCPU\_IO finishes its 1\,ms job and goes back to sleep while exhausting its budget at the same instant. vCPU\_CPU1 is selected because it is queued ahead of vCPU\_RT, consuming its full 3\,ms budget.

At $t=6$, vCPU\_CPU1 exhausts its budget. Because extratime = True, its priority level is demoted from 0 to 1 and its budget is refilled. The scheduler then selects vCPU\_RT, which executes from $t=6$ to $t=7$ before it returns to sleep.

From $t=7$ to $t=10$, vCPU\_CPU2 executes its 3\,ms allocation, exhausting its budget at $t=10$ and demoting to \code{priority\_level} = 1. At the same instant, vCPU\_RT is woken by the release of its second job, and the replenishment of vCPU\_CPU1 falls due. Since vCPU\_RT was blocked, it is not on the replenishment queue: its deadline and budget are updated on wake-up (\code{rt\_unit\_wake()} calls \code{rt\_update\_deadline()}), giving \code{cur\_deadline} = 20 and \code{cur\_budget} = 3. Because vCPU\_RT is woken while vCPU\_CPU1 still holds its demoted extra-time status (\code{priority\_level} = 1), vCPU\_RT is inserted ahead of vCPU\_CPU1 in the run queue. When vCPU\_CPU1's period is subsequently replenished, \code{deadline\_queue\_insert()} places it behind vCPU\_RT under FIFO tie-breaking for equal deadlines. Consequently, vCPU\_RT takes the pCPU and executes from $t=10$ to $t=12$. Both events occur in the same timer softirq. It is assumed that the wake-up of vCPU\_RT is processed first. In the opposite order, vCPU\_CPU1 would run first ($t=10$ to 12) and vCPU\_RT would complete at $t=16$, still meeting its deadline.

At $t=12$, an I/O interrupt wakes vCPU\_IO. The scheduler compares vCPU\_IO (deadline 18) against the running vCPU\_RT (deadline 20). Because its deadline is earlier than vCPU\_RT, vCPU\_IO preempts vCPU\_RT and executes from $t=12$ to $t=13$.

At $t=13$, vCPU\_CPU1 is selected and executes its full 3\,ms budget from $t=13$ to $t=16$.

At $t=15$, vCPU\_CPU2 undergoes period replenishment.

At $t=16$, vCPU\_CPU1 exhausts its budget and is demoted to \code{priority\_level} = 1 (extratime = True); simultaneously, an I/O wake event occurs for vCPU\_IO, but because its period replenishment is scheduled for $t=18$, it remains depleted. The scheduler selects vCPU\_RT, which executes its remaining 1\,ms budget from $t=16$ to $t=17$, finishes its work, and goes back to sleep.

At $t=17$, as vCPU\_RT finishes its work and vCPU\_CPU1 is demoted to extra-time status, vCPU\_CPU2 takes the pCPU and executes from $t=17$ to $t=18$.

At $t=18$, vCPU\_IO's period is replenished. Since its deadline (24) is earlier than vCPU\_CPU2's (30), vCPU\_IO preempts vCPU\_CPU2 at $t=18$ and executes from $t=18$ to $t=19$.

Finally, at $t=19$, vCPU\_IO completes its work and blocks, allowing vCPU\_CPU2 to resume and execute.

\subsection{Priority-based scheduler}\label{sec:priority}

The Priority-based scheduler is a scheduling algorithm designed to enforce vCPU execution hierarchy based on static priority levels. This scheduling model is natively implemented in the ACRN hypervisor~\cite{acrn2026prio}.

\subsubsection{Parameters}

\textbf{Priority (}\code{params->prio}\textbf{)}: A static integer assigned to each vCPU at creation time, defining its execution precedence relative to competing vCPUs on the same pCPU. Although the C structure stores an integer, ACRN's build-time configuration schema (\code{types.xsd}) exposes only two discrete levels: \code{PRIO\_LOW} and \code{PRIO\_HIGH}.

\noindent\textbf{Per-pCPU priority queue (}\code{INIT\_LIST\_HEAD(\&prio\_ctl->prio\_queue)}\textbf{)}: A doubly linked list maintained per pCPU that stores runnable vCPUs sorted in descending order of priority (with FIFO ordering among vCPUs of equal priority).

\subsubsection{Selection criterion}

The Priority-based scheduler applies a priority-driven policy. On each pCPU, the scheduler inspects its local priority run-queue and selects the vCPU situated at the head of the list, which corresponds to the runnable vCPU with the highest priority. If the run-queue is empty, the scheduler falls back to selecting the system IDLE vCPU.

When a vCPU wakes up, it is inserted into the run-queue, via a linear traversal, ahead of the first vCPU encountered with a strictly lower priority.

\subsubsection{Triggers}

Scheduling decisions and context switches are driven exclusively by two triggers:

\begin{itemize}

\item \textbf{Wake-up event (}\code{sched\_prio\_wake}\textbf{):} An asynchronous hardware interrupt or event wakes a sleeping vCPU, triggering its insertion into the local priority queue based on its priority level and prompting a preemption check against the currently running vCPU.

\item \textbf{Voluntary sleep / block / yield (}\code{sched\_prio\_sleep}\textbf{):} When the currently running vCPU completes its execution, yields, or blocks waiting for I/O, it is removed from the local run-queue, prompting the scheduler to pick the next highest-priority candidate at the head of the list.

\end{itemize}

\subsubsection{Admission control}

The Priority-based scheduler does not perform admission control or workload feasibility checks. While the build-time configuration schema (\code{types.xsd}) restricts configured priorities to two levels (\code{PRIO\_LOW} and \code{PRIO\_HIGH}), the C initialization routine in \code{sched\_prio.c} directly assigns the priority integer without checking for priority collisions or evaluating total pCPU capacity. It imposes no restrictions on the number of vCPUs assigned to a given priority tier or on aggregate CPU utilization. Consequently, the hypervisor accepts any valid vCPU configuration without rejecting overcommitted workloads, relying entirely on the system architect to manually bound execution demands offline to prevent the starvation of lower-priority vCPUs.

\subsubsection{Pseudo-code}

\begin{lstlisting}
"""
(*\textbf{Parameters}*):
    priority: Static integer assigned to each vCPU at creation time.
    prio_queue: Per-pCPU doubly linked list storing runnable vCPUs sorted in descending order of priority.
"""

(*\textbf{pick\_next}*)(prio_queue):   # sched_prio_pick_next()
    # Front-of-queue selection (runnable vCPU with highest priority)
    if not prio_queue_is_empty(prio_queue):
        return prio_queue_get_head(prio_queue)
    return IDLE_VCPU

(*\textbf{on\_wake}*)(vcpu, current_vcpu):   # sched_prio_wake()
    """Trigger: Asynchronous wake-up event / higher-priority preemption."""
    vcpu.state = RUNNABLE
    prio_queue_insert_sorted(prio_queue, vcpu)   # prio_queue_add()
    if vcpu.priority > current_vcpu.priority:
        trigger_preempt()

(*\textbf{on\_sleep}*)(vcpu):   # sched_prio_sleep()
    """Trigger: Voluntary yield / block / sleep."""
    vcpu.state = SLEEP
    prio_queue_remove(prio_queue, vcpu)   # prio_queue_remove()
    return pick_next(prio_queue)
\end{lstlisting}

\subsubsection{Pros}

The Priority-based scheduler ensures that critical vCPUs assigned to higher priority levels execute immediately without preemption from lower-priority workloads. However, any \textit{real-time guarantee} assumes that the system integrator bounds the execution demand of every vCPU at the same or higher priority levels. For interactive workloads, \textit{I/O responsiveness} is particularly good when assigned high priorities, enabling asynchronous wake-ups to preempt lower-priority entities instantly. As a \textit{work-conserving} algorithm, it promotes efficient resource utilization. A pCPU is never left idle as long as at least one runnable vCPU exists in the local queue. Regarding \textit{scheduler efficiency}, the algorithm exhibits minimal execution overhead during context switches, relying on straightforward head-of-queue extractions and integer comparisons. Finally, it provides \textit{partitioned} \textit{multicore support} by maintaining independent run-queues per pCPU.

\subsubsection{Cons}

Conversely, the primary drawback of this scheduler is its total lack of \textit{fair CPU sharing}, as an unconstrained high-priority vCPU can cause execution starvation for lower-priority vCPUs. This preemption model severely degrades \textit{temporal isolation} because a high-priority vCPU that misbehaves or runs indefinitely is never preempted or limited. Consequently, \textit{mixed-criticality} workloads remain susceptible to timing interference. System \textit{configurability} is limited: priority levels are static and assigned at vCPU creation, preventing dynamic priority adjustments during runtime without hypervisor reboots. Furthermore, priority assignment creates a trade-off between real-time guarantees and I/O responsiveness: elevating an interactive vCPU to improve I/O latency directly degrades the deadline guarantees of any lower-priority vCPU it preempts.

\subsubsection{Timeline}

The schedule under the Priority-based scheduler was derived from the rules presented in Section~\ref{sec:priority}. In accordance with ACRN's configuration schema (\code{types.xsd}), which exposes only two priority levels, vCPU\_RT and vCPU\_IO are both configured with \code{PRIO\_HIGH}, while vCPU\_CPU1 and vCPU\_CPU2 share \code{PRIO\_LOW}.

\begin{figure}[!htbp]
\centering
\includegraphics[width=\linewidth]{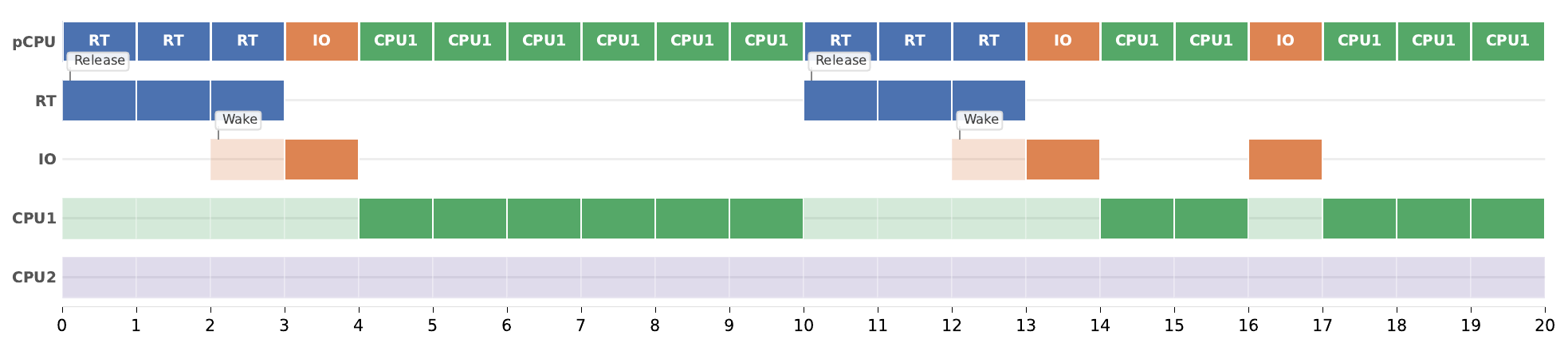}
\caption{Example timeline for Priority-based scheduler}\label{fig:7}
\end{figure}

At $t=0$, vCPU\_RT releases its first period and is selected as the highest-priority runnable vCPU.

At $t=2$, an I/O interrupt wakes vCPU\_IO. vCPU\_IO is inserted behind vCPU\_RT, since \code{prio\_queue\_add()} places a vCPU before the first one of strictly lower priority. The rescheduling requested by \code{wake\_thread()} selects the head of the queue, which is still vCPU\_RT.

At $t=3$, vCPU\_RT completes its 3\,ms execution requirement and transitions to sleep. The scheduler then selects vCPU\_IO from the head of the run-queue. vCPU\_IO executes from $t=3$ to $t=4$ and returns to sleep.

At $t=4$, with both higher-priority vCPUs asleep, the scheduler evaluates the remaining ready vCPUs. Because vCPU\_CPU1 was inserted prior to vCPU\_CPU2 among the equal-priority background vCPUs, vCPU\_CPU1 is selected at the head of the queue and executes continuously from $t=4$ to $t=10$. Because the algorithm lacks round-robin quantum expiration or time-slicing among equal-priority vCPUs, vCPU\_CPU1 never yields the pCPU, causing vCPU\_CPU2 to suffer total starvation during this interval.

At $t=10$, vCPU\_RT releases its second period and wakes up. Because vCPU\_RT has a higher priority than vCPU\_CPU1, it preempts vCPU\_CPU1. vCPU\_RT executes from $t=10$ to $t=13$.

At $t=12$, an I/O interrupt wakes vCPU\_IO. Since vCPU\_RT is already running at \code{PRIO\_HIGH}, no preemption occurs, and vCPU\_IO is queued behind vCPU\_RT.

At $t=13$, vCPU\_RT finishes its execution and sleeps. The scheduler selects vCPU\_IO, which executes from $t=13$ to $t=14$ before sleeping.

At $t=14$, control returns to vCPU\_CPU1, which executes from $t=14$ to $t=16$.

At $t=16$, a third I/O interrupt wakes vCPU\_IO. Because vCPU\_IO holds a higher priority than vCPU\_CPU1, vCPU\_IO preempts vCPU\_CPU1 and executes from $t=16$ to $t=17$ before returning to sleep.

Finally, at $t=17$, control returns once again to vCPU\_CPU1, which executes from $t=17$ to $t=20$, leaving vCPU\_CPU2 completely starved across the entire 20\,ms horizon.

\subsection{Borrowed Virtual Time (BVT) scheduler}\label{sec:bvt}

The BVT scheduler~\cite{duda1999bvt,acrn2026bvt} uses virtual time to provide weighted fair sharing of pCPU time, while a temporal “warp” mechanism enables low-latency dispatch for latency-sensitive workloads. Crucially, a discrepancy exists between the theoretical model and its actual codebase: while the configuration data structures define warp parameters, the ACRN kernel implementation leaves the warp calculation as an unfulfilled “TODO” stub.

\subsubsection{Parameters}

\textbf{Weight (}\code{data->weight}\textbf{):} A VM-level parameter (configured at build time, bounded between 1 and 128) defining the proportional share of pCPU time allocated to a vCPU relative to others. Higher weights make virtual time elapse slower.

\noindent\textbf{Actual Virtual Time (AVT)} (\texttt{data->avt}): A 64-bit counter tracking the cumulative execution time consumed by a vCPU, advanced proportionally to its inverse weight. AVT is updated exclusively during \texttt{sched\_bvt\_pick\_next()}.

\noindent\textbf{Scheduler Virtual Time (SVT)} (\texttt{svt}): A per-pCPU reference equal to the AVT of the run-queue head, recorded at the last invocation of \texttt{sched\_bvt\_pick\_next()} (\texttt{update\_svt()}). Since $\text{EVT} = \text{AVT}$ in the implementation, it is the minimum AVT among the runnable vCPUs of that pCPU at that instant. It is used to align the AVT of waking vCPUs after a sleep period and prevent CPU monopolization.

\noindent\textbf{Effective Virtual Time (EVT) (}\code{data->evt}\textbf{):} The primary metric used for scheduling decisions. Theoretically, $\mathit{EVT} = \mathit{AVT} - \mathit{warp}$. In the analyzed ACRN codebase, because the warp feature is marked as a “TODO”, EVT is assigned directly equal to AVT.

\noindent\textbf{Warp parameters (}\code{warp\_value}, \code{warp\_limit}, \code{unwarp\_period}\textbf{):} Theoretical parameters intended to allow latency-sensitive vCPUs to temporarily lower their EVT (warp) to gain prompt preemption, subject to a maximum warp execution limit and a mandatory unwarp recovery window. These mechanisms are unimplemented in ACRN's default BVT scheduler.

\noindent\textbf{Minimum Charging Unit (MCU) (}\code{BVT\_MCU\_MS}\textbf{):} The base hardware tick resolution (configured to 1\,ms) used to measure pCPU consumption and convert between physical time and virtual time.

\noindent\textbf{Context Switch Allowance (CSA) (}\code{BVT\_CSA\_MCU}\textbf{):} A configuration constant (default 5 MCU) that acts as a virtual time threshold (SVT - CSA) to limit how much virtual credit a sleeping vCPU can accumulate upon wake-up.

\subsubsection{Selection criterion}

The scheduler maintains a local per-pCPU run-queue sorted in ascending order of EVT. When selecting the next vCPU to execute, the scheduler picks the runnable vCPU positioned at the head of the list, which possesses the smallest EVT.

To prevent context-switch thrashing between vCPUs with close EVT values, a dynamic hardware timer is calculated based on the EVT difference between the head candidate and the second candidate in line, augmented by the CSA.

\subsubsection{Triggers}

Scheduling decisions and context switches in the BVT scheduler are driven by three triggers:

\begin{itemize}

\item \textbf{Wake-up event (}\code{sched\_bvt\_wake}\textbf{):} An asynchronous hardware interrupt or event wakes a sleeping vCPU. The scheduler updates the vCPU's AVT against SVT (as recorded at the last \code{pick\_next()}). If the AVT lags behind SVT – CSA, it is resynchronized to SVT to prevent long-sleeping vCPUs from claiming an excessive CPU share accumulated while inactive. It then inserts the vCPU into the run-queue (behind vCPUs with equal EVT) and issues a reschedule request.

\item \textbf{Voluntary sleep / block / yield (}\code{sched\_bvt\_sleep}\textbf{):} The blocking vCPU is removed from the run-queue and a rescheduling is requested.

\item \textbf{Countdown expiry (} \code{sched\_tick\_handler}\textbf{):} \code{sched\_bvt\_pick\_next()} arms a one-shot timer that expires after $(\mathit{EVT}_{\mathit{second}} - \mathit{EVT}_{\mathit{head}}) / \mathit{vt\_ratio}_{\mathit{head}} + \mathit{CSA}$~MCUs; its handler only requests a rescheduling.

\end{itemize}

\subsubsection{Admission control}

The BVT scheduler does not implement workload admission control or CPU capacity validation. Operating as a virtual-time-based fair-share algorithm, it allows an arbitrary number of vCPUs to be scheduled without checking aggregate CPU demand against physical execution capacity. During vCPU initialization, parameter validation is limited to sanitizing the assigned weight, which is bounded to a valid range between 1 and 128. Invalid or out-of-bounds weight inputs are silently adjusted to safe operational boundaries without raising error codes or rejecting vCPU instantiation, relying entirely on static build-time configuration to manage resource distribution.

\subsubsection{Pseudo-code}

\begin{lstlisting}
"""
(*\textbf{Parameters}*):
    weight: Proportional share weight (1 to 128) determining virtual time progression rate.
    MCU: 1 ms; CSA: 5 ms
    avt: Actual Virtual Time recording consumed pCPU time weighted by VM share.
    evt: Effective Virtual Time used as runqueue sorting key.
    svt: per-pCPU, avt of the run-queue head at the last decision.
"""

(*\textbf{pick\_next}*)(runqueue):   # sched_bvt_pick_next()
    if current is not IDLE:
        update_vt(current)   # update_vt()
        if current is runnable:
            reinsert_by_evt(current)
    cancel(countdown_timer)
    if not runqueue : return IDLE_VCPU
    update_svt()   #update_svt()
    first = runqueue.head
    if len(runqueue)>1:
        gap = runqueue[1].evt - first.evt
        arm_one_shot(countdown_timer, now + (gap / first.vt_ratio + CSA) * MCU)
    return first

(*\textbf{on\_countdown\_expiry}*) ():   # sched_tick_handler()
    """ Trigger: One-shot countdown timer expiration (quantum end)."""
    request_reschedule()

(*\textbf{on\_wake}*)(vcpu):   # sched_bvt_wake()
    """Trigger: Asynchronous I/O wake-up event."""
    vcpu.state = RUNNABLE
    if vcpu.avt <= svt - CSA:
        vcpu.avt = svt
    vcpu.evt = vcpu.avt
    #TODO: Add warp limit enforcement logic here
    #vcpu.evt = vcpu.avt - (vcpu.warp_value if vcpu.is_warped else 0)
    runqueue_insert_by_evt(vcpu)
    request_reschedule()

(*\textbf{on\_sleep}*)(vcpu):   # sched_bvt_sleep()
    """Trigger: Voluntary yield / block / sleep."""
    vcpu.state = SLEEP
    runqueue_remove(vcpu)
    request_reschedule()
\end{lstlisting}

\subsubsection{Pros}

First, the BVT scheduler supports \textit{fair CPU sharing} by distributing pCPU time in proportion to assigned weights through virtual time accounting. As a work-conserving algorithm, it contributes to effective \textit{resource utilization}, reallocating unconsumed pCPU cycles to other runnable vCPUs. Theoretically, the “warp” mechanism improves \textit{I/O responsiveness} by enabling preemption, while resetting AVT to SVT on wake-up prevents long-sleeping vCPUs from starving others. Finally, it provides native \textit{multicore support} through independent per-pCPU run-queues.

\subsubsection{Cons}

Conversely, the principal limitation of the BVT scheduler is its \textit{efficiency}, as maintaining a run-queue sorted by EVT requires $\mathcal{O}(N)$ linear traversal on every wake-up or re-insertion, compounded by frequent virtual-to-physical arithmetic conversions. Because scheduling is governed by virtual time decay rather than explicit deadlines, the algorithm cannot provide \textit{real-time guarantees}. Consequently, \textit{temporal isolation} and \textit{mixed-criticality support} are compromised, as interactive or computationally heavy vCPUs can introduce execution jitter into critical workloads. Furthermore, a major implementation discrepancy exists in current ACRN code: the warp mechanism is incomplete and explicitly marked as a “TODO”, forcing $\mathit{evt} = \mathit{avt}$. As a result, the code implementation fails to deliver the theoretical \textit{I/O responsiveness} gains provided by warping, causing BVT in practice to function as a standard weighted fair-share virtual time scheduler. Finally, vCPU weights are statically fixed at build time (via \code{config.xsd}), with no interface provided for dynamic runtime adjustment, meaning the system lacks runtime \textit{configurability}.

\subsubsection{Timeline}

The schedule under the BVT scheduler was derived from the rules presented in Section~\ref{sec:bvt}. Unlike the other algorithms, the timeline covers the window from $t=20$ to $t=40$ rather than $t=0$ to $t=20$, and the events of the reference workload are shifted accordingly: vCPU\_RT is released at $t=20$ and $t=30$, and the I/O interrupts for vCPU\_IO occur at $t=22$, $t=32$ and $t=36$. At initialization ($t=0$), all vCPUs start with an identical virtual time balance (AVT = EVT = 0). Since execution from $t=0$ exhibits an initial transient behavior that differs significantly from steady-state operation once virtual time drift has accumulated over several cycles, the illustration focuses on the $t=20$ to $t=40$ window. At $t=20$, cumulative execution has driven the SVT baseline to 160, yielding the initial virtual time states EVT\_CPU1 = 192, EVT\_CPU2 = 160, and EVT\_RT = 160 (clamped to SVT upon its release at $t=20$), while vCPU\_IO is sleeping. The first period of this window ($t=20$ to 30) still reflects the symmetric start-up, in which vCPU\_CPU1 and vCPU\_CPU2 have equal virtual times (160 at $t=19$). This tie, resolved in FIFO order, delays vCPU\_RT at $t=20$ and vCPU\_IO at $t=22$. From $t=30$ on, every waking vCPU is dispatched immediately, which is the steady-state behavior.

BVT manages vCPUs using virtual time tracking, where AVT advances during execution by \texttt{vt\_ratio} virtual-time units per ms of execution (with $\texttt{vt\_ratio} = 1024 / \text{weight}$, or normalized proportionally) and EVT determines run-queue ordering. Higher weights are assigned to vCPU\_RT and vCPU\_IO (weight = 128, yielding \code{vt\_ratio}= 8), while the background vCPUs receive lower weights (weight = 32, yielding \code{vt\_ratio}= 32). As a result, the background vCPUs advance their virtual time four times as fast as the real-time and I/O vCPUs when executing. Furthermore, when a vCPU at the head of the queue is selected, its execution quantum (in ms) is computed as $(\mathit{EVT}_{\mathit{next}} - \mathit{EVT}_{\mathit{head}}) / \mathit{vt\_ratio}_{\mathit{head}} + \mathit{CSA}$ where CSA=5\,ms.

\begin{figure}[!htbp]
\centering
\includegraphics[width=\linewidth]{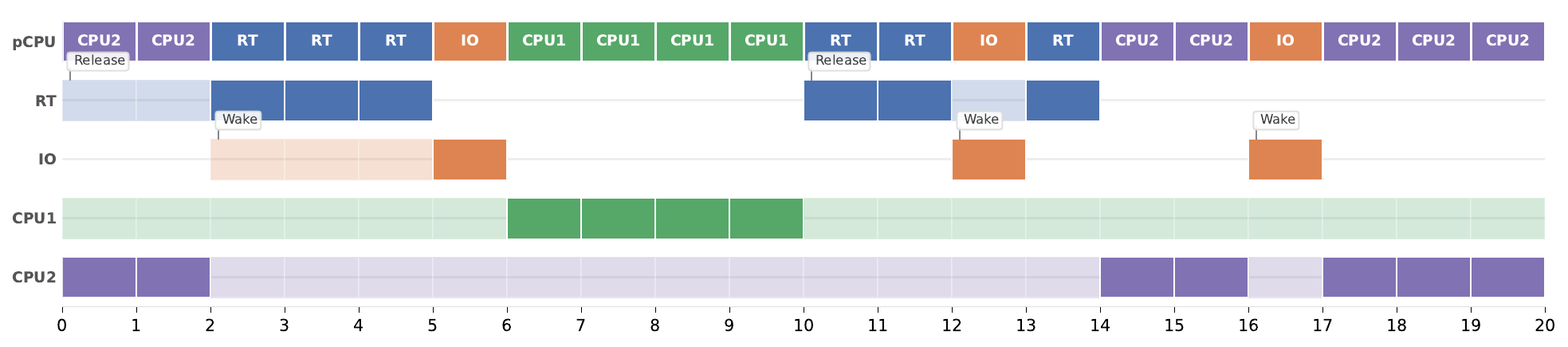}
\caption{Example timeline for BVT scheduler}\label{fig:8}
\end{figure}

At $t=20$, vCPU\_RT is released. Since it was blocked, the wake is effective: its AVT (48) is below the threshold, so it is clamped to the baseline $\mathit{SVT} = 160$. The run-queue is ordered vCPU\_CPU2 (160), RT (160), vCPU\_CPU1 (192). With the EVT tie at 160, vCPU\_CPU2 precedes vCPU\_RT under the FIFO insertion rule. vCPU\_CPU2 is elected for a quantum of 5\,ms.

At $t=22$, an interrupt wakes vCPU\_IO. Its AVT (24) is below $\mathit{SVT} - 5 = 155$, so it is clamped to $\mathit{SVT} = 160$. The run-queue becomes vCPU\_RT (160), vCPU\_IO (160), vCPU\_CPU1 (192), vCPU\_CPU2 (224). vCPU\_RT holds FIFO precedence over vCPU\_IO and is elected with a quantum of 5\,ms. It runs its full 3\,ms job from $t=22$ to $t=25$, advancing its EVT to 184 before sleeping.

At $t=25$, vCPU\_RT sleeps. vCPU\_IO holds the minimum EVT (160) and receives the pCPU. It executes for 1\,ms and goes back to sleep at $t=26$ with an EVT of 168.

At $t=26$, only the background vCPUs remain: vCPU\_CPU1 (EVT 192) and vCPU\_CPU2 (EVT 224). SVT is updated to 192, and vCPU\_CPU1, holding the minimum EVT, is elected with a quantum of 6\,ms.

At $t=30$, vCPU\_RT is released. It was blocked, so the wake is effective: its AVT (184) is not above SVT $-$ 5 = 187, so it is clamped to SVT = 192. The runqueue is vCPU\_RT (192), vCPU\_CPU2 (224), vCPU\_CPU1 (320), so vCPU\_RT is selected with a quantum of 9\,ms.

At $t=32$, an interrupt wakes vCPU\_IO, whose AVT (168) is clamped to SVT = 192. Since IO (192) now has a lower EVT than RT (208), vCPU\_IO preempts vCPU\_RT and runs. vCPU\_RT still has 1\,ms of work pending.

At $t=33$, vCPU\_IO completes its 1\,ms of work and sleeps with an EVT of 200. vCPU\_RT, now at the head of the runqueue (EVT 208), resumes and finishes its last millisecond at $t=34$, sleeping with an EVT of 216.

At $t=34$, only the background vCPUs remain. SVT is updated to 224 and vCPU\_CPU2, which holds the minimum EVT (224 vs. 320 for vCPU\_CPU1), is selected with a quantum of 8\,ms.

At $t=36$, an interrupt wakes vCPU\_IO. Its AVT (200) is below SVT $-$ 5 = 219, so it is clamped to SVT = 224. It holds the minimum EVT, preempting vCPU\_CPU2. It runs 1\,ms and sleeps at $t=37$ with an EVT of 232.

At $t=37$, vCPU\_CPU2 (EVT 288) is again at the head of the runqueue and resumes execution, with a quantum of 6\,ms.

\subsection{I/O sensitive Round Robin (IORR) scheduler}\label{sec:iorr}

The IORR scheduler is a hybrid scheduling algorithm designed to improve latency for I/O-bound vCPUs while maintaining egalitarian time-sharing for CPU-bound vCPUs. This scheduling model is implemented in the ACRN hypervisor~\cite{acrn2026iorr}.

\subsubsection{Parameters}

\textbf{Time slice (}\code{CONFIG\_SLICE\_MS}\textbf{):} The default execution quantum (hardcoded to 10\,ms) allocated to each vCPU before Round-Robin preemption and tail rotation occur.

\noindent\textbf{Per-vCPU execution counters (}\code{data->left\_cycles} , \code{data->last\_cycles}\textbf{):} Internal tracking variables maintained per vCPU to measure remaining execution time within the active time slice.

\noindent\textbf{Per-pCPU run-queue (}\code{INIT\_LIST\_HEAD(\&iorr\_ctl->runqueue)}\textbf{):} A doubly linked list maintained independently per pCPU that stores runnable vCPUs.

\subsubsection{Selection criterion}

The IORR scheduler uses a hybrid wake-up policy that places awakened vCPUs at the front of the queue and rotates the remaining vCPUs in Round-Robin order. The scheduler inspects the local per-pCPU run-queue and selects the runnable vCPU situated at the head of the list. If the run-queue is empty, the scheduler falls back to selecting the system IDLE vCPU.

To differentiate between I/O-bound and CPU-bound workloads without requiring manual priority configuration, the scheduler modifies queue insertion points: waking vCPUs are inserted directly at the front of the run-queue, whereas currently running vCPUs, whether preempted by a waking vCPU or having exhausted their 10\,ms time slice, are rotated to the tail of the list.

\subsubsection{Triggers}

Scheduling decisions and context switches in the IORR scheduler are driven by three triggers:

\begin{itemize}

\item \textbf{Wake-up event (}\code{sched\_iorr\_wake}\textbf{):} An asynchronous hardware interrupt or event wakes a sleeping vCPU. The scheduler inserts the waking vCPU directly at the head of the local run-queue and issues a reschedule request to execute it promptly.

\item \textbf{Voluntary sleep / block / yield (}\code{sched\_iorr\_sleep}\textbf{):} When the currently running vCPU completes its execution or blocks on I/O, it is removed from the run-queue, prompting the scheduler to pick the next candidate waiting at the head of the list.

\item \textbf{Time slice expiration / tick handler (}\code{sched\_tick\_handler}\textbf{):} A periodic per-pCPU timer interrupt (1\,ms tick) decrements the running vCPU's left cycles. When it becomes equal to 0, the scheduler replenishes its allocation, removes it from the head, and moves it to the tail of the run-queue.

\end{itemize}

\subsubsection{Admission control}

The IORR scheduler does not perform admission control or workload feasibility checks. During vCPU initialization, the scheduler ignores external configuration parameters and statically sets every vCPU's execution quantum to a hardcoded compile-time value (\code{CONFIG\_SLICE\_MS}=10\,ms). It enforces no validation checks on vCPU parameters, imposes no limits on the number of active vCPUs, and does not evaluate aggregate CPU utilization. Consequently, the hypervisor accepts any vCPU configuration without raising errors or rejecting overcommitted workloads, relying entirely on static system design to bound execution requirements offline.

\subsubsection{Pseudo-code}

\begin{lstlisting}
"""
(*\textbf{Parameters}*):
    time_slice: Default execution quantum allocated to each vCPU before Round-Robin tail rotation (10 ms).
    left_cycles: Remaining execution time counter maintained per vCPU for the active time slice.
    runqueue: Per-pCPU doubly linked list storing active runnable vCPUs.
"""

(*\textbf{pick\_next}*)(runq, current_vcpu):   # sched_iorr_pick_next()
    # Front-of-queue selection; rotate current vCPU to tail if time slice exhausted
    if current_vcpu and current_vcpu.is_runnable():
        if current_vcpu.left_cycles <= 0:
            current_vcpu.left_cycles += current_vcpu.time_slice
        runq_move_to_tail(runq, current_vcpu)
    if not runq_is_empty(runq):
        next_vcpu = runq_get_head(runq)
        if next_vcpu.left_cycles <= 0:
            next_vcpu.left_cycles += next_vcpu.time_slice
        return next_vcpu
    return IDLE_VCPU

(*\textbf{on\_tick}*)(current_vcpu, delta_time):   # sched_tick_handler()
    """Trigger: Per-pCPU timer tick handler (1 ms)."""
    current_vcpu.left_cycles -= delta_time
    if current_vcpu.left_cycles <= 0:
        return pick_next(runq, current_vcpu)

(*\textbf{on\_wake}*)(vcpu):   # sched_iorr_wake()
    """Trigger: Asynchronous I/O wake-up event."""
    vcpu.state = RUNNABLE
    runqueue_add_head(runq, vcpu)   # runqueue_add_head()
    return pick_next(runq, current_vcpu)

(*\textbf{on\_sleep}*)(vcpu):   # sched_iorr_sleep()
    """Trigger: Voluntary yield / block / sleep."""
    vcpu.state = SLEEP
    runqueue_remove(runq, vcpu)   # runqueue_remove()
    return pick_next(runq, None)
\end{lstlisting}

\subsubsection{Pros}

The IORR scheduler achieves high \textit{I/O responsiveness} for event-driven workloads by inserting waking vCPUs directly at the head of the queue for preemption. As a \textit{work-conserving} algorithm, it supports efficient resource utilization by instantly reassigning idle cycles. Regarding \textit{scheduler efficiency}, it operates with constant-time complexity via simple pointer manipulations without complex arithmetic. Finally, it provides \textit{partitioned} \textit{multicore support} through independent per-pCPU run-queues.

\subsubsection{Cons}

Conversely, IORR lacks \textit{temporal isolation}, as any vCPU that sleeps and wakes up ~is automatically placed at the head of the queue regardless of its actual importance, allowing frequent I/O-bound vCPUs to monopolize the pCPU. This unconstrained preemption eliminates \textit{real-time guarantees}. Consequently, \textit{mixed-criticality} workloads remain susceptible to interference from I/O-bound vCPUs, potentially affecting the execution of safety-critical workloads. Furthermore, the algorithm lacks \textit{fair CPU sharing} mechanisms, making proportional pCPU time distribution impossible. Finally, system \textit{configurability} is restricted because the 10\,ms time slice parameter is hardcoded in the hypervisor source code and cannot be modified dynamically at runtime.

\subsubsection{Timeline}

The schedule under the IORR scheduler was derived from the rules presented in Section~\ref{sec:iorr}. No algorithm-specific parameters are required for this trace.

\begin{figure}[!htbp]
\centering
\includegraphics[width=\linewidth]{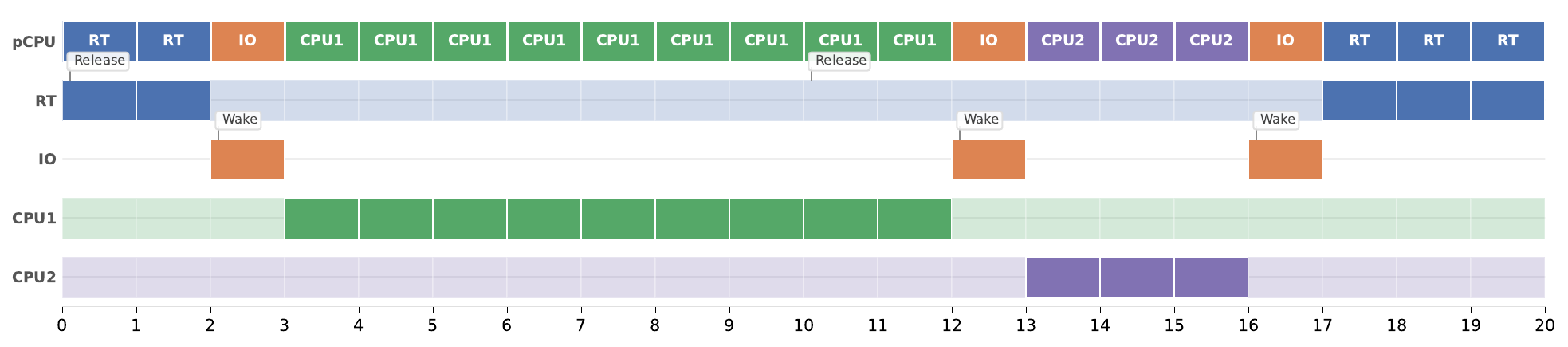}
\caption{Example timeline for IORR scheduler}\label{fig:9}
\end{figure}

At $t=0$, vCPU\_RT releases its first period, takes the pCPU, and executes from $t=0$ to $t=2$.

At $t=2$, an asynchronous I/O interrupt wakes vCPU\_IO. The scheduler places vCPU\_IO at the head of the runqueue and preempts vCPU\_RT. The preempted vCPU\_RT is appended to the tail of the runqueue. vCPU\_IO executes from $t=2$ to $t=3$ and transitions to a sleeping state.

At $t=3$, the scheduler selects the vCPU at the head of the run-queue. Because vCPU\_RT was moved to the tail at $t=2$, vCPU\_CPU1 resides at the head and takes the pCPU. Then, it executes from $t=3$ through $t=10$.

The job released at $t=0$ has not completed by its implicit deadline $t=10$\textit{.} Because vCPU\_RT is currently runnable in the run-queue rather than sleeping, no wake-up boost is granted, and vCPU\_RT remains near the tail of the run-queue without preempting vCPU\_CPU1. Furthermore, because vCPU\_CPU1's round-robin time slice (\code{CONFIG\_SLICE\_MS} = 10\,ms) has not expired, vCPU\_CPU1 continues executing from $t=10$ to $t=12$.

At $t=12$, a second I/O interrupt wakes vCPU\_IO, which preempts vCPU\_CPU1 and moves vCPU\_CPU1 to the tail of the runqueue. vCPU\_IO executes from $t=12$ to $t=13$ before returning to sleep.

At $t=13$, the scheduler selects vCPU\_CPU2 from the head of the runqueue, which executes from $t=13$ to $t=16$. At $t=16$, a third I/O interrupt wakes vCPU\_IO, preempting vCPU\_CPU2 and appending vCPU\_CPU2 to the tail of the runqueue. vCPU\_IO executes from $t=16$ to $t=17$ before returning to sleep.

Finally, at $t=17$, vCPU\_RT first completes the remaining 1\,ms of job 1, which had already missed its deadline at $t=10$. It then starts job 2, which has only received 2 of its 3\,ms by $t=20$, so job 2 also misses its deadline.

\subsection{PRR (Priority Round Robin) scheduler}\label{sec:prr}

The PRR scheduler is a hybrid scheduling algorithm that combines a static priority policy with a Round-Robin mechanism for vCPUs operating at the same priority level. This scheduling model is implemented in the Xvisor hypervisor~\cite{xvisor2026repo,xvisor2026prr}.

\subsubsection{Parameters}

\textbf{Priority (}\code{vcpu->priority}\textbf{):} A static integer assigned to each vCPU at creation time, ranging from 0 to 7, defining its relative execution tier.

\noindent\textbf{Time slice (}\code{vcpu->time\_slice}\textbf{):} A per-vCPU temporal execution quantum defined at setup that dictates the maximum continuous runtime allowed before Round-Robin preemption and tail rotation.

\noindent\textbf{Priority run-queue (}\code{rqi->list}\textbf{):} An array of doubly linked list heads maintained per run-queue, indexed from 0 to 7, where each entry holds the active runnable vCPUs belonging to the associated priority level.

\subsubsection{Selection criterion}

The PRR scheduler applies a priority policy with Round-Robin head extraction. When a pCPU requires a new vCPU, the scheduler iterates through its priority run-queue array starting from the maximum priority index down to 0 to locate the first non-empty list.

Upon identifying the highest active priority tier, the scheduler extracts the first vCPU waiting at the head of the list and assigns it an execution timer matching its time slice. If all priority lists are empty, the lookup fails, prompting the system to select the system IDLE vCPU.

\subsubsection{Triggers}

Scheduling decisions and context switches in the PRR scheduler are driven by four triggers:

\begin{itemize}

\item \textbf{Wake-up event (}\code{vmm\_schedalgo\_rq\_enqueue}\textbf{):} An asynchronous hardware interrupt or event wakes a sleeping vCPU. The scheduler enqueues the vCPU at the tail of the list matching its priority level and triggers a preemption check.

\item \textbf{Voluntary sleep / block / yield (}\code{vmm\_schedalgo\_rq\_detach}\textbf{):} When a running vCPU completes its task or blocks on I/O, it is removed from its priority list, prompting the scheduler to select the next highest-priority candidate.

\item \textbf{Time slice expiration:} A hardware timer interrupt fires when the currently running vCPU exhausts its assigned time slice. The vCPU is preempted, detached from the head, and re-enqueued at the tail of its priority list, forcing a Round-Robin rotation to the next vCPU of identical priority.

\item \textbf{Preemption trigger (}\code{vmm\_schedalgo\_rq\_prempt\_needed}\textbf{):} When a higher-priority vCPU wakes up, the scheduler detects that a non-empty list exists at a priority level strictly greater than that of the currently running vCPU, forcing preemption.

\end{itemize}

\subsubsection{Admission control}

The PRR scheduler does not perform a workload feasibility test or evaluate aggregate CPU utilization, allowing an arbitrary number of vCPUs to be assigned to any priority level. At the algorithm level (\code{vmm\_schedalgo\_prr.c}), initialization checks are confined to defensive null-pointer and heap allocation verifications, returning an error code only if memory allocation fails. Parameters are, however, sanitized by \code{vmm\_manager.c} when a vCPU is created from the device tree: the priority is clamped to [0, 7], a null time slice is replaced by the default (10\,ms), the deadline is raised to at least the time slice, and the periodicity to at least the deadline (default deadline 100\,ms, default periodicity 1s). Inconsistent parameters are thus silently adjusted rather than rejected, leaving the responsibility for preventing priority-tier starvation to offline system integration.

\subsubsection{Pseudo-code}

\begin{lstlisting}
"""
(*\textbf{Parameters}*):
    priority: Static integer assigned to each vCPU at creation time (0 to 7).
    time_slice: Per-vCPU execution quantum dictating maximum continuous runtime before Round-Robin tail rotation.
    priority_rq: Array of priority runqueues indexed from 0 to 7.
"""

(*\textbf{pick\_next}*)(priority_rq):   # vmm_schedalgo_rq_dequeue()
    # Highest active priority tier lookup with Round-Robin head extraction
    for p in range(MAX_PRIORITY, -1, -1):
        if not priority_rq_is_empty(priority_rq[p]):
            return priority_rq_get_head(priority_rq[p])
    return IDLE_VCPU

(*\textbf{on\_tick}*)(current_vcpu, delta_time):   # vmm_schedalgo_rq_dequeue()
    """Trigger: Time slice expiration."""
    current_vcpu.time_slice_left -= delta_time
    if current_vcpu.time_slice_left <= 0:
        current_vcpu.time_slice_left = current_vcpu.time_slice
        priority_rq_move_to_tail(current_vcpu)
        return pick_next(priority_rq)

(*\textbf{on\_wake}*)(vcpu, current_vcpu):   # vmm_schedalgo_rq_enqueue()
    """Trigger: Asynchronous wake-up event."""
    vcpu.state = RUNNABLE
    priority_rq_add_tail(vcpu)
    if vcpu.priority > current_vcpu.priority:
        trigger_preempt()

(*\textbf{on\_sleep}*)(vcpu):   # vmm_schedalgo_rq_detach()
    """Trigger: Voluntary yield / block / sleep."""
    vcpu.state = SLEEP
    priority_rq_remove(vcpu)
    return pick_next(priority_rq)
\end{lstlisting}

\subsubsection{Pros}

The PRR scheduler ensures that higher-priority vCPUs execute without preemption from lower-priority ones. However, any \textit{real-time guarantee} assumes that the system integrator bounds the execution demand of every vCPU at the same or higher priority levels, as the scheduler itself does not enforce budget caps or temporal isolation. For interactive workloads configured at high priority tiers, \textit{I/O responsiveness} is strong, as asynchronous interrupt wake-ups place vCPUs into their priority queue and quickly trigger preemption against lower-priority running vCPUs. Regarding \textit{scheduler efficiency}, the algorithm achieves constant time list manipulations for insertion and removal, utilizing direct array indexing that avoids complex list sorting. Its \textit{work-conserving} nature also helps keep pCPUs busy whenever runnable vCPUs are available. Finally, it provides \textit{partitioned} \textit{multicore support} by maintaining independent per-pCPU priority run-queue structures.

\subsubsection{Cons}

Conversely, the PRR scheduler lacks \textit{fair CPU sharing} across different priority tiers, as it lacks global weight or credit mechanisms to distribute pCPU cycles in fixed proportions. Its preemption model compromises \textit{temporal isolation} because a high-priority vCPU that misbehaves or runs indefinitely can monopolize the pCPU. Consequently, \textit{mixed-criticality} workloads remain susceptible to timing interference unless they are carefully partitioned or their priority assignments are appropriately managed by the system architect. Furthermore, static priority assignment creates a design trade-off between real-time guarantees and I/O responsiveness: elevating an interactive vCPU to a high priority tier to minimize I/O latency directly degrades the deadline guarantees of any lower-priority critical workloads it preempts. Finally, system \textit{configurability} is restricted because scheduling parameters are static and assigned at vCPU creation time, preventing dynamic adjustments during hypervisor runtime without system reboots.

\subsubsection{Timeline}

The schedule under the PRR scheduler was derived from the rules presented in Section~\ref{sec:prr}. The vCPUs are assigned to specific priority levels and individual Round-Robin time slices: vCPU\_RT has the highest priority (\code{time\_slice} = 4), vCPU\_IO the second highest (\code{time\_slice} = 1), and vCPU\_CPU1 and vCPU\_CPU2 share an identical, lower background priority level (\code{time\_slice} = 2).

\begin{figure}[!htbp]
\centering
\includegraphics[width=\linewidth]{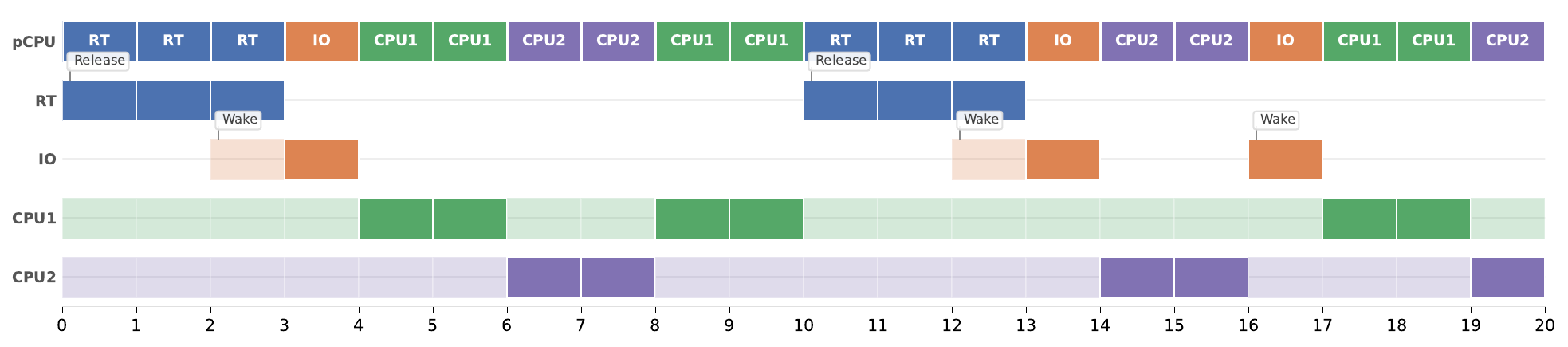}
\caption{Example timeline for PRR scheduler}\label{fig:10}
\end{figure}

At $t=0$, vCPU\_RT releases its first period, is enqueued, and is picked as the highest-priority runnable vCPU.

At $t=2$, an I/O interrupt wakes vCPU\_IO. Because vCPU\_RT holds a higher priority than vCPU\_IO, vCPU\_IO remains queued.

At $t=3$, vCPU\_RT completes its 3\,ms requirement and sleeps. The scheduler selects vCPU\_IO from the highest non-empty priority queue. vCPU\_IO executes from $t=3$ to $t=4$ to handle its I/O event before sleeping.

At $t=4$, with both higher-priority vCPUs asleep, the scheduler selects from the background priority queue. vCPU\_CPU1 is selected and executes for its 2\,ms time slice from $t=4$ to $t=6$.

At $t=6$, vCPU\_CPU1's quantum expires. It is enqueued at the tail of the background queue, allowing vCPU\_CPU2 (now at the head) to execute for its 2\,ms quantum from $t=6$ to $t=8$.

At $t=8$, vCPU\_CPU2's quantum expires, and vCPU\_CPU1 resumes for another 2\,ms time slice from $t=8$ to $t=10$.

At $t=10$, vCPU\_RT releases its second period. Because vCPU\_RT's priority is the highest, it triggers preemption. vCPU\_RT executes from $t=10$ to $t=13$.

At $t=12$, an I/O interrupt wakes vCPU\_IO, but no preemption occurs as vCPU\_RT maintains higher priority.

At $t=13$, vCPU\_RT completes its execution and sleeps. The scheduler selects vCPU\_IO, which executes from $t=13$ to $t=14$ before returning to sleep.

At $t=14$, control returns to the background priority queue. Because vCPU\_CPU2 was at the head of the queue, it is selected and executes its 2\,ms quantum from $t=14$ to $t=16$.

At $t=16$, a third I/O interrupt wakes vCPU\_IO. Because vCPU\_IO holds a higher priority than the background vCPUs, it preempts and executes from $t=16$ to $t=17$ before returning to sleep.

At $t=17$, control returns to the background queue; since vCPU\_CPU2's quantum expired at $t=16$, vCPU\_CPU1 is now at the head of the queue and executes its 2\,ms quantum from $t=17$ to $t=19$.

Finally, at $t=19$, vCPU\_CPU1's quantum expires, allowing vCPU\_CPU2 to resume execution.

\subsection{PRM (Priority Rate Monotonic) scheduler}\label{sec:prm}

The PRM scheduler is a hybrid scheduling algorithm that combines a priority policy with a Rate Monotonic mechanism for vCPUs sharing the same priority level. This scheduling model is implemented in the Xvisor hypervisor~\cite{xvisor2026repo,xvisor2026prm}.

\subsubsection{Parameters}

\textbf{Priority (}\code{vcpu->priority}\textbf{):} A static integer assigned to each vCPU at creation time (ranging from 0 to 7), defining its primary execution tier.

\noindent\textbf{Periodicity (}\code{vcpu->periodicity}\textbf{):} A per-vCPU temporal parameter representing the vCPU's periodicity duration, used to order vCPUs within the same priority level according to the Rate Monotonic principle (shorter periods yield higher precedence).

\noindent\textbf{Time Slice (}\code{vcpu->time\_slice}\textbf{):} A temporal execution quantum allocated to a vCPU before preemption and re-insertion into its priority structure occur.

\noindent\textbf{Per-priority Red-Black trees (}\code{rqi->root}\textbf{):} An array of Red-Black tree roots maintained per run-queue, indexed by priority level (from 0 to 7), where each tree stores active runnable vCPUs sorted in ascending order of their periodicity.

\subsubsection{Selection criterion}

The PRM scheduler applies a highest-priority lookup with Rate Monotonic Red-Black tree extraction. When a pCPU requires a new vCPU, the scheduler iterates through its priority array starting from the maximum priority index down to 0 to locate the first non-empty priority tier.

Once the highest active priority tier is identified, the scheduler extracts the leftmost node from the corresponding Red-Black tree, which represents the runnable vCPU with the shortest periodicity. If all priority trees are empty, the lookup fails, prompting the system to select the system IDLE vCPU.

\subsubsection{Triggers}

Scheduling decisions and context switches in the PRM scheduler are driven by four triggers:

\begin{itemize}

\item \textbf{Wake-up event (}\code{vmm\_schedalgo\_rq\_enqueue}\textbf{):} An asynchronous hardware interrupt or event wakes a sleeping vCPU. The scheduler inserts the vCPU into the Red-Black tree corresponding to its priority level based on its periodicity and triggers a preemption check.

\item \textbf{Voluntary sleep / block / yield (}\code{vmm\_schedalgo\_rq\_detach}\textbf{):} When a running vCPU completes its task or blocks waiting for I/O, it is removed from its priority tree, prompting the scheduler to select the next highest-priority, shortest-periodicity candidate.

\item \textbf{Time slice expiration:} A hardware timer fires when the currently running vCPU exhausts its assigned time slice. The vCPU is preempted and re-inserted into its Red-Black tree, forcing a context switch to the next eligible vCPU.

\item \textbf{Preemption trigger (}\code{vmm\_schedalgo\_rq\_prempt\_needed}\textbf{):} When a vCPU wakes up, the algorithm evaluates if preemption is required. Due to the codebase implementation gap (\code{/* TODO: check lowest periodicity of highest priority with vcpu */}), preemption is strictly triggered only when a non-empty tree exists at a priority tier strictly greater than that of the currently running vCPU. Preemption is not triggered for vCPUs at the same priority level, even if the newly awakened vCPU possesses a shorter periodicity.

\end{itemize}

\subsubsection{Admission control}

PRM does not perform a workload feasibility test: although grounded in Rate Monotonic scheduling theory, PRM omits classical schedulability checks (e.g., verifying that cumulative CPU utilization satisfies Liu and Layland's bound~\cite{liu1973scheduling}). At the algorithm level (\code{vmm\_schedalgo\_prm.c}), vCPU setup validation is confined to defensive null-pointer and dynamic memory allocation checks. Parameters are, however, sanitized by \code{vmm\_manager.c} when a guest vCPU is created from the device tree: the priority is clamped to [0, 7], a null time slice is replaced by the default (10\,ms), the deadline is raised to at least the time slice, and the periodicity to at least the deadline (default deadline 100\,ms, default periodicity 1s). Inconsistent parameters are thus silently adjusted rather than rejected, shifting the full responsibility for timing feasibility and starvation prevention to offline system integration.

\subsubsection{Pseudo-code}

\begin{lstlisting}
"""
(*\textbf{Parameters}*):
    priority: Static integer assigned to each vCPU at creation time (0 to 7), defining its primary execution tier.
    periodicity: Temporal parameter representing vCPU periodicity duration; shorter periods give higher precedence (Rate Monotonic).
    time_slice: Temporal execution quantum allocated to a vCPU before preemption and tree re-insertion.
    priority_rbtree: Array of Red-Black tree roots indexed from 0 to 7, storing runnable vCPUs sorted by ascending periodicity.
"""

(*\textbf{pick\_next}*)(priority_rbtree):   # vmm_schedalgo_rq_dequeue()
    # Highest priority tier lookup with Rate Monotonic (shortest periodicity) tree extraction
    for p in range(MAX_PRIORITY, -1, -1):
        if not rbtree_is_empty(priority_rbtree[p]):
            return rbtree_get_min_periodicity(priority_rbtree[p])
    return IDLE_VCPU

(*\textbf{on\_tick}*)(current_vcpu, delta_time):   # vmm_scheduler_next()
    """Trigger: Time slice expiration / hardware timer interrupt."""
    rbtree_reinsert_by_periodicity(current_vcpu)
    return pick_next(priority_rbtree)

(*\textbf{on\_wake}*)(vcpu, current_vcpu):   # vmm_schedalgo_rq_enqueue()
    """Trigger: Asynchronous wake-up event / higher-priority preemption."""
    vcpu.state = RUNNABLE
    rbtree_insert_by_periodicity(vcpu)
    if vcpu.priority > current_vcpu.priority:   # vmm_schedalgo_rq_prempt_needed()
        trigger_preempt()

(*\textbf{on\_sleep}*)(vcpu):   # vmm_schedalgo_rq_detach()
    """Trigger: Voluntary yield / block / sleep."""
    vcpu.state = SLEEP
    rbtree_remove(vcpu)
    return pick_next(priority_rbtree)
\end{lstlisting}

\subsubsection{Pros}

First, PRM orders vCPUs within each priority level by periodicity, systematically prioritizing vCPUs with shorter periods. However, any \textit{real-time guarantee} assumes that the system integrator bounds the execution demand of every vCPU at the same or higher priority levels, as the scheduler itself does not enforce budget caps or vCPU isolation. By utilizing per-priority Red-Black trees, the algorithm achieves a logarithmic time complexity for insertion and removal operations, maintaining bounded \textit{scheduler efficiency}. For interactive workloads, \textit{I/O responsiveness} can be strong provided they are assigned high-priority tiers. As a \textit{work-conserving} policy, it keeps pCPUs occupied whenever runnable vCPUs are available. Finally, it provides \textit{partitioned} \textit{multicore support} by maintaining independent per-pCPU run-queue structures.

\subsubsection{Cons}

One principal limitation of the PRM scheduler is its lack of \textit{fair CPU sharing} across priority tiers. Also, preemption compromises \textit{temporal isolation}, because a high-priority vCPU or a vCPU with an extremely short periodicity can monopolize the pCPU and cause execution starvation for lower-priority vCPUs. Consequently, \textit{mixed-criticality} workloads remain susceptible to interference caused by misbehaving vCPUs. Furthermore, because the scheduler leaves same-priority periodicity evaluation as an unfulfilled TODO stub, a newly awakened vCPU with a shorter periodicity cannot preempt a running vCPU at the same priority tier, degrading real-time responsiveness for peer periodic vCPUs. Static priority and periodicity assignments also create a design trade-off between real-time guarantees and I/O responsiveness: elevating an interactive vCPU to a high priority tier directly degrades the deadline guarantees of lower-priority critical workloads. Finally, system \textit{configurability} is restricted because scheduling parameters are static and assigned at vCPU creation time, preventing adjustments during hypervisor runtime without system reboots.

\subsubsection{Timeline}

The schedule under the PRM scheduler was derived from the rules presented in Section~\ref{sec:prm}. The vCPUs are assigned to specific priority levels, periods, and time slices: vCPU\_RT has the highest static priority (deadline = periodicity = 10, \code{time\_slice} = 4), vCPU\_IO the second highest (deadline = periodicity = 6, \code{time\_slice} = 1), and vCPU\_CPU1 (deadline = periodicity = 10, \code{time\_slice} = 2) and vCPU\_CPU2 (deadline = periodicity = 15, \code{time\_slice} = 2) share an identical, lower priority level.

\begin{figure}[!htbp]
\centering
\includegraphics[width=\linewidth]{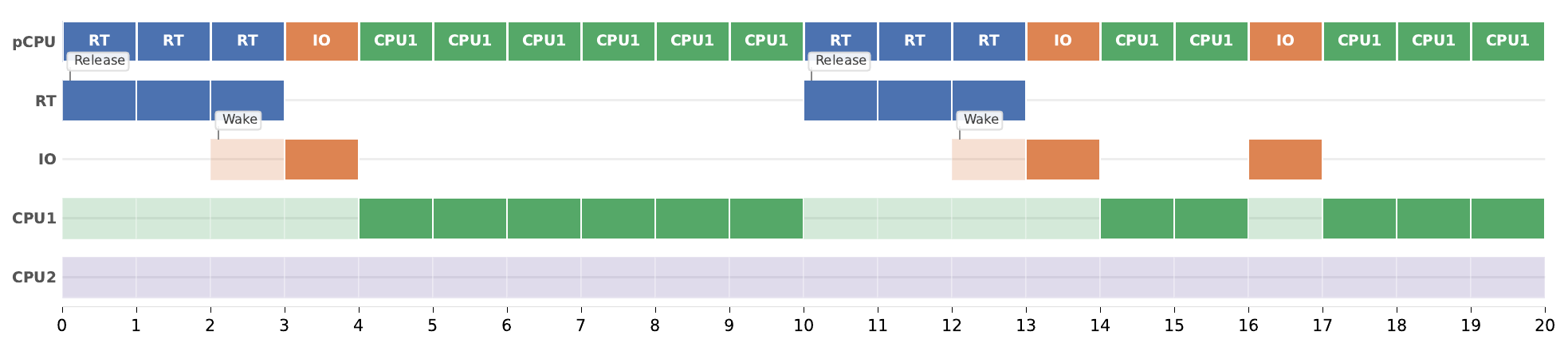}
\caption{Example timeline for PRM scheduler}\label{fig:11}
\end{figure}

At $t=0$, vCPU\_RT releases its first period, is inserted into the run-queue, and is selected as the highest priority vCPU.

At $t=2$, an I/O interrupt wakes vCPU\_IO. Because vCPU\_RT holds a higher priority than vCPU\_IO, vCPU\_IO is inserted into its priority level's red-black tree.

At $t=3$, vCPU\_RT completes its 3\,ms execution requirement and sleeps. The scheduler selects vCPU\_IO from the highest non-empty priority tree. vCPU\_IO executes from $t=3$ to $t=4$ before sleeping.

At $t=4$, the scheduler extracts the leftmost node of the background tree, vCPU\_CPU1 (periodicity 10). Each time its time slice expires, vCPU\_CPU1 is re-inserted into the tree according to its periodicity and, being shorter than that of vCPU\_CPU2 (15), it is again the leftmost node and is immediately re-selected. Then, vCPU\_CPU2 starves for as long as vCPU\_CPU1 remains runnable.

At $t=10$, vCPU\_RT releases its second period. Because vCPU\_RT's priority is the highest, it triggers preemption. vCPU\_RT executes from $t=10$ to $t=13$.

At $t=12$, an I/O interrupt wakes vCPU\_IO, but no preemption occurs as vCPU\_RT maintains higher priority.

At $t=13$, vCPU\_RT completes its 3\,ms execution time and sleeps. The scheduler selects vCPU\_IO, which executes from $t=13$ to $t=14$ before sleeping.

At $t=14$, vCPU\_CPU1 (periodicity = 10) is selected once again over vCPU\_CPU2 (periodicity = 15) due to its shorter periodicity in the red-black tree sorting, executing from $t=14$ to $t=16$.

At $t=16$, a third I/O interrupt wakes vCPU\_IO. Because vCPU\_IO holds a higher priority than vCPU\_CPU1, vCPU\_IO preempts vCPU\_CPU1 and executes from $t=16$ to $t=17$ before returning to sleep. Finally, at $t=17$, control returns once again to vCPU\_CPU1, which executes continuously from $t=17$ to $t=20$ due to its shorter periodicity in the red-black tree, leaving vCPU\_CPU2 completely starved over the entire 20\,ms horizon.

\subsection{Synthesis of the reference schedules}\label{sec:synthesis}

Table~\ref{tab:3} summarizes the reference schedules derived above over 20\,ms. These values are not performance measurements: they result from a single workload, specific parameter choices and given initial state. Their purpose is to make visible, on a common example, the mechanisms identified in the code: reserved windows left idle (ARINC653), the starvation of a vCPU sharing a priority level (Priority-based, PRM), the precedence given to any waking vCPU (IORR), the deferrable-server latency of a vCPU waking with an exhausted budget (RTDS), and the dependence of wake-up latency on accumulated state (Credit2, BVT).

\begin{table}[!htbp]
\caption{Analytical comparison of the reference schedules derived under the workload scenario described in Section~\ref{sec:reference-workload}. Metrics report real-time job deadline fulfillment (Done/Missed), wake-up latencies for sporadic I/O events, total execution time for background vCPUs (CPU1, CPU2), and total pCPU idle time (all latencies and execution times are expressed in ms).}\label{tab:3}
\begin{tabular*}{\tblwidth}{@{}LCCCCCCCC@{}}
\toprule
 & \multicolumn{2}{c}{RT job deadline} & \multicolumn{3}{c}{IO wake-up latency} & \multicolumn{2}{c}{Execution time} & \\
\cmidrule{2-3}\cmidrule{4-6}\cmidrule{7-8}
Scheduler & Job 1 & Job 2 & Wake-up 1 & Wake-up 2 & Wake-up 3 & CPU1 & CPU2 & Idle time \\
\midrule
ARINC653 & Done & Done & 2 & 2 & 8 & 4 & 4 & 4 \\
Credit & Done & Done & 1 & 1 & 0 & 9 & 2 & 0 \\
Credit2 & Done & Done & 1 & 0 & 0 & 4 & 7 & 0 \\
RTDS & Done & Done & 0 & 0 & 2 & 6 & 5 & 0 \\
Priority-based & Done & Done & 1 & 1 & 0 & 11 & 0 & 0 \\
BVT & Done & Done & 3 & 0 & 0 & 4 & 7 & 0 \\
IORR & Missed & Missed & 0 & 0 & 0 & 9 & 3 & 0 \\
PRR & Done & Done & 1 & 1 & 0 & 6 & 5 & 0 \\
PRM & Done & Done & 1 & 1 & 0 & 11 & 0 & 0 \\
\bottomrule
\end{tabular*}
\end{table}

Table~\ref{tab:3} highlights the trade-offs between the schedulers on this workload. Only IORR misses the real-time deadlines, whereas all other schedulers meet them. Under IORR, vCPU\_RT is preempted once at $t=2$ by the waking vCPU\_IO and rotated to the tail of the run-queue behind vCPU\_CPU1 and vCPU\_CPU2. Consequently, vCPU\_CPU1 claims the pCPU at $t=3$ for its 10\,ms slice; when the second job of vCPU\_RT is released at $t=10$, vCPU\_RT is still runnable in the queue rather than transitioning from blocked to ready, so it never benefits from head-of-queue wake-up insertion. In exchange, IORR provides the best I/O responsiveness, with no wake-up latency (0\,ms) for the three I/O events.

ARINC653 shows the opposite behavior. Its time windows protect vCPU\_RT, which meets both deadlines, but I/O requests must wait for the window of vCPU\_IO: the first two wake-ups wait 2\,ms, and the third waits 8\,ms. It is also the only scheduler that leaves the pCPU idle (4\,ms), since a window is not reassigned when its owner has nothing to run. This is the cost of temporal partitioning: strong isolation and predictability, but lower reactivity and lower utilization.

RTDS achieves a balanced trade-off on this workload between deadline preservation, reactivity, and utilization. Both real-time deadlines are met, albeit with a longer response time for vCPU\_RT (7\,ms for job 1, versus 3\,ms under fixed-priority schedulers, due to EDF tie-breaking with vCPU\_CPU1), while the first two I/O events are served immediately and the third waits 2\,ms for budget replenishment. Moreover, the pCPU is never left idle, as unreserved time is reclaimed by the background vCPUs thanks to the extratime flag.

The remaining schedulers meet the real-time deadlines but differ in how they allocate the remaining 11\,ms of pCPU time between the background vCPUs. Priority-based and PRM give unbalanced shares (11\,ms for vCPU\_CPU1 and 0\,ms for vCPU\_CPU2): in the Priority-based scheduler, equal-priority vCPUs (\code{PRIO\_LOW}) have no time slice so the first background vCPU runs without yielding, whereas in PRM, even when vCPU\_CPU1's time slice expires, its shorter periodicity (10\,ms vs. 15\,ms) places it back at the leftmost position of the Red-Black tree, starving vCPU\_CPU2. PRR avoids this starvation thanks to its 2\,ms Round-Robin time slices (6\,ms vs. 5\,ms). Conversely, the asymmetric shares observed for Credit (9\,ms vs. 2\,ms) and BVT (4\,ms vs. 7\,ms) result from the 20\,ms observation window and the initial ordering rather than long-term fairness properties, whereas the asymmetry observed for Credit2 (4\,ms vs. 7\,ms) is the intended effect of a parameter choice. In Credit, fairness is enforced over 30\,ms accounting periods (\code{csched\_acct}) that lie beyond the 20\,ms trace. In BVT, vCPU\_CPU2 enters the $t=20$ window with a lower accumulated virtual time (160 vs. 192). In Credit2, vCPU\_CPU1 is explicitly restricted to 4\,ms (2\,ms per 10\,ms period) by its 20\% utilization cap, which leaves the remaining 7\,ms to the uncapped vCPU\_CPU2. Finally, for I/O responsiveness, Credit, Credit2, Priority-based, PRR, and PRM serve all I/O events with at most 1\,ms of latency, whereas BVT delays the first I/O event ($t=22$) by 3\,ms because vCPU\_RT (clamped to the same baseline SVT = 160 inherited from the initial CPU1/CPU2 progression) precedes vCPU\_IO in FIFO order and executes from $t=22$ to $t=25$.

\paragraph{Fault scenario (overrun of vCPU\_RT).}

\begin{table}[!htbp]
\caption{Containment of vCPU\_RT overruns under each scheduling policy}
\label{tab:confinement}
\footnotesize
\begin{tabular}{p{2.6cm}p{3.2cm}p{2.6cm}p{\dimexpr\textwidth-8.4cm-8\tabcolsep\relax}}
\toprule
\textbf{Policy} & \textbf{IO~1 latency (nominal $\rightarrow$ fault)} & \textbf{Overrun contained?} & \textbf{Mechanism} \\
\midrule
ARINC653 & $2 \rightarrow 2$ & Yes & RT is confined to its 4~ms window; IO, CPU1, and CPU2 are unaffected; RT only misses its own deadlines. \\
\addlinespace
Credit & $1 \rightarrow 6$ & No & RT keeps \texttt{BOOST} until a tick finds it running ($t = 10$); IO, also in \texttt{BOOST}, waits behind RT in FIFO order. \\
\addlinespace
Credit2 & $1 \rightarrow 1$ & Partial & Proportional sharing: CPU1 and CPU2 still receive their slices, but RT has no hard bound (unless a cap is set). \\
\addlinespace
RTDS & $0 \rightarrow 0$ & Yes & Budget of 3~ms per 10~ms: RT is exhausted at $t = 7$ and again at $t = 14$; IO and background vCPUs are unaffected. \\
\addlinespace
Priority-based & $1 \rightarrow 6$ & No & RT at \texttt{PRIO\_HIGH} runs for 8~ms; IO (same priority level) is queued behind it. \\
\addlinespace
BVT & $3 \rightarrow 5$ & Partial & Proportional sharing; the RT quantum (CSA $= 5$~ms) delays IO. \\
\addlinespace
IORR & $0 \rightarrow 0$ & Yes (by rotation) & RT is placed at the tail of the queue regardless of its needs; RT itself is not protected. \\
\addlinespace
PRR & $1 \rightarrow 6$ & No & The 4~ms time slice expires, but RT is alone at its level and resumes immediately. \\
\addlinespace
PRM & $1 \rightarrow 6$ & No & Same as PRR. \\
\bottomrule
\end{tabular}
\end{table}

To illustrate temporal isolation on the reference timeline, a fault scenario is evaluated in which the first job of vCPU\_RT released at the start of the observation window requires $8$~ms of execution instead of $3$~ms, while all other parameters remain unchanged. As confirmed by the executable models (Table~\ref{tab:confinement}), only three policies prevent the overrun from degrading vCPU\_IO and background execution: ARINC653 (which confines vCPU\_RT to its $4$~ms minor frame per $10$~ms major frame), RTDS (which suspends vCPU\_RT when its $3$~ms budget is exhausted, at $t = 7$ and again at $t = 14$), and IORR (where the wake-up of vCPU\_IO at $t = 2$~ms preempts vCPU\_RT and rotates it to the tail of the run-queue). Conversely, Credit2 and BVT bound the overrun proportionally through credit consumption and virtual-time advancement, whereas Credit, Priority-based, PRR, and PRM propagate the full $5$~ms overrun to vCPU\_IO, increasing its first wake-up latency from $1$~ms to $6$~ms.

Beyond the normal-operation schedules of Table~\ref{tab:3}, the code-level analysis conducted in Sections~2.4 to~2.13 reveals several discrepancies between published algorithm descriptions or hypervisor documentation and the actual kernel implementations. Table~\ref{tab:rq2_deviations} consolidates these findings by mapping each documented behavior to its exact source-code location at the frozen commits of Table~\ref{tab:2} and stating its practical consequence for the system integrator.

\begin{table}[!htbp]
\caption{Deviations between documented and implemented behaviors across the surveyed scheduling policies (addressing RQ2). Code locations refer to file, function, and line numbers at the frozen commits listed in Table~\ref{tab:2}.}
\label{tab:rq2_deviations}
\footnotesize
\begin{tabular}{p{2.1cm}p{3.8cm}p{5.2cm}p{\dimexpr\textwidth-11.1cm-8\tabcolsep\relax}}
\toprule
\textbf{Policy} & \textbf{Documented behavior} & \textbf{Implemented behavior (code location)} & \textbf{Consequence for the integrator} \\
\midrule
BVT (ACRN) & Temporal warp lowers the EVT of latency-sensitive vCPUs~\cite{duda1999bvt,acrn2026bvt}; warp parameters are defined in data structures. & \texttt{sched\_bvt.c}: warp disabled at initialization (\texttt{warp\_on = false}, l.\,213); warp subtraction left as a comment in \texttt{update\_vt()} and \texttt{sched\_bvt\_wake()} (\texttt{/* TODO */}, l.\,248, 327), forcing $\text{EVT} = \text{AVT}$. & Warp parameters have no effect; BVT acts as a standard weighted fair-share scheduler without low-latency preemption, despite being ACRN's default (\texttt{config.xsd}, l.\,56). \\
\addlinespace
Priority-based (ACRN) & Static priority-based vCPU scheduling~\cite{acrn2026prio}. & \texttt{types.xsd} (l.\,170--179) exposes only two schema values (\texttt{PRIO\_LOW}, \texttt{PRIO\_HIGH}), though \texttt{sched\_prio.c} (l.\,14, 35) stores an \texttt{int}. \texttt{prio\_queue\_add()} (l.\,52--74) inserts in FIFO order among equals without time-slicing. & Only two priority tiers are usable without modifying the configuration schema; the first runnable vCPU in a tier executes until it blocks, starving same-tier peers. \\
\addlinespace
IORR (ACRN) & Round-Robin with I/O sensitivity~\cite{acrn2026iorr}. & \texttt{sched\_iorr.c}: \texttt{CONFIG\_SLICE\_MS} is hardcoded to 10\,ms (l.\,12) and forced at initialization (l.\,162); every wake-up inserts directly at the run-queue head (\texttt{sched\_iorr\_wake()}, l.\,212--214). & No tunable parameters exist; any frequently waking vCPU unconditionally preempts the running vCPU and can delay periodic workloads. \\
\addlinespace
PRM (Xvisor) & Rate Monotonic periodicity ordering within each priority tier~\cite{xvisor2026prm}. & \texttt{vmm\_schedalgo\_prm.c}, \texttt{vmm\_schedalgo\_rq\_prempt\_needed()} (l.\,190--214): preemption is triggered only for a strictly higher priority tier; same-tier periodicity check is unimplemented (\texttt{/* TODO */}, l.\,211). & A waking vCPU with a shorter period cannot preempt a running vCPU at the same priority level and must wait for it to block or exhaust its slice. \\
\addlinespace
PRR, PRM (Xvisor) & Priority, time slice, deadline, and periodicity are configurable per-vCPU parameters~\cite{xvisor2026repo}. & \texttt{vmm\_manager.c}: priority is clamped to $[0,7]$ (l.\,1303--1308), a null slice is replaced by 10\,ms (l.\,1396--1398), deadline is raised to at least the slice (l.\,1403--1405), and periodicity to at least the deadline (l.\,1410--1412). & Inconsistent timing parameters in the device tree are silently rewritten rather than rejected with an error. \\
\addlinespace
Credit (Xen) & Waking I/O vCPUs receive a \texttt{BOOST} priority elevation~\cite{xen2026creditwiki, xen2026creditcode}. & \texttt{credit.c}, \texttt{csched\_unit\_wake()} (l.\,1164--1169): \texttt{BOOST} is granted only if the vCPU is \texttt{UNDER}; \texttt{csched\_unit\_acct()} (l.\,957--961) clears \texttt{BOOST} only when the 10\,ms tick samples the vCPU running; \texttt{\_\_runq\_tickle()} (l.\,405, 455) preempts only on strictly higher priority. & \texttt{OVER} vCPUs receive no wake-up boost; a boosted vCPU retains \texttt{BOOST} across sleep/wake cycles until a tick catches it running; multiple \texttt{BOOST} vCPUs are served in FIFO order. \\
\addlinespace
Credit2 (Xen) & Sleeping vCPUs preserve their credits to favor I/O responsiveness~\cite{xen2026credit2wiki, xen2026creditcode}. & \texttt{credit2.c}, \texttt{reset\_credit()} (l.\,1783--1839): when the run-queue head reaches $\le 0$ credits, all vCPUs (including sleeping ones) are reset and clipped to \texttt{CREDIT\_INIT + CARRYOVER\_MAX} (10.5\,ms, l.\,1838--1839). & Credit preservation for sleeping I/O vCPUs is bounded by the next global credit reset; utilization caps act as periodic parking budgets rather than deferrable servers. \\
\addlinespace
RTDS (Xen) & Budget/period reservations under Global EDF~\cite{xi2014realtime, xen2026rtdswiki, xen2026rtdscode}. & \texttt{rt.c}: \texttt{rt\_try\_set\_utilization()} (l.\,706--721) rejects configurations exceeding total capacity with \texttt{-ENOSPC} (l.\,1584, 1653). Added in commits \texttt{5772eaa} and \texttt{04584da} (21/09/2026); absent from release 4.22.0. & Online admission control checks only $\sum \text{budget}/\text{period} \le M$, which is a necessary but insufficient schedulability condition under Global EDF with deferrable servers. \\
\addlinespace
ARINC653 (Xen) & Static cyclic time-table scheduling~\cite{xen2026arinc653wiki, xen2026arinc653code}. & \texttt{arinc653.c}, \texttt{arinc653\_sched\_set()} (l.\,235--270) checks $\sum \text{runtime} \le \text{major\_frame}$; \texttt{a653sched\_pick\_resource()} assigns all units to the first pCPU of the pool. & Enforces frame-fit admission control only; execution is restricted to a single pCPU per scheduling pool. \\
\bottomrule
\end{tabular}
\end{table}

\subsection{Threats to validity}
\paragraph{Construct validity.} The nine qualitative criteria defined in Section~2.2 involve architectural trade-offs, and summarizing them in Table~\ref{tab:4} requires structured judgment. To mitigate subjectivity, Section~3 defines explicit operational rules for each rating symbol (\ding{51}, (\ding{51}), \ding{55}), and every rating is traced directly to the source-code mechanisms analyzed in Sections~2.4 to~2.13.

\paragraph{Internal validity.} The reference schedules of Sections~2.5 to~2.13 were derived from the source code under explicit simplifying assumptions: zero hypervisor scheduling and context-switch overhead, a 1\,ms time resolution, a periodic tick phase aligned with $t=0$ (for Credit and IORR), and simultaneous event processing prior to rescheduling (Section~2.3). In Credit, this phase assumption is critical: with the accounting tick shifted by 1~ms ($t = 1 + 10k$), the tick at $t = 1$~ms clears the \texttt{BOOST} priority of vCPU\_RT before the first wake-up of vCPU\_IO at $t = 2$~ms; vCPU\_IO then preempts vCPU\_RT, which is requeued behind the two background vCPUs and misses both deadlines on the reference workload (whereas tick phases $0, 2, 4$, or $5$~ms preserve the schedule of Figure~4). Where the kernel code leaves the ordering of simultaneous timer events to runtime queue traversal (such as at $t=10$ in Credit2 and RTDS), the chosen ordering and its alternative are stated explicitly in the text. Both the code reading and the schedule derivations were cross-checked against the frozen source files by the authors. Each schedule in Figures~3--11 and the metrics in Table~\ref{tab:3} were re-derived and cross-checked with an executable Python model of the scheduling rules, provided as a public research artifact~\cite{artifact_zenodo}. Because these models encode the same code reading as the paper, they guarantee reproducibility and internal consistency rather than independent empirical validation. An independent validation would require execution traces collected on a running hypervisor (e.g., via \texttt{xentrace} on Xen or ACRN under QEMU), which is left to future work.

\paragraph{External validity.} All code-level findings apply to the frozen repository commits listed in Table~\ref{tab:2}, and behavioral differences with the latest official releases are documented in Section~2.1. Furthermore, the reference workload is an analytical illustration on a single pCPU rather than an empirical benchmark: end-to-end real-time performance on physical targets remains subject to hardware interrupt latency, shared cache/memory contention, and guest OS scheduling.

\section{Qualitative comparison and selection guide}\label{sec:qualitative-comparison}

\begin{figure}[!htbp]
\centering
\resizebox{\textwidth}{!}{%
\begin{tikzpicture}[
    >=Stealth,
    q/.style={
        draw,
        rounded corners=3pt,
        fill=blue!8,
        align=center,
        text width=4.3cm,
        minimum height=1.15cm,
        inner sep=4pt,
        font=\small
    },
    leaf/.style={
        draw,
        rounded corners=3pt,
        fill=green!12,
        align=center,
        text width=3.5cm,
        minimum height=1.75cm,
        inner sep=4pt,
        font=\small\bfseries
    },
    edge/.style={->, thick},
    lab/.style={
        font=\scriptsize\bfseries,
        fill=white,
        inner sep=2pt
    }
]

\node[q] (q1) at (-3.5, 0)
    {1. Can each vCPU be given\\a dedicated pCPU?};

\node[leaf, text width=3.8cm] (static) at (-7.5, -3.5)
    {Static partitioning\\
    {\scriptsize\normalfont\itshape Multicore: partitioned}\\
    {\scriptsize\normalfont\itshape $\mathcal{O}(1)$}};

\node[q, text width=4.8cm] (q2) at (1.5, -2.8)
    {2. Must critical vCPUs receive a\\guaranteed pCPU share regardless of\\others (temporal isolation)?};

\node[q, text width=4.5cm] (q3) at (-4.2, -6.0)
    {3. Can unused time\\be lost?};

\node[q, text width=4.2cm] (q4) at (6.2, -6.0)
    {4. Is a weighted share of\\the pCPU required?};

\node[leaf, text width=3.4cm] (arinc) at (-7.2, -9.5)
    {ARINC653\\
    {\scriptsize\normalfont\itshape Uniprocessor}\\
    {\scriptsize\normalfont\itshape $\mathcal{O}(1)$}};

\node[leaf, text width=4.1cm] (rtds) at (-2.2, -9.5)
    {RTDS\\[1pt]
    {\scriptsize\normalfont\itshape Multicore: global}\\
    {\scriptsize\normalfont\itshape $\mathcal{O}(N)$}};

\node[q, text width=4.0cm] (q5) at (3.2, -8.3)
    {5. Must parameters\\be changed at runtime?};

\node[q, text width=4.0cm] (q6) at (11.7, -8.3)
    {6. Is there a static order of\\importance between\\vCPUs?};

\node[leaf, text width=3.7cm] (credit) at (1.1, -11.5)
    {Credit, Credit2\\[1pt]
    {\scriptsize\normalfont\itshape Multicore: global}\\
    {\scriptsize\normalfont\itshape $\mathcal{O}(N)$}};

\node[leaf, text width=3.4cm] (bvt) at (5.1, -11.5)
    {BVT\\[1pt]
    {\scriptsize\normalfont\itshape Multicore: partitioned}\\
    {\scriptsize\normalfont\itshape $\mathcal{O}(N)$}};

\node[leaf, text width=4.3cm] (prio) at (9.5, -11.5)
    {Priority, PRR, or PRM\\[1pt]
    {\scriptsize\normalfont\itshape Multicore: partitioned}\\
    {\scriptsize\normalfont\itshape $\mathcal{O}(N)$ | $\mathcal{O}(1)$ | $\mathcal{O}(\log N)$}};

\node[leaf, text width=3.5cm] (iorr) at (13.9, -11.5)
    {IORR\\[1pt]
    {\scriptsize\normalfont\itshape Multicore: partitioned}\\
    {\scriptsize\normalfont\itshape $\mathcal{O}(1)$}};

\draw[edge] (q1.south) -- ++(0,-0.45) -| node[pos=0.75, above, lab] {Yes} (static.north);
\draw[edge] (q1.south) -- ++(0,-0.45) -| node[pos=0.75, above, lab] {No}  (q2.north);

\draw[edge] (q2.south) -- ++(0,-0.45) -| node[pos=0.75, above, lab] {Yes} (q3.north);
\draw[edge] (q2.south) -- ++(0,-0.45) -| node[pos=0.75, above, lab] {No}  (q4.north);

\draw[edge] (q3.south) -- ++(0,-0.45) -| node[pos=0.75, above, lab] {Yes} (arinc.north);
\draw[edge] (q3.south) -- ++(0,-0.45) -| node[pos=0.75, above, lab] {No}  (rtds.north);

\draw[edge] (q4.south) -- ++(0,-0.45) -| node[pos=0.75, above, lab] {Yes} (q5.north);
\draw[edge] (q4.south) -- ++(0,-0.45) -| node[pos=0.75, above, lab] {No}  (q6.north);

\draw[edge] (q5.south) -- ++(0,-0.45) -| node[pos=0.75, above, lab] {Yes} (credit.north);
\draw[edge] (q5.south) -- ++(0,-0.45) -| node[pos=0.75, above, lab] {No}  (bvt.north);

\draw[edge] (q6.south) -- ++(0,-0.45) -| node[pos=0.75, above, lab] {Yes} (prio.north);
\draw[edge] (q6.south) -- ++(0,-0.45) -| node[pos=0.75, above, lab] {No}  (iorr.north);

\end{tikzpicture}%
}
\caption{Decision tree for selecting a vCPU scheduler. Each leaf gives the simplest policies that match the answers along the path; answering ``No'' to a requirement does not exclude policies from other leaves that also provide the property. The multicore organization and asymptotic scheduling complexity are indicated within each leaf.}
\label{fig:12}
\end{figure}
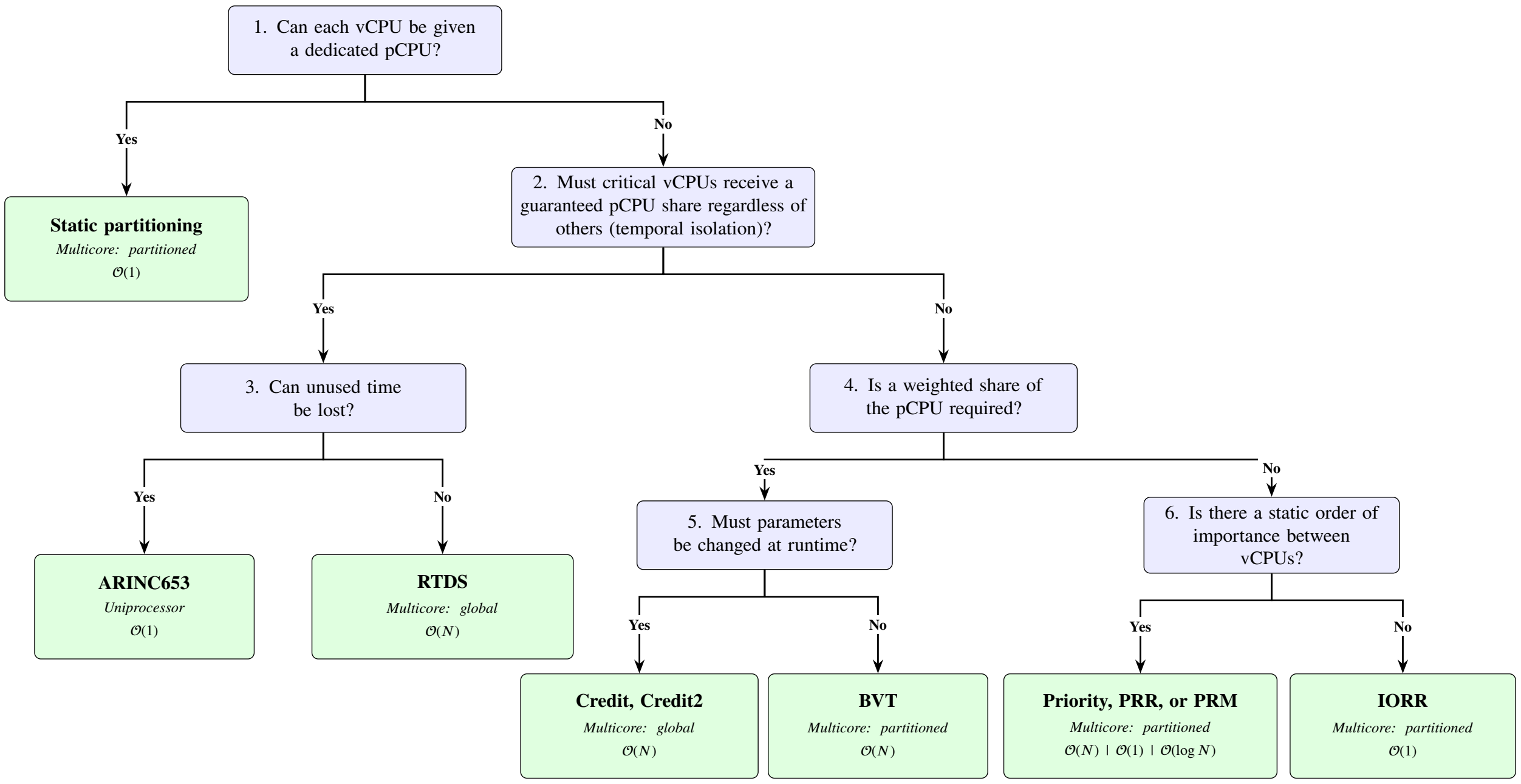

The previous section analyzed each algorithm separately and compared their behavior on a single reference workload. This section generalizes these observations. Table~\ref{tab:4} summarizes how each algorithm satisfies the nine evaluation criteria defined in Section~\ref{sec:criteria-definitions}, based on the code-level mechanisms described in Sections~\ref{sec:static-partitioning} to~\ref{sec:prm}. For the six behavioral criteria, each symbol follows an operational rating rule: a tick (\ding{51}) indicates that the scheduling mechanism natively guarantees the property for all admissible parameter configurations by construction; a parenthesized tick ((\ding{51})) indicates conditional support, meaning that the property holds only under specific offline configuration or schedulability assumptions placed on the system integrator (detailed in the notes below Table~\ref{tab:4} and anchored in scheduling theory in Table~\ref{tab:sched_theory}); a cross (\ding{55}) indicates that no mechanism in the scheduler can provide the property; and \textit{n/a} means that the criterion is not applicable. The ratings were derived directly from the code-level mechanisms identified in Sections~2.4 to~2.13 and cross-reviewed by all authors to ensure consistency across hypervisors. Table~\ref{tab:4} is then used to construct a decision tree (Figure~\ref{fig:12}) that guides the system architect in selecting a scheduler according to application requirements.

\begin{table}[!htbp]
\caption{Comparative evaluation of scheduling policies across qualitative criteria. Symbols denote full support (\ding{51}), conditional support ((\ding{51})), and lack of support (\ding{55}).}\label{tab:4}
\footnotesize\renewcommand{\arraystretch}{1.15}
\begin{tabular*}{\tblwidth}{@{}l@{\extracolsep{\fill}}cccccc>{\raggedright\arraybackslash}p{3.05cm}>{\raggedright\arraybackslash}p{2.2cm}l@{}}
\toprule
Policy & \rh{Temporal\\isolation} & \rh{RT\\guarantees} & \rh{I/O\\responsiveness} & \rh{Work-\\conserving} & \rh{Fair\\sharing} & \rh{Mixed-criticality\\support} & \makecell[lb]{Efficiency and\\scalability} & Configurability & Multicore \\
\midrule
Static partitioning & \ding{51} & \ding{51} & \ding{51} & \ding{55} & n/a & \ding{51} & $\mathcal{O}(1)$, no scheduler & build/boot; runtime for Xen & partitioned \\
ARINC653 & \ding{51} & (\ding{51}) & \ding{55} & \ding{55} & \ding{55} & \ding{51} & $\mathcal{O}(1)$ lookup & runtime & uniprocessor \\
Credit & \ding{55} & \ding{55} & (\ding{51}) & \ding{51} & (\ding{51}) & \ding{55} & $\mathcal{O}(1)$ pick, $\mathcal{O}(N)$ insert/acct & runtime & global \\
Credit2 & \ding{55} & \ding{55} & (\ding{51}) & \ding{51} & (\ding{51}) & \ding{55} & $\mathcal{O}(1)$ pick, $\mathcal{O}(N)$ insert/EWMA & runtime & global \\
RTDS & (\ding{51}) & (\ding{51}) & (\ding{51}) & (\ding{51}) & \ding{55} & (\ding{51}) & $\mathcal{O}(1)$ pick, $\mathcal{O}(N)$ insert & runtime & global \\
Priority & \ding{55} & (\ding{51}) & (\ding{51}) & \ding{51} & \ding{55} & (\ding{51}) & $\mathcal{O}(1)$ pick, $\mathcal{O}(N)$ insert & build & partitioned \\
BVT & \ding{55} & \ding{55} & (\ding{51}) & \ding{51} & \ding{51} & \ding{55} & $\mathcal{O}(N)$ pick/insert/VT update & build & partitioned \\
IORR & \ding{55} & \ding{55} & \ding{51} & \ding{51} & (\ding{51}) & \ding{55} & $\mathcal{O}(1)$ pick/insert & none & partitioned \\
PRR & \ding{55} & (\ding{51}) & (\ding{51}) & \ding{51} & (\ding{51}) & (\ding{51}) & $\mathcal{O}(1)$ pick/insert & boot & partitioned \\
PRM & \ding{55} & (\ding{51}) & (\ding{51}) & \ding{51} & \ding{55} & (\ding{51}) & $\mathcal{O}(\log N)$ pick/insert & boot & partitioned \\
\bottomrule
\end{tabular*}
\end{table}

Notes on conditional support (\ding{51}) in Table~\ref{tab:4}:

Temporal isolation: In RTDS, budget policing prevents any vCPU from exceeding its reservation, but guaranteeing that each vCPU receives its full budget in every period requires the set of deferrable servers to be schedulable under global EDF. However, the hypervisor admission check only verifies that total utilization does not exceed platform capacity (in the Xen development branch only, Section~2.1; absent from Xen 4.22.0), which is a necessary but insufficient condition (due to Dhall's effect~\cite{dhall1978real} and back-to-back deferrable server execution~\cite{strosnider1995deferrable}).

RT guarantees: In ARINC 653, guarantees hold if the integrator aligns minor/major frames offline with vCPU periods and WCETs. In RTDS, guarantees hold if the reservation set satisfies global EDF schedulability bounds with deferrable servers, which the integrator must verify offline as the online admission test only checks total capacity. In Priority, PRR, and PRM, guarantees for high-priority vCPUs hold only if the integrator strictly bounds the execution demand of all vCPUs at the same or higher priority tiers (and additionally in PRM, shares no priority tier among periodic vCPUs due to the unimplemented same-priority preemption TODO).

I/O responsiveness: In Credit, Credit2, and BVT, low wake-up latency holds only if the waking I/O vCPU has not exhausted its credits (UNDER state in Credit, competitive credit balance in Credit2) or has slept long enough to be clamped to SVT in BVT (since the warp mechanism is left as an unimplemented TODO). In RTDS, immediate preemption occurs only if the waking I/O vCPU has remaining budget in its current period and an earlier deadline. In Priority, PRR, and PRM, low dispatch latency requires assigning the I/O vCPU to a high static priority tier.

Work-conserving: RTDS is work-conserving only when the integrator enables the optional extratime flag on vCPUs. Otherwise, depleted vCPUs are strictly throttled even if the pCPU goes idle.

Fair sharing: IORR and PRR only provide egalitarian Round-Robin time-slicing (equal fixed quanta rather than weighted shares), and in PRR this sharing is restricted to vCPUs operating within the exact same priority tier. Credit and Credit2 are rated (\ding{51}) because proportional sharing holds only for uncapped VMs (configuring a hard cap overrides proportional weight allocation) and only over multiple accounting periods (Credit) or credit resets (Credit2), with short-term distortions caused by \texttt{BOOST} wake-ups or credit saturation. In BVT, proportional sharing holds once the initial wake-up transient has elapsed.

Mixed-criticality support: None of the analyzed hypervisors implements Vestal-style criticality mode changes. Mixed-criticality support here refers to safely co-locating workloads of different criticality levels without temporal interference. In RTDS, it is supported because per-vCPU budgets prevent overruns and extratime demotes best-effort execution to a lower priority tier, provided the tier-0 reservation set is shown schedulable offline. In Priority, PRR, and PRM, safe co-location holds only if critical vCPUs are assigned strictly higher priorities and their execution demand is bounded offline by the integrator.

To anchor the conditional guarantees ((\ding{51})) of Table~\ref{tab:4} in real-time scheduling theory, Table~\ref{tab:sched_theory} summarizes the formal schedulability analysis frameworks applicable to each policy for offline verification by the system integrator.

\begin{table}[!htbp]
\caption{Applicable real-time schedulability analysis frameworks for establishing the conditional guarantees ((\ding{51})) of Table~\ref{tab:4}.}
\label{tab:sched_theory}
\footnotesize
\begin{tabular}{p{3.2cm}p{\dimexpr\textwidth-6.2cm-6\tabcolsep\relax}p{3.0cm}}
\toprule
\textbf{Policy} & \textbf{Applicable schedulability and isolation analysis} & \textbf{Key references} \\
\midrule
Static partitioning & No hypervisor-level scheduling; each guest OS is analyzed as on bare metal, with worst-case execution time (WCET) bounds inflated for shared cache and memory-bus contention. & \cite{martins2023shedding, axer2014building} \\
\addlinespace
ARINC653 & Compositional/hierarchical scheduling using Periodic or Explicit Deadline Periodic (EDP) resource models: each vCPU partition supplies $(Q, P)$ capacity, against which guest tasks are validated via supply-bound functions. & \cite{shin2003periodic, easwaran2007compositional} \\
\addlinespace
RTDS & 
Each deferrable server is modeled as a sporadic task of period $P$, execution demand $Q$, and release jitter $P - Q$, to which a global EDF schedulability test or virtual-clustering analysis is applied. & \cite{dhall1978real, strosnider1995deferrable, baker2003multiprocessor, shin2008hierarchical} \\
\addlinespace
Priority-based, PRR, PRM & Fixed-priority Response-Time Analysis (RTA) at the vCPU level, assuming the integrator strictly bounds the execution demand and minimum inter-arrival time of all higher- or equal-priority vCPUs. Intra-tier FIFO (Priority), time-slicing (PRR), or non-preemptive periodicity order (PRM) must be modeled as blocking terms. These bounds must hold by construction of the guests, since none of the three schedulers enforces a budget (Sections~2.9.4, 2.12.4, and 2.13.4). & \cite{audsley1993applying, davis2011survey} \\
\addlinespace
Credit, Credit2, BVT & Proportional-share lag bounds (GPS/PGPS) characterize long-term bandwidth allocation, but no established hard real-time schedulability analysis covers wake-up \texttt{BOOST} (Credit), global credit reset clipping (Credit2), or SVT clamping without warp (BVT). & \cite{parekh1993generalized, stoica1996proportional} \\
\addlinespace
IORR & None: unconditional head-of-queue insertion upon every wake-up prevents analytical bounding of worst-case response times for runnable vCPUs. & --- \\
\bottomrule
\end{tabular}
\end{table}

Table~\ref{tab:4} confirms the observations of Section~\ref{sec:synthesis}: no single algorithm satisfies all criteria simultaneously, as structural trade-offs oppose strict isolation, resource utilization, fairness, and runtime overhead. Based on these trade-offs, the ten policies fall into three distinct architectural groups:

\textbf{Strong isolation, deterministic mixed-criticality, low utilization.} Static partitioning and ARINC653 are the only mechanisms that provide strict temporal isolation by construction, either through dedicated physical cores or fixed cyclic time windows. Consequently, they natively support mixed-criticality workloads with $\mathcal{O}(1)$ scheduling efficiency and minimal runtime overhead, which simplifies the safety certification argument. However, this determinism carries a severe utilization cost: neither policy is work-conserving, as unused pCPU cycles within an idle core or an early-completing minor frame are permanently lost. Furthermore, while Static partitioning achieves immediate I/O responsiveness through direct hardware assignment, ARINC653 delays asynchronous I/O events until the vCPU's next scheduled minor frame, as demonstrated in the timeline of Section~\ref{sec:arinc653}. These policies are therefore suited to hard real-time and safety-critical workloads with stable, well-characterized execution demands.

\textbf{Proportional resource sharing without strict real-time or mixed-criticality guarantees.} Credit, Credit2, BVT, and IORR are work-conserving policies designed to maximize pCPU utilization, with the first three distributing CPU time proportionally to vCPU weights. However, none of them provides real-time guarantees or safe mixed-criticality support, because they lack deadline awareness and prioritize interactive fairness over deterministic isolation. While Credit and Credit2 allow administrators to limit a vCPU's consumption via execution caps, this mechanism only enforces an upper utilization bound over a time window and does not prevent short-term preemption jitter from waking vCPUs. Their I/O responsiveness relies on best-effort heuristics (BOOST priority in Credit, sleep credit preservation in Credit2, SVT alignment in BVT since warp remains unimplemented, and head-of-queue wake-up insertion in IORR). In terms of scheduler efficiency, IORR uses $\mathcal{O}(1)$ pointer manipulations (albeit driven by a 1\,ms periodic timer tick), whereas Credit, Credit2, and BVT add periodic accounting (Credit), EWMA load tracking (Credit2), or virtual-time arithmetic (BVT) alongside $\mathcal{O}(N)$ sorted queue insertions. These algorithms are suited to general-purpose or soft real-time workloads where average throughput and fairness outweigh strict deadline guarantees.

\textbf{Priority and deadline-driven scheduling.} Priority, PRR, PRM, and RTDS establish an explicit execution hierarchy that favors critical or urgent workloads. The three fixed-priority schedulers (Priority, PRR, and PRM) deliver fast I/O responsiveness and predictable execution to high-priority vCPUs, with high scheduler efficiency, ranging from $\mathcal{O}(1)$ array/list extractions in PRR to $\mathcal{O}(\log N)$ Red-Black tree lookups in PRM, and purely event-driven $\mathcal{O}(1)$ head selection (albeit $\mathcal{O}(N)$ wake-up insertion) in ACRN's Priority scheduler. However, because they lack budget enforcement, they support real-time and mixed-criticality workloads only under the assumption that the system integrator bounds the execution demand of higher-priority vCPUs offline. Otherwise, a misbehaving high-priority vCPU will monopolize the pCPU and starve lower tiers. Moreover, Priority and PRM provide no fair sharing among equal-priority or background vCPUs, whereas PRR enforces egalitarian Round-Robin rotation within each priority tier. Conversely, RTDS is a dynamic policy that enforces per-vCPU budgets while offering work-conserving slack reclamation (through extratime), which supports mixed-criticality co-location and real-time guarantees provided the reservation set is shown schedulable under global EDF with deferrable servers (as its admission test only checks total utilization). However, RTDS does not provide fair sharing among extratime vCPUs, and its multicore scalability is constrained by a global spinlock protecting an $\mathcal{O}(N)$ deadline-ordered run-queue across all pCPUs.

\textbf{Practical deployment constraints: Configurability and Multicore scalability.} Beyond pure scheduling logic, the \textit{Configurability}, \textit{Scheduler efficiency}, and \textit{Multicore} columns introduce decisive architectural constraints. Several schedulers bind their parameters statically at build time (Priority, BVT) or boot time (PRR, PRM), and IORR offers no configurable parameters at all. Only Xen's schedulers (ARINC653, Credit, Credit2, RTDS, and \code{null}) support online runtime reconfiguration via hypercalls. Finally, multicore architectures dictate how scheduling overhead scales with core count: partitioned schedulers (Static partitioning, Priority, BVT, IORR, PRR, PRM) instantiate independent per-pCPU run-queues, completely eliminating cross-core lock contention and vCPU migration overheads at the cost of static core-to-vCPU bin-packing. In contrast, global schedulers (Credit, Credit2, RTDS) dynamically balance load across pCPUs via work-stealing, topology-aware EWMA migration, or a shared global EDF queue, maximizing hardware utilization at the expense of higher synchronization overhead and cache invalidation penalties. Importantly, the ``global'' and ``partitioned'' labels in Table~\ref{tab:4} describe each scheduler's default design rather than its only possible deployment mode. In Xen, the integrator can restrict any scheduler to a disjoint subset of pCPUs via \texttt{cpupools} and pin vCPUs to specific cores via hard affinity: for instance, instantiating one single-pCPU RTDS \texttt{cpupool} per core transforms Global EDF into partitioned EDF, eliminating cross-core lock contention and migration. Similarly, Credit2 supports a boot-time run-queue granularity parameter (per-pCPU, core, socket, NUMA node, or global), allowing clustered scheduling. When global scheduling is retained, integrators must account for well-known multiprocessor pitfalls, including cross-core migration and cache-reload delays~\cite{davis2011survey}, utilization loss under Global EDF (Dhall's effect~\cite{dhall1978real}), and back-to-back deferrable server execution across pCPUs~\cite{strosnider1995deferrable}.

These observations lead to the decision tree in Figure~\ref{fig:12}, which guides the system architect from hardware and application constraints to the simplest matching policies. Importantly, answering “No” to a question means that a property is not required by the application, not that policies providing it are forbidden. Therefore, each leaf highlights the most parsimonious schedulers satisfying the path. The first question (Step 1) separates systems that can afford a dedicated pCPU per vCPU (where Static partitioning (Bao, Xen \code{null}, ACRN \code{NOOP}) eliminates scheduling overhead altogether) from consolidated systems where multiple vCPUs must share a pCPU. When pCPUs are shared, Step 2 asks whether critical vCPUs require strict temporal isolation (a guaranteed pCPU share regardless of competing workloads). If so (Step 3), the choice lies between ARINC653, when an offline table-driven schedule and non-work-conserving idle losses are acceptable, and RTDS, when EDF reservations must be combined with work-conserving slack reclamation via extratime. Conversely, if strict temporal isolation is not required, Step 4 distinguishes proportional weighted sharing from priority-based or latency-driven execution. When weighted sharing is needed (Step 5), Credit and Credit2 are required if weights or upper-bound caps must be adjusted at runtime, whereas BVT provides a static build-time alternative (alongside Credit and Credit2). Finally, when weighted sharing is not required (Step 6), systems with a static importance hierarchy between vCPUs map to ACRN's Priority-based scheduler (2 configured priority levels, no time-slicing within a level), Xvisor's PRR (8 priority levels with Round-Robin rotation within each level), or PRM (8 priority levels with Rate Monotonic periodicity ordering within each level). In the absence of any priority hierarchy, IORR provides minimal wake-up latency with egalitarian time-sharing and no real-time guarantees. Ultimately, this tree serves as a structural guide rather than a replacement for platform-specific evaluation: as indicated by the (\ding{51}) entries in Table~\ref{tab:4}, the actual guarantees of most algorithms depend on how the system integrator sizes and validates the scheduling parameters.

\paragraph{Illustrative case study.} Consider a consolidated automotive Electronic Control Unit (ECU) hosting three VMs on shared pCPUs: a safety-critical chassis control VM (10\,ms periodic loop), a driver-assistance VM with bursty camera processing, and a best-effort infotainment VM. Because pCPUs are shared across VMs (Step~1: No) and the chassis control VM requires strict temporal isolation from camera bursts and infotainment load (Step~2: Yes), while unused pCPU cycles must be reclaimed by the camera and infotainment workloads (Step~3: No), the decision tree leads directly to \textbf{RTDS}: the control vCPU is configured as a tier-0 reservation, and the driver-assistance and infotainment vCPUs are configured with \texttt{extratime} enabled, provided the reservation set is verified schedulable offline (Table~\ref{tab:sched_theory}). Note that the two \texttt{extratime} vCPUs then share the residual CPU capacity without proportional control (Section~2.8.7). If the camera VM must be favored over infotainment during overload, a dedicated tier-0 reservation sized for its average demand is required. Furthermore, the temporal isolation guaranteed by RTDS applies strictly at the CPU scheduler level. Bounding interference on shared caches, memory controllers, and I/O buses requires complementary hardware partitioning mechanisms.

\section{Related Work}\label{sec:related-work}

To position the contributions of this paper, this section reviews the existing literature across three areas: general surveys on embedded virtualization, theoretical mixed-criticality scheduling models, and empirical evaluations or extensions of specific hypervisor schedulers.

\subsection{Surveys on embedded systems}\label{sec:surveys}

Using hypervisors to run multiple applications on a single hardware platform reduces Size, Weight, Power, and Cost (SWaP-C) constraints. Several surveys have studied this approach in embedded systems. Early work by Gu and Zhao~\cite{gu2012survey} reviewed real-time virtualization challenges in Xen, KVM, and L4 microkernels. However, it was published before the introduction of modern static-partitioning hypervisors (such as Bao or Jailhouse), Xen's event-driven RTDS and Credit2 schedulers, and ACRN. More recently, Cinque et al.~\cite{cinque2022virtualizing} reviewed industrial trends and isolation issues in avionics, automotive, and railway systems. Similarly, Lozano et al.~\cite{lozano2023comprehensive} compared open-source and commercial hypervisors (including Xen, Bao, Jailhouse, and ACRN), focusing on safety certification and hardware isolation mechanisms. Earlier, Garc{\'i}a-Valls et al.~\cite{garciavalls2014challenges} identified the core challenges of real-time virtualization and vCPU scheduling from the perspective of predictable cloud and embedded computing.

While these surveys provide a good high-level overview of hypervisor architectures and isolation features, they do not analyze the internal logic of the vCPU schedulers. They usually list the available schedulers by name (for instance, stating that Xen includes ARINC653 and RTDS, or that ACRN includes BVT) without inspecting their source code. As a result, they do not detail the exact selection rules, triggers, and admission checks, and they miss important differences between theoretical algorithms and actual kernel code (such as the unimplemented warp feature in ACRN's BVT or the missing same-priority preemption check in Xvisor's PRM).

\subsection{Mixed-criticality scheduling models}\label{sec:mc-models}

Running applications with different safety levels on a shared processor is the main focus of Mixed-Criticality Systems (MCS) research. Classical real-time systems rely on static time tables or preemptive priority and deadline algorithms~\cite{liu1973scheduling,ramamritham1994scheduling}. For mixed-criticality workloads, Vestal~\cite{vestal2007preemptive} introduced a model where each task has multiple Worst-Case Execution Time (WCET) estimates depending on its criticality level. Since then, many theoretical studies have proposed mode-switching algorithms, such as Earliest Deadline First with Virtual Deadlines (EDF-VD)~\cite{baruah2012preemptive}, and slack management methods~\cite{zhang2025slack}. In these models, low-criticality tasks are dropped or delayed if a high-criticality task exceeds its optimistic budget. Recently, Reinforcement Learning (RL) has also been studied in simulation to adaptively schedule mixed-criticality workloads on multicore processors~\cite{liang2024adaptive,elmahdy2025improving}.

As noted by Burns and Davis~\cite{burns2017survey} and Ernst and Di Natale~\cite{ernst2016mixed}, there is a clear gap between theoretical MCS models and actual hypervisor implementations. Theoretical mode-switching and RL algorithms require complex runtime monitoring and task dropping mechanisms that are not implemented in current open-source hypervisors. Instead, as shown in Section~\ref{sec:algorithms}, embedded hypervisors handle mixed-criticality workloads using static core partitioning~\cite{martins2020bao,ramsauer2017jailhouse}, cyclic ARINC653 time windows~\cite{arinc653,vanderleest2010arinc}, fixed priorities~\cite{acrn2026prio,xvisor2026prr,xvisor2026prm}, or two-tier deferrable servers (extratime in Xen RTDS~\cite{xi2014realtime}). This paper bridges this gap by analyzing the scheduling algorithms that are actually implemented in open-source hypervisors and by stating the offline configuration assumptions they place on the system integrator.

\subsection{Evaluation and customization of hypervisor schedulers}\label{sec:evaluation-customization}

Several studies have experimentally tested or modified the schedulers of specific hypervisors. The closest precedent for comparing hypervisor schedulers is the early experimental study by Cherkasova et al.~\cite{cherkasova2007comparison}, which evaluated Xen's BVT, SEDF, and Credit schedulers in 2007. However, BVT and SEDF have since been removed from Xen, and Credit has been superseded by Credit2 as the default scheduler. In Xen, Ongaro et al.~\cite{ongaro2008scheduling} studied how the Credit scheduler impacts I/O latency. Masrur et al.~\cite{masrur2010vm} showed through an automotive case study that Credit does not provide hard real-time guarantees and proposed a custom Priority-Switched EDF (PSEDF) scheduler. The RT-Xen project by Xi et al.~\cite{xi2014realtime,xi2011rtxen} implemented and tested global and partitioned real-time schedulers in Xen, which led to the mainline RTDS scheduler. Li et al.~\cite{li2018multimode} later used RTDS for multi-mode soft real-time systems. More recently, Cinque et al.~\cite{cinque2024temporal} used a Design of Experiments (DoE) approach to test the temporal isolation of Xen's \code{null}, Credit2, and RTDS schedulers under heavy CPU load in a railway use case. For static partitioning hypervisors, Danielsson et al.~\cite{danielsson2019testing} and Jiménez et al.~\cite{jimenez2022specific} measured interrupt latency and isolation in Jailhouse. Martins and Pinto~\cite{martins2023shedding} compared Jailhouse and Bao on ARM platforms, showing that shared cache and memory bus contention can still delay vCPUs even with dedicated pCPUs. For Xvisor, Patel et al.~\cite{patel2015embedded} compared its general performance against KVM and Xen, Modica et al.~\cite{modica2018supporting} added cache coloring and memory bandwidth control, and Cui et al.~\cite{cui2026optimizing} extended Xvisor’s PRM scheduler with a contextual-bandit reinforcement-learning policy (eXvisor), reducing real-time task latency with respect to PRR. Outside dedicated embedded hypervisors, Abeni~\cite{abeni2024virtualized} recently evaluated real-time workloads in KVM virtual machines and containers scheduled by Linux's \texttt{SCHED\_DEADLINE}~\cite{abeni2019hierarchical}.

While these experimental works provide useful measurements on specific boards, they have two main limitations when comparing hypervisors. First, each study tests only one hypervisor (or compares two of them) on a specific hardware target and guest OS setup, which makes it difficult to compare scheduling algorithms across different hypervisors. Second, hardware benchmarks measure the combined effects of interrupts, cache contention, hypervisor context switches, and scheduling rules, which can hide specific behaviors of the scheduling algorithm itself. In contrast, this paper provides a unified, code-level analysis of the ten scheduling policies implemented across Bao, Xen, ACRN, and Xvisor, compares them on a common reference workload, and summarizes their trade-offs in a practical decision tree.

\section{Conclusion}\label{sec:conclusion}

This paper surveyed the ten CPU allocation policies offered by four actively maintained open-source hypervisors for embedded systems (Bao, Xen, ACRN, and Xvisor), analyzed at the source-code level on frozen repository states, without runtime measurement.

\textbf{RQ1.} Besides spatial partitioning (Bao, Xen \code{null}, ACRN \code{NOOP}), the nine time-sharing policies follow five decision rules: a static time table (ARINC653); fixed priorities, with three intra-level policies (Priority-based, PRR and PRM); EDF over budget–period servers (RTDS); service accounting by credits or virtual time (Credit, Credit2, BVT); and quantum rotation with wake-up promotion (IORR). Classifying policies by decision rule rather than by hypervisor makes mechanisms from different hypervisors directly comparable.

\textbf{RQ2.} Reading the source code rather than the documentation reveals deviations that matter to a system integrator (Table~\ref{tab:rq2_deviations}):

\begin{itemize}

\item The warp mechanism of BVT (the default scheduler in ACRN's configuration) is left as an unimplemented stub.

\item ACRN's Priority-based scheduler exposes only two priority levels (\code{PRIO\_LOW} and \code{PRIO\_HIGH}) in its configuration schema, despite using an integer priority in C.

\item In PRM, periodicity orders vCPUs within a priority level but never triggers same-priority preemption (left as a TODO), and Xvisor's management layer (\code{vmm\_manager.c}) silently adjusts inconsistent timing parameters instead of rejecting them.

\item Credit's wake-up BOOST is granted only to UNDER vCPUs and is cleared only once a 10\,ms evaluation tick finds the vCPU running. Its behavior is sensitive to the accounting tick phase: with the evaluation tick at $t = 1 + 10k$, $\text{vCPU}_{\text{RT}}$ loses \texttt{BOOST} before the first I/O wake-up, is requeued behind the background vCPUs, and misses both deadlines on the reference workload.

\item Credit2 preserves the credits of sleeping vCPUs only until the next credit reset, which clips all balances to a fixed ceiling.

\item Apart from the frame-fit check of \mbox{ARINC653}, among the surveyed hypervisor-level schedulers, only RTDS performs platform-capacity admission control (added to the Xen development branch in September 2026 and not yet included in release 4.22.0). However, it checks total utilization only ($\sum \text{budget}/\text{period} \le M$), which is a necessary but insufficient schedulability condition for global EDF with deferrable servers~\cite{dhall1978real, strosnider1995deferrable}.

\end{itemize}

\textbf{RQ3.} No surveyed policy holds an explicit criticality attribute or performs a Vestal-style mode change. What they offer for mixed-criticality integration is a bound (or no bound) on the pCPU time a vCPU can consume: by construction (Static partitioning, ARINC 653), by per-period budget (RTDS), by optional upper-bound cap (Credit, Credit2), or not at all (Priority-based, PRR, PRM, BVT, IORR). For accounting-based policies (Credit, Credit2, BVT), wake-up latency also depends on accumulated runtime state, making it difficult to bound analytically. In every case, offline schedulability analysis and the bounding of shared-resource interference remain the integrator's responsibility. Table~\ref{tab:4} and the decision tree (Figure~\ref{fig:12}) make these assumptions explicit.

Future work will extend this analysis to multicore mechanisms (migration, load balancing, and lock-holder preemption), to the fixed-priority round-robin scheduler of Gunyah once its open-source tree targets physical platforms, and to empirical trace validation on hardware targets. Also, some innovative algorithms could be designed and deployed to bridge the gap identified in Table~\ref{tab:4} between strict real-time isolation and fair slack sharing, for instance, by combining hierarchical budget enforcement for safety-critical vCPUs with low-overhead, weight-proportional slack reclamation for best-effort workloads, or by integrating lightweight runtime criticality-mode switches directly into open-source hypervisor kernels.

\section*{CRediT authorship contribution statement}
\noindent \textbf{Lisa Mignerey:} Conceptualization, Methodology, Investigation, Visualization, Software, Writing -- original draft, Writing -- review \& editing.\\
\noindent \textbf{Anthony Ferrari:} Methodology, Writing -- review \& editing, Supervision, Funding acquisition.\\
\noindent \textbf{Damien Sauveron:} Methodology, Writing -- review \& editing, Supervision, Funding acquisition.\\
\noindent \textbf{Karim Tamine:} Supervision, Funding acquisition.

\section*{Declaration of competing interest}
The authors declare the following financial interests/personal relationships which may be considered as potential competing interests: Lisa Mignerey and Anthony Ferrari report a relationship with Thales Cybersecurity \& Digital Identity that includes: employment. Damien Sauveron and Karim Tamine report that their laboratory (XLIM, University of Limoges) receives funding from Thales Cybersecurity \& Digital Identity under the CIFRE collaboration agreement supporting this work.

\section*{Funding}
This work was supported by Thales Cybersecurity \& Digital Identity and the French Ministry of Higher Education and Research (ANRT) [CIFRE PhD grant number 2024/1560] under a collaboration agreement between Thales Cybersecurity \& Digital Identity and XLIM. The French Ministry of Higher Education and Research had no involvement in the study design, in the analysis and interpretation, in the writing of the report, or in the decision to submit the article for publication. Thales Cybersecurity \& Digital Identity was involved in the study design, supervision, and manuscript review through the affiliated authors.

\section*{Data availability}
The analysed source code is publicly available in the repositories cited in the article, at the commits given in Table~\ref{tab:2}. The executable Python models used to derive the schedules and generate Figures~3--11 and Table~3 are publicly available on Zenodo at \url{https://doi.org/10.5281/zenodo.23687141}.

\section*{Declaration of generative AI and AI-assisted technologies in the manuscript preparation process}
During the preparation of this work, the authors used Claude (Anthropic) and Gemini (Google) to improve the language and readability of the manuscript and to cross-check the code-level descriptions and the hand-derived schedules against the source code. After using these tools, the authors reviewed and verified the content as needed and take full responsibility for the content of the published article.

\section*{Acknowledgments}
The authors would like to thank Pierre-François Bonnefoi for a number of very useful discussions.

\bibliographystyle{elsarticle-num-names}
\bibliography{refs}

\end{document}